\documentclass[twocolumn]{autart}    % Enable this line and disable the
\usepackage{graphicx}          % Include this line if your
\usepackage{subfigure}
\usepackage{amssymb}
\usepackage{amsmath}
\usepackage{mathrsfs}
\usepackage{hyperref}
\usepackage{colortbl,dcolumn}
\usepackage{booktabs}
\usepackage{xcolor}
\usepackage{cite}
\usepackage{threeparttable}

\usepackage{algorithm}
\usepackage[noend]{algpseudocode}

\usepackage{cases}

\usepackage{stmaryrd}

\newtheorem{assumption}{Assumption}
\newtheorem{definition}{Definition}
\newtheorem{lemma}{Lemma}
\newtheorem{theorem}{Theorem}
\newtheorem{proposition}{Proposition}
\newtheorem{corollary}{Corollary}
\newtheorem{remark}{Remark}

\newtheorem{problem}{Problem}

\newcommand{\aspref}[1]{Assumption~\ref{#1}}
\newcommand{\defref}[1]{Definition~\ref{#1}}
\newcommand{\lemref}[1]{Lemma~\ref{#1}}
\newcommand{\thmref}[1]{Theorem~\ref{#1}}
\newcommand{\propref}[1]{Proposition~\ref{#1}}
\newcommand{\corref}[1]{Corollary~\ref{#1}}
\newcommand{\figref}[1]{Fig.~\ref{#1}}
\newcommand{\tabref}[1]{Table~\ref{#1}}
\newcommand{\secref}[1]{Section~\ref{#1}}
\newcommand{\apxref}[1]{Appendix~\ref{#1}}
\newcommand{\probref}[1]{Problem~\ref{#1}}

\newcommand{\rekref}[1]{Remark~\ref{#1}}

\newenvironment{proof}{{\noindent\it Proof:}\quad}{\hfill $\square$\par}

\usepackage{mathtools} % \DeclareDelimiter etc

\usepackage{enumerate}
\usepackage{cuted}%\stripsep-3pt

\graphicspath{{pdfs/},{images/},{authorphotos/}}

\begin{document}

\begin{frontmatter}

\title{Stability of Linear Set-Membership Filters With Respect to Initial Conditions: An Observation-Information Perspective\thanksref{footnoteinfo}} % Title, preferably not more

\thanks[footnoteinfo]{
%This paper was not presented at any IFAC meeting.
Corresponding author: X.~Wang.
%Tel. +86-13975138479.
}

\author[NUDT]{Yirui~Cong}\ead{congyirui11@nudt.edu.cn},    % Add the
\author[NUDT]{Xiangke~Wang}\ead{xkwang@nudt.edu.cn},               % e-mail address
\author[ANU]{Xiangyun~Zhou}\ead{xiangyun.zhou@anu.edu.au}  % (ead) as shown

\address[NUDT]{College of Intelligence Science and Technology, National University of Defense Technology}  % Please supply
%\address[NUDT]{College of Intelligence Science and Technology, National University of Defense Technology}             % full addresses
\address[ANU]{School of Engineering, The Australian National University}        % here.

                             % keyword list or with the
                                          % help of the Automatica
                                          % keyword wizard

\begin{abstract}                          % Abstract of not more than 200 words.
The issue of filter stability with respect to (w.r.t.) the initial condition refers to the unreliable filtering process caused by improper prior information of the initial state. This paper focuses on analyzing and resolving the stability issue w.r.t. the initial condition of the classical Set-Membership Filters (SMFs) for linear time-invariance systems, which has not yet been well understood in the literature. To this end, we propose a new concept -- the Observation-Information Tower (OIT), which describes how the measurements affect the estimate in a set-intersection manner without relying on the initial condition. The proposed OIT enables a rigorous stability analysis, a new SMFing framework, as well as an efficient filtering algorithm. Specifically, based on the OIT, explicit necessary and sufficient conditions for stability w.r.t. the initial condition are provided for the classical SMFing framework. Furthermore, the OIT inspires a stability-guaranteed SMFing framework, which fully handles the stability issue w.r.t. the initial condition. Finally, with the OIT-inspired framework, we develop a fast and stable constrained zonotopic SMF, which significantly overcomes the wrapping effect.
%Stability with respect to (w.r.t.) the initial condition is fundamental to filter design. It reflects the sensitivity of the estimate when the initial condition does not match the true initial range of system states. In this paper, we focus on the stability of linear set-membership filters (SMFs), which has not yet been well understood. Considering linear time-invariant systems, we analyze the stability issue of the existing/classical SMFing framework and establish a new framework with guaranteed stability. First, we propose a new concept -- the Observation-Information Tower (OIT), which describes how the measurements affect the estimate in a set-intersection manner. The OIT enables a rigorous stability analysis of the classical linear SMFing framework, where explicit necessary and sufficient conditions for stability are provided. It turns out that the classical framework requires the knowledge of the true initial range of the system state to guarantee stability, which is hence sensitive to initial conditions. To handle this problem, we establish an OIT-inspired filtering framework such that the stability is unaffected by the initial conditions. Under this stability-guaranteed framework, we develop a stable and fast constrained zonotopic SMF.
\end{abstract}

\end{frontmatter}

\section{Introduction}\label{sec:Introduction}

%\subsection{Motivation and Related Work}\label{sec:Motivation and Related Work}

Set-Membership Filter (SMF) is a very important class of non-stochastic filters, which gives the optimal solution to the filtering problems with bounded noises whose probability distributions are unknown.
It is fundamentally equivalent to the Bayes filter~\cite{sarkkaS2013BOOK} under the non-stochastic Markov condition (caused by unrelated noises)~\cite{CongY2022TAC}.
For linear systems, it can be regarded as a non-stochastic counterpart of the well-known Kalman filter.
Therefore, the SMF has great potentials in many important fields like control systems, telecommunications, and navigation, as the Bayes filter does.
%%
%Under the non-stochastic Markov condition, the classical SMFing framework is optimal in the sense of deriving the smallest set/estimate that contains all possible states.\footnote{If the condition is violated, the classical SMFing framework simply provides an outer bound, and the conservativeness is determined by the relatedness of the noises~\cite{CongY2022TAC}.}
%%
%Thus, for linear systems, it can be regarded as a non-stochastic counterpart of the well-known Kalman filtering, and should have broad applications as well.
%%
However, the linear SMFing has not drawn an adequate amount of attention to realize its full potential.
One important issue is \emph{stability w.r.t. the initial condition}\footnote{Similarly to the true initial distribution/measure and the initial condition for stochastic filters~\cite{OconeD1996,vanHandel2006PhdThesis}, in this article: the term true initial set refers to the set of all possible initial states (objectively determined by the system), which could be a singleton when we focus on one trial;
the term initial condition (for SMFs) represents the initial set subjectively chosen in the design of a filter.}, which characterizes whether the resulting estimate, as time elapses, remains reliable when the filter is improperly initialized~\cite{OconeD1996,vanHandel2006PhdThesis}.

For Kalman filters, stability w.r.t. the initial condition is a central feature and well-established in an asymptotic manner: the posterior distribution, under some simple system assumptions, converges to the true conditional distribution of the system state (given all observed measurements) regardless of initial conditions~\cite{JazwinskiAH1970BOOK,KalmanR1961,LewisF2007BOOK}.
This means measurements in stable filters can help correct/forget the information from the initial condition.

For linear SMFs, in contrast, there is very limited knowledge of the stability w.r.t. the initial condition.
The most related work is on the uniform boundedness of the estimate, which guarantees that the size of estimate does not increase unboundedly with time.
In the literature, the uniform boundedness analysis is considered for two types of SMFs, i.e., the ellipsoidal and polytopic SMFs.\footnote{The uniform boundedness of interval observers~\cite{MazencF2011,XuF2020} is not included since their basic structures are linear observers.}
For ellipsoidal SMFs,~\cite{BecisAubryY2008} proposed an input-to-state stable algorithm that ensures the uniform boundedness of the estimate;
since in~\cite{BecisAubryY2008} the input-to-state stability required to solve a time-inefficient polynomial equation at each time step,~\cite{LiuY2016} developed an SMF with increased efficiency based on minimizing an important upper bound;
then,~\cite{ShenQ2018} provided a parallelotope-bounding technique to reduce the complexity, which was further improved by~\cite{BecisAubry2021}.
For polytopic SMFs,~\cite{CombastelC2015} proposed a zonotopic Kalman filter, where a sufficient condition (called robust detectability) for the uniform boundedness of the estimate was given;
in~\cite{WangY2019}, a zonotopic SMF was designed for linear parameter-varying systems, and an upper bound on the radius of the estimate was derived by solving linear matrix inequalities (LMIs);
based on~\cite{WangY2019}, the article~\cite{FeiZ2022} proposed a zonotopic SMF for switched linear systems, where an LMI-based upper bound was obtained for the radius of the estimate.

It is important to note that the uniform boundedness of the estimate does not imply reliable estimates when the initial condition is improperly chosen.
Therefore, existing linear SMFs still face the issue of \emph{stability w.r.t. the initial condition}, reflected in two aspects\footnote{Note that stability of SMFs w.r.t. their initial conditions is different from stability of control systems with estimation~\cite{ChenC1999,ChenJ2000,MilaneseM1996BOOK,MilaneseM1989BOOK}; for an unstable system, an stable SMF w.r.t. the initial condition can still be designed.}:
\begin{itemize}
\item   \textbf{Ill-posedenss:}
    In the literature, the initial condition of an SMF should include the true initial set of the system state.
    But if the initial condition does not contain the true initial set, the resulting estimate at some time steps can be an empty set;
    in this case, we say the SMF is ill-posed\footnote{In the SMF community, the ill-posedness is a case of the falsification of a priori assumption. To the best of our knowledge, this issue has not been studied in the literature.} or not well-posed, which is a key different feature from the stability of Kalman filters.
    Unfortunately, ill-posedness widely exists, which means the initial condition must be carefully chosen to guarantee non-empty estimates.
\item   \textbf{Unbounded estimation gap:}
    Even if the estimate is non-empty, perturbations of the initial condition can gradually amplify the difference (called the estimation gap) between the estimate and the true set of the system states.
    Consequently, the estimation gap can go unbounded with time:
    inherently, the classical SMFing framework cannot always ensure the bounded estimation gap (the boundedness condition is unknown);
    extrinsically, approximation/reduction methods result in the wrapping effect~\cite{KunhW1998} so that the estimation gap is unboundedly accumulated.
\end{itemize}

To directly tackle the stability issue, we should analyze when the classical linear SMFing framework is stable w.r.t. the initial condition, i.e., well-posed and with a bounded estimation gap;
then, it is necessary to propose a new SMFing framework with guaranteed stability.

In addition to the stability, a linear SMF should be \emph{fast and accurate}, which is also a challenging issue arising from the computational intractability of the optimal estimate.
In general, to improve the time efficiency of an existing SMF, one has to sacrifice the estimation accuracy, and vice versa.

A recent study in~\cite{AlthoffM2021} showed that the constrained zonotopic SMF proposed by~\cite{ScottJ2016} can give tighter estimates than other types of modern SMFs (e.g., ellipsoidal SMFs~\cite{LiuY2016,LoukkasN2017} and zonotopic SMFs~\cite{CombastelC2015,WangY2018Automatica}), which sheds new light on balancing the efficiency and the accuracy.
This is largely because constrained zonotopes are closed under Minkowski sums and set intersections in the prediction and update steps, respectively.
Nevertheless, the constrained zonotopic SMF still has a relatively large computation time due to the limitation of the existing approximation techniques.
Specifically, the reduction relies on the geometric properties of the constrained zonotopes at each single time step (i.e., without considering the ``time correlation'').
As a result, overestimation cumulates over time,
which means maintaining accuracy would lead to an unnecessary increase of complexity.
The situation would be even worse for high-dimensional systems.

Taking the above discussions into account, it is of great theoretical and practical interest to develop a fast constrained zonotopic SMF with guaranteed stability, which is also efficient for high-dimensional systems.

In this work, we focus on understanding and analyzing the stability w.r.t. the initial condition of SMFing for linear time-invariant systems and establish a stability-guaranteed filtering framework to develop a new SMF.
The main contribution is to put forward a concept of Observation-Information Tower (OIT).
It reflects the impact of the measurements on the estimate, independent of any reliance on the initial condition.
The OIT enables to provide a rigorous stability analysis, resolve the ill-posedness issue, and develop an efficient filtering algorithm.
More specifically:
\begin{itemize}
	\item	Applying the OIT to a projected system, we analyze the stability of the classical linear SMFing framework.
	An explicit stability criterion (sufficient condition) is given, which turns out to be surprisingly close to the necessity w.r.t. the bounded estimation gap.
	
	\item	The OIT inspires a new SMFing framework, through a rigorously proven invariance property.
	This framework completely fixes the ill-posedness problem of the existing linear SMFing, without relying on any information of the true initial set.
	
	\item	 With this new framework, we design a stable and fast constrained zonotopic SMF with uniform boundedness;
	different from the existing reduction methods based on geometric properties of constrained zonotopes, our method utilizes the properties of system dynamics via the OIT, which has high efficiency and significantly overcomes the wrapping effect.
\end{itemize}

In \secref{sec:System Model and Problem Statement}, the system model is given and three problems (on \emph{stability of classical linear SMFing framework}, \emph{new framework with guaranteed stability}, and \emph{stable and fast constrained zonotopic SMF}) are described.
We propose the OIT in \secref{sec:Observation-Information Tower} which is the key to solving those three problems, and the solutions are provided in Sections~\ref{sec:Stability Analysis of Optimal Linear Set-Membership Filtering Framework}--\ref{sec:A Stable and Fast Constrained Zonotopic SMF}, respectively.
\secref{sec:Numerical Examples} gives the simulation examples to corroborate our theoretical results.
Finally, the concluding remarks are given in \secref{sec:Conclusion}.

\textbf{Notation:} Throughout this paper, $\mathbb{R}$, $\mathbb{N}_0$, and $\mathbb{Z}_+$ denote the sets of real numbers, non-negative integers, and positive integers, respectively.
$\mathbb{R}^n$ stands for the $n$-dimensional Euclidean space.
For an uncertain variable $\mathbf{x}$ defined on a sample space $\Omega$, its range is $\llbracket\mathbf{x}\rrbracket = \{\mathbf{x}(\omega)\colon \omega \in \Omega\}$ and its realization is $x = \mathbf{x}(\omega)$~\cite{NairG2013,CongY2022TAC}.
For multiple uncertain variables with consecutive indices, we define $\mathbf{x}_{k_1:k_2} = \mathbf{x}_{k_1},\ldots,\mathbf{x}_{k_2}$ (with their realizations $x_{k_1:k_2} = x_{k_1},\ldots,x_{k_2}$) where $k_2 \geq k_1$.
Given two sets $\mathcal{S}_1$ and $\mathcal{S}_2$ in a Euclidean space, the operation $\oplus$ stands for the Minkowski sum of $\mathcal{S}_1$ and $\mathcal{S}_2$, i.e., $\mathcal{S}_1 \oplus \mathcal{S}_2 = \{s_1 + s_2\colon s_1 \in \mathcal{S}_1, s_2 \in \mathcal{S}_2\}$.
The summation $\sum_{i = a}^b \mathcal{S}_i$ represents $\mathcal{S}_a \oplus \mathcal{S}_{a+1} \oplus \cdots \oplus \mathcal{S}_b$ for $a \leq b$, but returns $\{0\}$ for $a > b$.\footnote{This is different from the mathematical convention, but can keep many expressions compact in this paper.}
We use $\|\cdot\|$ to represent the Euclidean norm (of a vector) or the spectral norm (of a matrix).
The set $\mathcal{S}_k$ is \emph{uniformly bounded}\footnote{A mathematically rigorous statement should be ``the indexed family $(\mathcal{S}_k)_{k \in \mathcal{K}}$ is uniformly bounded''.
However, this minor abuse of notation considerably simplifies the presentation.}
(w.r.t. $k \in \mathcal{K}$) if there exists $\bar{d} > 0$ such that $d(\mathcal{S}_k) \leq \bar{d}$ for all $k \in \mathcal{K}$, where $d(\mathcal{S}_k) = \sup_{s, s' \in \mathcal{S}_k} \|s - s'\|$ is the diameter of $\mathcal{S}_k$.
The interval hull of a bounded set $\mathcal{S} \ni s = (s^{(1)},\ldots,s^{(n)})$ is $\overline{\mathrm{IH}}(\mathcal{S}) = \prod_{i=1}^{n}[\underline{s}^{(i)}, \overline{s}^{(i)}]$, where $\underline{s}^{(i)} = \inf_{s^{(i)}} \{s \in \mathcal{S}\}$ and $\overline{s}^{(i)} = \sup_{s^{(i)}} \{s \in \mathcal{S}\}$.
The limit superior is denoted by $\varlimsup$.
Given a matrix $M \in \mathbb{R}^{m\times n}$, the Moore-Penrose inverse is $M^+$; the range space and the kernel (null space) are denoted by $\mathrm{ran}(M)$ and $\ker(M)$, respectively.
The notation $\circ$ stands for the composition of two maps.

\section{System Model and Problem Description}\label{sec:System Model and Problem Statement}

\subsection{Linear SMF with Inaccurate Initial Condition}\label{sec:System Model}

Consider the following discrete-time linear system described by uncertain variables:
\begin{align}
\mathbf{x}_{k+1} &= A \mathbf{x}_k + B \mathbf{w}_k,\label{eqn:State Equation}\\
\mathbf{y}_k &= C \mathbf{x}_k + \mathbf{v}_k,\label{eqn:Measurement Equation}
\end{align}
at time $k \in \mathbb{N}_0$, where~\eqref{eqn:State Equation} and~\eqref{eqn:Measurement Equation} are called the state and measurement equations, respectively, where $A \in {\mathbb R}^{n\times n}$, $B \in {\mathbb R}^{n\times p}$, and $C \in \mathbb{R}^{m\times n}$.
The state equation describes how the system state $\mathbf{x}_k$ (with its realization $x_k \in \llbracket\mathbf{x}_k\rrbracket \subseteq \mathbb{R}^n$) changes over time, where the \emph{true initial set} $\llbracket\mathbf{x}_0\rrbracket$ is non-empty and bounded;
$\mathbf{w}_k$ is the process/dynamical noise (with its realization $w_k \in \llbracket\mathbf{w}_k\rrbracket \subseteq \mathbb{R}^p$), and there exists a constant $d_w > 0$ such that $d(\llbracket\mathbf{w}_k\rrbracket) \leq d_w$ for all $k \in \mathbb{N}_0$.
The measurement equation gives how the system state is measured from the measurement $\mathbf{y}_k$ (with its realization, called observed measurement, $y_k \in \llbracket\mathbf{y}_k\rrbracket \subseteq \mathbb{R}^m$);
$\mathbf{v}_k$ (with its realization $v_k \in \llbracket\mathbf{v}_k\rrbracket \subseteq \mathbb{R}^m$) stands for the measurement noise, and there exists a constant $d_v > 0$ such that $d(\llbracket\mathbf{v}_k\rrbracket) \leq d_v$ for all $k \in \mathbb{N}_0$.
The process noises, the measurement noises, and the initial state are unrelated~\cite{NairG2013}, given in \aspref{asp:Unrelated Noises and Initial State}.
This makes the system satisfy a non-stochastic hidden Markov model, which guarantees the optimality of the classical SMFing framework~\cite{CongY2022TAC}.

\begin{assumption}\textbf{\emph{(Unrelated Noises and Initial State)}}\label{asp:Unrelated Noises and Initial State}
$\forall k \in \mathbb{N}_0$, $\mathbf{w}_{0:k}, \mathbf{v}_{0:k},\mathbf{x}_0$ are unrelated.
\end{assumption}

Unless otherwise stated, the results and discussions in the rest of this paper are under \aspref{asp:Unrelated Noises and Initial State}.

The classical linear SMFing framework~\cite{CongY2022TAC} is given in Algorithm~\ref{alg:Classical Linear Set-Membership Filtering}, where an initial condition $\llbracket\hat{\mathbf{x}}_0\rrbracket$ is required.
The existing linear SMFs are based on this framework.
We highlight the prediction and update steps in~\eqref{eqn:Prediction - Optimal Linear Set-Membership Filter} and~\eqref{eqn:Update - Optimal Linear Set-Membership Filter}, respectively:
\begin{align}
\llbracket\hat{\mathbf{x}}_k| y_{0:k-1}\rrbracket &= A \llbracket\hat{\mathbf{x}}_{k-1}|y_{0:k-1}\rrbracket \oplus B \llbracket\mathbf{w}_{k-1}\rrbracket, \label{eqn:Prediction - Optimal Linear Set-Membership Filter}\\
\llbracket\hat{\mathbf{x}}_k| y_{0:k}\rrbracket &= \mathcal{X}_k(C, y_k, \llbracket\mathbf{v}_k\rrbracket) \bigcap \llbracket\hat{\mathbf{x}}_k| y_{0:k-1}\rrbracket,\label{eqn:Update - Optimal Linear Set-Membership Filter}
\end{align}
where we define $\llbracket\hat{\mathbf{x}}_0\rrbracket =: \llbracket\hat{\mathbf{x}}_0| y_{0:-1}\rrbracket$ for consistency, and
\begin{equation}\label{eqn:Specific Form of the Solution Set X}
\begin{split}
\!\!\mathcal{X}_k(C, y_k, \llbracket\mathbf{v}_k\rrbracket) &= \left\{x_k\colon y_k = C x_k + v_k,~v_k \in \llbracket\mathbf{v}_k\rrbracket\right\}\\
&= \ker(C) \oplus C^+ (\left\{y_k\right\} \oplus \llbracket-\mathbf{v}_k\rrbracket).
\end{split}
\end{equation}

\begin{algorithm}
\begin{footnotesize}
\caption{Classical Linear SMFing Framework}\label{alg:Classical Linear Set-Membership Filtering}
\begin{algorithmic}[1]
\State  \textbf{Initialization:} $\llbracket\hat{\mathbf{x}}_0\rrbracket \subset \mathbb{R}^n$;\label{line:Optimal Linear Set-Membership Filtering - Initialization}
\If {$k > 0$}\label{line:OIT-Inspired Filtering - Prediction - Start}
    \State  \textbf{Prediction:}~$\llbracket\hat{\mathbf{x}}_k| y_{0:k-1}\rrbracket \leftarrow$~\eqref{eqn:Prediction - Optimal Linear Set-Membership Filter}; \% Returns the prior set.\label{line:Optimal Linear Set-Membership Filtering - Prediction}
\EndIf\label{line:Optimal Linear Set-Membership Filtering - Prediction -End}
\State  \textbf{Update:}~$\llbracket\hat{\mathbf{x}}_k| y_{0:k}\rrbracket \leftarrow$~\eqref{eqn:Update - Optimal Linear Set-Membership Filter}; \% Returns the estimate/the posterior set.\label{line:Optimal Linear Set-Membership Filtering - Update}
\end{algorithmic}
\end{footnotesize}
\end{algorithm}

If the true initial set $\llbracket\mathbf{x}_0\rrbracket$ can be perfectly known, i.e., $\llbracket\hat{\mathbf{x}}_0\rrbracket = \llbracket\mathbf{x}_0\rrbracket$,
Algorithm~\ref{alg:Classical Linear Set-Membership Filtering} returns the true set $\llbracket\mathbf{x}_k| y_{0:k}\rrbracket$ as the estimate $\llbracket\hat{\mathbf{x}}_k| y_{0:k}\rrbracket$.
In this case, Algorithm~\ref{alg:Classical Linear Set-Membership Filtering} is optimal\footnote{The optimality is in the sense that no filters can give a smaller set containing all possible states~\cite{CongY2022TAC}.}.
Under this classical SMFing framework, existing SMFs employ different set-based descriptions to represent or outer bound $\llbracket\hat{\mathbf{x}}_k| y_{0:k}\rrbracket$, as pointed out in~\cite{CongY2022TAC};
that means the estimate $\mathcal{Z}_k$ of a linear SMF, at $k \in \mathbb{N}_0$, satisfies $\mathcal{Z}_k \supseteq \llbracket\hat{\mathbf{x}}_k| y_{0:k}\rrbracket$.

However, the initial condition is usually not accurate in practice, i.e., $\llbracket\hat{\mathbf{x}}_0\rrbracket \neq \llbracket\mathbf{x}_0\rrbracket$, which can cause stability issue as stated in \secref{sec:Introduction}.
To rigorously derive the theoretical results in the sequel of this paper, we introduce the filtering map as follows.

\begin{definition}\textbf{\emph{(Filtering Map)}}\label{def:Filtering Map}
At time $k \in \mathbb{Z}_+$, the prediction-update map under $\llbracket\mathbf{w}_{k-1}\rrbracket$ and $\llbracket\mathbf{v}_k\rrbracket$ is
\begin{equation}\label{eqn:Prediction-Update Map}
f_k: \mathcal{S} \mapsto \mathcal{X}_k(C, y_k, \llbracket\mathbf{v}_k\rrbracket) \bigcap (A \mathcal{S} \oplus B \llbracket\mathbf{w}_{k-1}\rrbracket),
\end{equation}
where $\mathcal{S} \subseteq \mathbb{R}^n$.
The filtering map from time $i \in \mathbb{Z}_+$ to $k \geq i$ is $F_{k,i}$ with the following form (where $0 \leq i \leq k$)
\begin{equation}\label{eqn:Filtering Map}
\!\!\!\!\!F_{k,i}(\mathcal{S}) \!=\!
\begin{cases}
\!f_k \!\circ \cdots \circ \! f_{i+1} (\mathcal{X}_i(C, y_i, \llbracket\mathbf{v}_i\rrbracket) \bigcap \mathcal{S}), &\!\!\! k > i,\\
\!\mathcal{X}_i(C, y_i, \llbracket\mathbf{v}_i\rrbracket) \bigcap \mathcal{S}, &\!\!\! k = i.
\end{cases}
\end{equation}
\end{definition}

By \defref{def:Filtering Map}, Algorithm~\ref{alg:Classical Linear Set-Membership Filtering} can be described by a compact form by the filtering map:
\begin{equation}\label{eqn:The Filtering Map Description of Optimal SMFing}
\llbracket\hat{\mathbf{x}}_k| y_{0:k}\rrbracket = F_{k,0}(\llbracket\hat{\mathbf{x}}_0\rrbracket),\quad k \in \mathbb{N}_0.
\end{equation}

\subsection{Problem Description: Analysis and Synthesis of Stability w.r.t. Initial Condition}\label{sec:Problem Description}

In this work, we focus on understanding and handling the stability issue of linear SMFs, which includes three parts:
(i) analyzing the stability of the classical linear SMFing framework (see \probref{prob:Stability of Optimal Linear SMF}),
(ii) establishing a new linear SMFing framework with guaranteed stability (see \probref{prob:Stability-Guaranteed Framework}),
(iii) designing an efficient linear SMF under the proposed framework (see \probref{prob:Stability-Guaranteed Zonotopic SMF}).

To analyze the stability of existing linear SMFs, we define the stability in \defref{def:Stability of SMF}, based on the well-posedness and the bounded estimation gap.

\begin{definition}\textbf{\emph{(Well-Posedness)}}\label{def:Well-Posedness}
An SMF with the initial condition $\llbracket\hat{\mathbf{x}}_0\rrbracket$ is well-posed, if $\mathcal{Z}_k \neq \emptyset$ for all $k \in \mathbb{N}_0$.
\end{definition}

\begin{definition}\textbf{\emph{(Bounded Estimation Gap)}}\label{def:Estimation Gap}
At $k \in \mathbb{N}_0$, the estimation gap is the Hausdorff distance between the estimate $\mathcal{Z}_k$ and the true set $\llbracket\mathbf{x}_k|y_{0:k}\rrbracket$
%(which measures how far two sets are from each other)
%
\begin{equation}\label{eqn:Estimation Gap}
d_k^{\mathrm{g}}(\mathcal{Z}_k) := d_{\mathrm{H}}(\mathcal{Z}_k, \llbracket\mathbf{x}_k|y_{0:k}\rrbracket),
\end{equation}
where
\begin{equation}\label{eqn:Hausdorff Distance}
\!\!\!d_{\mathrm{H}}(\mathcal{S},\mathcal{T}) = \max\Big\{\adjustlimits\sup_{s \in \mathcal{S}} \inf_{t \in \mathcal{T}}\|s - t\|, \adjustlimits\sup_{t \in \mathcal{T}} \inf_{s \in \mathcal{S}}\|s - t\|\Big\}.
\end{equation}
The estimation gap is bounded, if there exists a $\bar{d} > 0$ such that $d_k^{\mathrm{g}}(\mathcal{Z}_k) \leq \bar{d}$ for all $k \in \mathbb{N}_0$.
\end{definition}

\begin{definition}\textbf{\emph{(Stability of SMF)}}\label{def:Stability of SMF}
An SMF is stable w.r.t. its initial condition, if for all bounded $\llbracket\hat{\mathbf{x}}_0\rrbracket \subset \mathbb{R}^n$, the SMF is well-posed and the estimation gap is bounded.
\end{definition}

\begin{remark}
The stability of an SMF reflects the insensitivity of $\mathcal{Z}_k$ to the initial conditions.
Note that the well-posedness is a necessary condition for the stability, which is different from that of Kalman filters.
\end{remark}

With \defref{def:Stability of SMF}, we are ready to study the stability of the classical linear SMFing framework (that the existing linear SMFs are based on).
For the framework described in Algorithm~\ref{alg:Classical Linear Set-Membership Filtering}:
the initial condition $\llbracket\hat{\mathbf{x}}_0\rrbracket$ must be carefully chosen in case of ill-posedness;\footnote{For example, consider the linear system with parameters $A = 1$, $B = 1$, $C = 1$, $\llbracket\mathbf{w}_k\rrbracket = [-1, 1]$, and $\llbracket\mathbf{v}_k\rrbracket = [0, 1]$.
If $\llbracket\mathbf{x}_0\rrbracket = [-1, 1]$ and $\llbracket\hat{\mathbf{x}}_0\rrbracket = [0, 2]$, with~\eqref{eqn:Update - Optimal Linear Set-Membership Filter} we have $\llbracket\mathbf{x}_0|y_0\rrbracket = \mathcal{X}_0(C, y_0, \llbracket\mathbf{v}_0\rrbracket) \bigcap \llbracket\mathbf{x}_0\rrbracket = [-2, -1] \bigcap [-1, 1] = \{-1\}$ while $\llbracket\hat{\mathbf{x}}_0|y_0\rrbracket = \mathcal{X}_0(C, y_0, \llbracket\mathbf{v}_0\rrbracket) \bigcap \llbracket\hat{\mathbf{x}}_0\rrbracket = [-2, -1] \bigcap [0, 2] = \emptyset$.
From~\eqref{eqn:Prediction - Optimal Linear Set-Membership Filter}, we know that $\llbracket\hat{\mathbf{x}}_1|y_0\rrbracket = \emptyset$ and thus~\eqref{eqn:Update - Optimal Linear Set-Membership Filter} gives $\llbracket\hat{\mathbf{x}}_1|y_{0:1}\rrbracket = \emptyset$.
Proceeding forward, we have $\llbracket\hat{\mathbf{x}}_k|y_{0:k}\rrbracket = \emptyset$ for $k \geq 0$.}
furthermore, the estimation gap $d_k^{\mathrm{g}}(\llbracket\hat{\mathbf{x}}_k| y_{0:k}\rrbracket)$ can go unbounded as $k \to \infty$.
Thus, it is necessary to analyze the stability condition of Algorithm~\ref{alg:Classical Linear Set-Membership Filtering}; see \probref{prob:Stability of Optimal Linear SMF}.

\begin{problem}\textbf{\emph{(Stability of Classical Linear SMFing Framework)}}\label{prob:Stability of Optimal Linear SMF}
Under what condition is the framework in Algorithm~\ref{alg:Classical Linear Set-Membership Filtering} stable w.r.t. the initial condition?
\end{problem}

Since the well-posedness is sensitive to initial conditions, the classical framework faces the ill-posedness issue.
This motivates us to propose a new SMFing framework always with guaranteed stability (regardless of perturbations w.r.t. the initial condition);
see \probref{prob:Stability-Guaranteed Framework}.

\begin{problem}[Stability-Guaranteed Framework]\label{prob:Stability-Guaranteed Framework}
How to establish a new SMFing framework such that for any bounded $\llbracket\hat{\mathbf{x}}_0\rrbracket \subseteq \mathbb{R}^n$, the stability is guaranteed?
\end{problem}

Under the new framework, one should develop a filtering algorithm with low complexity and high accuracy.
Inspired by prior work (e.g.,~\cite{ScottJ2016}), we consider a constrained zonotopic SMF for the filter design as follows.

\begin{problem}\textbf{\emph{(Stable and Fast Constrained Zonotopic SMF)}}\label{prob:Stability-Guaranteed Zonotopic SMF}
How to design a constrained zonotopic SMF satisfying the stability-guaranteed framework while the complexity and accuracy can be well handled?
\end{problem}

To solve these three problems, we propose the concept of Observation-Information Tower (OIT) in \secref{sec:Observation-Information Tower}.
It plays a pivotal role in understanding the stability of linear SMFing and inspiring stability-guaranteed designs.
Then, the solutions to Problems~\ref{prob:Stability of Optimal Linear SMF}-\ref{prob:Stability-Guaranteed Zonotopic SMF} are studied in Sections~\ref{sec:Stability Analysis of Optimal Linear Set-Membership Filtering Framework}--\ref{sec:A Stable and Fast Constrained Zonotopic SMF}, respectively.

\section{The Observation-Information Tower}\label{sec:Observation-Information Tower}

Before defining the OIT, let us first analyze how a single observed measurement affects the estimate as time elapses.\footnote{This is quite similar to analyzing the impulse response of a control system.}
At $k = 0$, from the update step~\eqref{eqn:Update - Optimal Linear Set-Membership Filter} in Algorithm~\ref{alg:Classical Linear Set-Membership Filtering} we know that the estimate $\llbracket\hat{\mathbf{x}}_0| y_0\rrbracket$ is the intersection of $\mathcal{X}_0(C, y_0, \llbracket\mathbf{v}_0\rrbracket) =: \mathcal{O}_{0,0}$ and $\llbracket\hat{\mathbf{x}}_0\rrbracket =: \mathcal{E}_0$, which describes how the observed measurement $y_0$ affects the estimate at $k = 0$.
When we ignore all the successive observed measurements (like the impulse response does), only the prediction step in Algorithm~\ref{alg:Classical Linear Set-Membership Filtering} works for $k \geq 1$.
Based on ``set intersection under Minkowski sum'': for sets $\mathcal{S}_1,\ldots,\mathcal{S}_I$, and $\mathcal{T}$, we have\footnote{The proof is as follows.
$\forall l \in \big(\bigcap_{i=1}^{I} \mathcal{S}_i\big) \oplus \mathcal{T}$, $\exists s \in \bigcap_{i=1}^{I} \mathcal{S}_i$ and $t \in \mathcal{T}$ such that $l = s + t$.
Since $s \in \bigcap_{i=1}^{I} \mathcal{S}_i$, we have $s \in \mathcal{S}_i$ for all $i \in \{1,\ldots,I\}$.
Thus, $\forall i \in \{1,\ldots,I\}$, $l = s + t \in \mathcal{S}_i \oplus \mathcal{T}$, which implies $l \in \bigcap_{i=1}^{I} (\mathcal{S}_i \oplus \mathcal{T})$, and we get~\eqref{eqn:Set Intersection Under Minkowski Sum}.
}
\begin{equation}\label{eqn:Set Intersection Under Minkowski Sum}
	\bigg(\bigcap_{i=1}^{I} \mathcal{S}_i\bigg) \oplus \mathcal{T} \subseteq \bigcap_{i=1}^{I} (\mathcal{S}_i \oplus \mathcal{T}),
\end{equation}
and hence the estimate at $k = 1$ is outer bounded by the intersection of the following two sets:
\begin{align*}
A \mathcal{O}_{0,0} \oplus B \llbracket\mathbf{w}_0\rrbracket &= A \mathcal{X}_0(C, y_0, \llbracket\mathbf{v}_0\rrbracket) \oplus B \llbracket\mathbf{w}_0\rrbracket =: \mathcal{O}_{1,0},\\
A \mathcal{E}_0 \oplus B \llbracket\mathbf{w}_0\rrbracket &= A \llbracket\hat{\mathbf{x}}_0\rrbracket \oplus B \llbracket\mathbf{w}_0\rrbracket =: \mathcal{E}_1,
\end{align*}
i.e., $\llbracket\hat{\mathbf{x}}_1| y_0\rrbracket \subseteq \mathcal{O}_{1,0} \bigcap \mathcal{E}_1$,
which indicates how $y_0$ affects the estimate at $k = 1$.
As such, we can analyze the effect of $y_i$ on the estimate at time $k$.
It defines the observation-information and state-evolution sets as follows.

\begin{definition}\textbf{\emph{(Observation-Information and State-Evolution Sets)}}\label{def:Observation-Information and State-Evolution Sets}
The observation-information set at time $k \geq i$ contributed by $y_i$ (i.e, the observed measurement at time $i$) is
\begin{equation}\label{eqn:Observation-Information Set}
\mathcal{O}_{k,i} := A^{k-i} \mathcal{X}_i(C,y_i,\llbracket \mathbf{v}_i\rrbracket) \oplus \sum_{r=i}^{k-1} A^{k-1-r} B \llbracket\mathbf{w}_r\rrbracket.
\end{equation}
The state-evolution set at time $k$ is
\begin{equation}\label{eqn:State-Evolution Set}
\mathcal{E}_k := A^k \llbracket\hat{\mathbf{x}}_0\rrbracket \oplus \sum_{r=0}^{k-1} A^{k-1-r} B \llbracket\mathbf{w}_r\rrbracket.
\end{equation}
\end{definition}

\begin{remark}
%If there is only one observed measurement (say $y_i$), the posterior range at time $k$ is outer bounded by the intersection of $\mathcal{O}_{k,i}$ and $\mathcal{E}_k$.
%%
If we consider all the observed measurements $y_{0:k}$, the intersection of the observation-information sets and the state-evolution set forms an outer bound on the estimate (see \propref{prop:Set-Intersection-Based Outer Bound}), where the conservativeness comes from the converse of~\eqref{eqn:Set Intersection Under Minkowski Sum} is not true in general.
\end{remark}

\begin{proposition}\textbf{\emph{(Set-Intersection-Based Outer B-ound)}}\label{prop:Set-Intersection-Based Outer Bound}
At $k \in \mathbb{N}_0$, an outer bound on $\llbracket\hat{\mathbf{x}}_k|y_{0:k}\rrbracket$ in Algorithm~\ref{alg:Classical Linear Set-Membership Filtering} is
\begin{equation}\label{eqn:Set-Intersection-Based Outer Bound}
\!\!\!\!\!\!\llbracket\hat{\mathbf{x}}_k|y_{0:k}\rrbracket = F_{k,0}(\llbracket\hat{\mathbf{x}}_0\rrbracket) \subseteq \!\bigcap_{i=0}^{k} \mathcal{O}_{k,i} \bigcap \mathcal{E}_k,~\forall \llbracket\hat{\mathbf{x}}_0\rrbracket \subseteq \mathbb{R}^n,
\end{equation}
which means $\llbracket\hat{\mathbf{x}}_k|y_{0:k}\rrbracket \subseteq \bigcap_{i=0}^{k} \mathcal{O}_{k,i}$ and $\llbracket\hat{\mathbf{x}}_k|y_{0:k}\rrbracket \subseteq \mathcal{E}_k$.
\end{proposition}

\begin{proof}
See \apxref{apx:Proof of prop:Set-Intersection-Based Outer Bound}.
\end{proof}

\begin{figure}[ht]
\centering
\includegraphics [width=0.8\columnwidth]{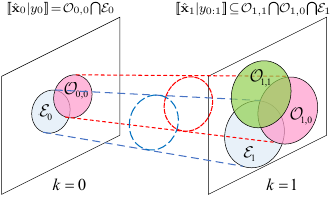}
\caption{Illustration of set-intersection-based outer bound.
For $k = 0$, the only observation-information set is $\mathcal{O}_{0,0}$;
the state-evolution set is $\mathcal{E}_0$;
the intersection of $\mathcal{O}_{0,0}$ and $\mathcal{E}_0$ gives the estimate $\llbracket\hat{\mathbf{x}}_0|y_0\rrbracket$.
For $k = 1$, $\mathcal{O}_{0,0}$ and $\mathcal{E}_0$ become $\mathcal{O}_{1,0}$ and $\mathcal{E}_1$, respectively;
the new observation-information set is $\mathcal{O}_{1,1}$;
the intersection of $\mathcal{O}_{1,1}$, $\mathcal{O}_{1,0}$, and $\mathcal{E}_1$ forms an outer bound on the estimate $\llbracket\hat{\mathbf{x}}_1|y_{0:1}\rrbracket$ as presented in \propref{prop:Set-Intersection-Based Outer Bound}.
}
\label{fig:Observation-information and state-evolution sets evolving with time.}
\end{figure}

A pictorial illustration of \propref{prop:Set-Intersection-Based Outer Bound} is given in \figref{fig:Observation-information and state-evolution sets evolving with time.}.

Note that
\begin{equation}\label{eqn:Introducing OIT}
\bigcap_{i=0}^{k} \mathcal{O}_{k,i} \subseteq \bigcap_{i=k-\delta}^{k} \mathcal{O}_{k,i}
\end{equation}
holds for any integer $\delta \in [0,k]$,\footnote{In this paper, when an integer is in an interval $[a, b]$, this interval denotes $\{i \in \mathbb{Z}\colon a \leq i \leq b\}$.}
we define the Right-Hand Side (RHS) of~\eqref{eqn:Introducing OIT} as the OIT (see \defref{def:Observation-Information Tower}).

\begin{definition}[Observation-Information Tower]\label{def:Observation-Information Tower}
The OIT at time $k \in \mathbb{N}_0$ is the intersection of the observation-information sets over $[k-\delta, k]$: $\bigcap_{i=k-\delta}^{k} \mathcal{O}_{k,i}$.
\end{definition}

Now, we provide a sufficient condition for the uniform boundedness (see \secref{sec:Introduction}) of the OIT as follows, which is fundamental to the results derived in the rest of this paper.

\begin{theorem}[Uniform Boundedness of OIT]\label{thm:Uniform Boundedness of OIT}
The OIT defined by \defref{def:Observation-Information Tower} is uniformly bounded w.r.t. $k \in \{k \geq \delta\colon \delta \geq \mu - 1\}$ (or simply $k \geq \delta \geq \mu - 1$), where $\mu$ is the observability index (see Chapter~6.3.1 in~\cite{ChenC1999}), if the pair $(A, C)$ is observable and $A$ is non-singular.
Furthermore, the diameter of the OIT is upper bounded by
\begin{equation}\label{eqn:An Upper Bound on the Diameter of the OIT}
\begin{split}
\!\!d\bigg(\bigcap_{i=k-\delta}^{k} \mathcal{O}_{k,i}\bigg) &\leq
\frac{\sqrt{\displaystyle\sum_{j=0}^{\delta} \bigg[d_v + \sum_{l=1}^{\delta-j} \|C A^{-l} B\| d_w\bigg]^{2}}}
{\sigma_{\min} (O_{\delta})}\\
&=: \bar{d}_{\delta}(A, B, C),
\end{split}
\end{equation}
where $\sigma_{\min}(O_{\delta})$ returns the smallest singular value of $O_{\delta} = [(A^{-\delta})^{\mathrm{T}}C^{\mathrm{T}} \ldots C^{\mathrm{T}}]^{\mathrm{T}}$.
\end{theorem}

\begin{proof}
See \apxref{apx:Proof of thm:Uniform Boundedness of OIT}.
\end{proof}

Note that the uniform boundedness of the OIT indicates the diameter of $\bigcap_{i=k-\delta}^{k} \mathcal{O}_{k,i}$ ($\forall \delta \geq \mu - 1$) cannot go unbounded as $k \to \infty$, which is upper bounded by~\eqref{eqn:An Upper Bound on the Diameter of the OIT}.

\section{Stability Analysis of Classical Linear Set-Membership Filtering Framework}\label{sec:Stability Analysis of Optimal Linear Set-Membership Filtering Framework}

In this section, we study the stability issue w.r.t. the initial condition of the classical linear SMFing framework to address \probref{prob:Stability of Optimal Linear SMF}, based on the OIT.

Since the uniform boundedness of the OIT requires observability, for general $(A, C)$ we introduce the observability decomposition:
there exists a nonsingular $P \in \mathbb{R}^{n\times n}$ such that the equivalence transformation $\tilde{\mathbf{x}}_k = [(\tilde{\mathbf{x}}_k^o)^\mathrm{T} (\tilde{\mathbf{x}}_k^{\bar{o}})^\mathrm{T}]^\mathrm{T} = P \mathbf{x}_k$ transforms~\eqref{eqn:State Equation} and~\eqref{eqn:Measurement Equation} into\footnote{The observability decomposition follows Theorem~6.O6 in~\cite{ChenC1999}: $P = [P_o^{\mathrm{T}}~P_{\bar{o}}^{\mathrm{T}}]^{\mathrm{T}}$ and $U = P^{-1} = [U_o~U_{\bar{o}}]$, where $P_o \in \mathbb{R}^{n_o \times n}$, $P_{\bar{o}}  \in \mathbb{R}^{n_{\bar{o}} \times n}$, $U_o \in \mathbb{R}^{n \times n_o}$, and $U_{\bar{o}} \in \mathbb{R}^{n \times n_{\bar{o}}}$, such that $\tilde{A}_o = P_o A U_o$, $\tilde{A}_{21} = P_{\bar{o}} A U_o$, $\tilde{A}_{\bar{o}} =P_{\bar{o}} A U_{\bar{o}}$, $\tilde{B}_o = P_o B$, $\tilde{B}_{\bar{o}} = P_{\bar{o}} B$, and $\tilde{C}_o = C U_o$.
Note that the eigenvalues of $\tilde{A}_{\bar{o}}$ only depends on $(A, C)$ but independent of $P$.}
\begin{equation}\label{eqn:Observability Decomposition}
\begin{split}
\begin{bmatrix}
\tilde{\mathbf{x}}_{k+1}^o \\
\tilde{\mathbf{x}}_{k+1}^{\bar{o}}
\end{bmatrix}
&=
\begin{bmatrix}
\tilde{A}_o & 0\\
\tilde{A}_{21} & \tilde{A}_{\bar{o}}
\end{bmatrix}
\begin{bmatrix}
\tilde{\mathbf{x}}_k^o \\
\tilde{\mathbf{x}}_k^{\bar{o}}
\end{bmatrix}
+
\begin{bmatrix}
\tilde{B}_o \\
\tilde{B}_{\bar{o}}
\end{bmatrix}
\mathbf{w}_k,\\
\mathbf{y}_k &= \tilde{C}_o \tilde{\mathbf{x}}_k^o + \mathbf{v}_k,
\end{split}
\end{equation}
where $\tilde{x}_k^o \in \mathbb{R}^{n_o}$ and $\tilde{x}_k^{\bar{o}} \in \mathbb{R}^{n_{\bar{o}}}$;
the pair $(\tilde{A}_o, \tilde{C}_o)$ is observable with the observability index $\mu_o$;
it is well-known that $(A, C)$ is detectable if and only if $\rho(\tilde{A}_{\bar{o}}) < 1$, where $\rho(\cdot)$ returns the spectral radius of a matrix.

Now, we propose a stability condition of the classical linear SMFing framework as follows, where we define $\llbracket\hat{\tilde{\mathbf{x}}}_0^o\rrbracket$ and $\llbracket\hat{\tilde{\mathbf{x}}}_0^{\bar{o}}\rrbracket$ as the projections of $\llbracket\hat{\tilde{\mathbf{x}}}_0\rrbracket = \llbracket P \hat{\mathbf{x}}_0\rrbracket$ to the subspaces w.r.t. $\tilde{x}_0^o$ and $\tilde{x}_0^{\bar{o}}$, respectively.

\begin{theorem}[Stability Criterion]\label{thm:Stability Criterion}
The classical SM-Fing framework in Algorithm~\ref{alg:Classical Linear Set-Membership Filtering} is stable w.r.t. its initial condition, if the following conditions hold:
\begin{itemize}
\item[(i)]  $\llbracket\hat{\tilde{\mathbf{x}}}_0^o\rrbracket \supseteq \llbracket\tilde{\mathbf{x}}_0^o\rrbracket$;
\item[(ii)] $\lim_{k\to\infty}\big(\tilde{A}_{\bar{o}}^k + \sum_{i=0}^{k-1} \tilde{A}_{\bar{o}}^{k-1-i} \tilde{A}_{21}\big) < \infty$.
\end{itemize}
\end{theorem}

\begin{proof}
See \apxref{apx:Proof of thm:Stability Criterion}.
\end{proof}

\begin{remark}\label{rek:Stability Criterion}
\thmref{thm:Stability Criterion} tells that when conditions~(i) and~(ii) hold, the classical SMFing framework is stable, which solves \probref{prob:Stability of Optimal Linear SMF}.
More precisely, these two conditions ensure the well-posedness
and the bounded estimation gap, respectively.
However, under condition~(i), we cannot always keep $\llbracket\hat{\mathbf{x}}_k|y_{0:k}\rrbracket \supseteq \llbracket\mathbf{x}_k|y_{0:k}\rrbracket$ (which is different from \propref{prop:Stability of Observable Systems} in \apxref{apx:Proof of thm:Stability Criterion} for observable systems), i.e., the outer boundedness breaks.
\end{remark}

As mentioned in \rekref{rek:Stability Criterion}, condition~(ii) in \thmref{thm:Stability Criterion} is a sufficient condition for the boundedness of the estimation gap.
To evaluate the difference between the sufficiency and the necessity, we provide a necessary condition for the bounded estimation gap in \propref{prop:Converse for Bounded Estimation Gap}.

\begin{proposition}\textbf{\emph{(Necessary Condition for Bounded Estimation Gap)}}\label{prop:Converse for Bounded Estimation Gap}
For all bounded $\llbracket\hat{\mathbf{x}}_0\rrbracket \subset \mathbb{R}^n$ with $\llbracket\hat{\tilde{\mathbf{x}}}_0^o\rrbracket \supseteq \llbracket\tilde{\mathbf{x}}_0^o\rrbracket$, if the estimation gap $d_k^{\mathrm{g}}(\llbracket\hat{\mathbf{x}}_k|y_{0:k}\rrbracket)$ of Algorithm~\ref{alg:Classical Linear Set-Membership Filtering} is bounded,
$\tilde{A}_{\bar{o}}$ must be marginally stable\footnote{In this paper, it refers to that $\tilde{A}_{\bar{o}}$ is the system matrix of a marginally stable discrete-time system, i.e., all eigenvalues of $\tilde{A}_{\bar{o}}$ have magnitudes less or equal to one and those equal to one are non-defective~\cite{ChenC1999}.}.
\end{proposition}

\begin{proof}
See \apxref{apx:Proof of prop:Converse for Bounded Estimation Gap}.
\end{proof}

\begin{remark}\label{rek:Gap Between the Criterion and the Converse}
For all bounded $\llbracket\hat{\mathbf{x}}_0\rrbracket \subset \mathbb{R}^n$ with $\llbracket\hat{\tilde{\mathbf{x}}}_0^o\rrbracket \supseteq \llbracket\tilde{\mathbf{x}}_0^o\rrbracket$, the proposed sufficient condition for the bounded estimation gap [i.e., condition~(ii) in \thmref{thm:Stability Criterion}] is very close to the necessary condition given in \propref{prop:Converse for Bounded Estimation Gap}.
When $\tilde{A}_{21} = 0$, condition~(ii) means $\tilde{A}_{\bar{o}}$ is marginally stable;
in that case, it becomes a necessary and sufficient condition.
When $\tilde{A}_{21} \neq 0$, in general we need $\rho(\tilde{A}_{\bar{o}}) < 1$ (which is close to marginal stability of $\tilde{A}_{\bar{o}}$), i.e., $(A, C)$ is detectable, to provide the bounded-input bounded-output stability to guarantee the bounded estimation gap.
\end{remark}

The following corollary gives a sufficient condition for simultaneously guaranteeing the stability w.r.t. the initial condition and the uniform boundedness of the estimate.

\begin{corollary}[Egregium]\label{cor:Egregium}
For bounded $\llbracket\hat{\mathbf{x}}_0\rrbracket \subset \mathbb{R}^n$ with $\llbracket\hat{\tilde{\mathbf{x}}}_0^o\rrbracket \supseteq \llbracket\tilde{\mathbf{x}}_0^o\rrbracket$, the classical SMFing framework is stable w.r.t. the initial condition and has the uniformly bounded estimate w.r.t. $k \in \mathbb{N}_0$ if $(A, C)$ is detectable.
\end{corollary}

\begin{proof}
See \apxref{apx:Proof of cor:Egregium}.
\end{proof}

\begin{remark}
\corref{cor:Egregium} provides an explicit condition for the stability and the uniform boundedness\footnote{Even though the detectability is usually assumed for the uniform boundedness~\cite{CombastelC2003}, to be best of our knowledge, there are no existing results on rigorously proving it.}.
This result is consistent with the discrete-time Kalman filter that detectable $(A, C)$ implies a stable\footnote{For Kalman filters, to achieve asymptotic stability w.r.t. the initial condition needs an additional condition on the reachability w.r.t. process noises.
For linear SMFs, even though the numerical results in \secref{sec:Classical and Stability-Guaranteed Filtering Frameworks} show the asymptotic stability, proving it requires to introduce probability measures, which is beyond the scope of this work.} filter~\cite{LewisF2007BOOK}.
Thus, it builds an important bridge between stochastic and non-stochastic linear filters on their stability w.r.t. initial conditions.
It should also be highlighted that our result is independent of the types of the noises;
while for stochastic filtering very few results are on the stability of discrete-time linear optimal filters with non-Gaussian noises.
\end{remark}

\begin{remark}\textbf{\emph{(Instability Caused by Improper Initial Conditions)}}\label{rek:Instability Caused by Improper Initial Conditions}
Different from Kalman filtering, the classical linear SMFing framework is not stable for all bounded initial conditions.
If the designer has little information on $\llbracket\mathbf{x}_0\rrbracket$, it would be hard to choose a proper $\llbracket\hat{\mathbf{x}}_0\rrbracket$ to guarantee the stability w.r.t. the initial condition (see \thmref{thm:Stability Criterion}).
This motivates us to propose a stability-guaranteed filtering framework without using the knowledge of $\llbracket\mathbf{x}_0\rrbracket$, presented in \secref{sec:Stability-Guaranteed Filtering Inspired by Observation-Information Tower}.
\end{remark}

\section{Stability-Guaranteed Filtering Framework}\label{sec:Stability-Guaranteed Filtering Inspired by Observation-Information Tower}

In this section, we establish a stability-guaranteed filtering framework, called OIT-inspired filtering (see Algorithm~\ref{alg:OIT-Inspired Filtering}), to solve \probref{prob:Stability-Guaranteed Framework}.

From \thmref{thm:Stability Criterion}, we know that the initial condition $\llbracket\hat{\mathbf{x}}_0\rrbracket$ should satisfy $\llbracket\hat{\tilde{\mathbf{x}}}_0^o\rrbracket \supseteq \llbracket\tilde{\mathbf{x}}_0^o\rrbracket$ to ensure the stability of Algorithm~\ref{alg:Classical Linear Set-Membership Filtering} w.r.t. the initial condition.
This motivates us to choose a sufficiently large $\llbracket\hat{\tilde{\mathbf{x}}}_0^o\rrbracket \supseteq \llbracket\tilde{\mathbf{x}}_0^o\rrbracket$ so that the ill-posedness can be fully handled.
However, it is difficult to choose such $\llbracket\hat{\tilde{\mathbf{x}}}_0^o\rrbracket$ due to the following two issues:
\begin{itemize}
	\item	Since $\llbracket\tilde{\mathbf{x}}_0^o\rrbracket$ is unlikely to know exactly, we can hardly $100\%$ guarantee $\llbracket\hat{\tilde{\mathbf{x}}}_0^o\rrbracket \supseteq \llbracket\tilde{\mathbf{x}}_0^o\rrbracket$.
	\item	Larger $\llbracket\hat{\tilde{\mathbf{x}}}_0^o\rrbracket$ can increase the possibility of the inclusion but brings more conservativeness.
	%, and it is not known whether the resulting estimate $\llbracket\hat{\mathbf{x}}_k| y_{0:k}\rrbracket$ goes unbounded with $\llbracket\hat{\tilde{\mathbf{x}}}_0^o\rrbracket$.
\end{itemize}
In the existing SMFing framework, these two issues cannot be effectively resolved.
Thus, we propose a new SMFing framework inspired by OIT, with the help of the following lemma.

\begin{lemma}[OIT-Inspired Invariance]\label{lem:OIT-Inspired Invariance}
	For all $k \geq \max\{\mu_o-1+n_{\lambda_0^o}, 1\} =: k_*$, where $n_{\lambda_0^o}$ is the index of eigenvalue $0$ of $\tilde{A}_o$ (if $\tilde{A}_o$ does not contain $0$ eigenvalues, $n_{\lambda_0^o} = 0$), define the family of nested sets $\{\mathcal{B}_{\theta_k}^{\infty}[\hat{c}_0^o]\}_{\theta_k \in [0,\infty)}$, in which $\mathcal{B}_{\theta_k}^{\infty}[\hat{c}_0^o]$ is a closed $n_o$-cube of edge length $2 \theta_k$ centered at $\hat{c}_0^o \in \mathbb{R}^{n_o}$.
	Then, $\exists \bar{\theta}_k \geq 0$ s.t. $\forall \theta'_k \leq \bar{\theta}_k \leq \theta''_k$,~\eqref{eqn:OIT-Inspired Invariance} holds, where $P_o$ is a submatrix formed by the first $n_o$ rows of $P$.
\end{lemma}

\begin{figure*}[!t]
	
	\normalsize
	
	\begin{equation}\label{eqn:OIT-Inspired Invariance}
		P_o F_{k,0}(P^{-1}(\mathcal{B}_{\theta'_k}^{\infty}[\hat{c}_0^o] \times \llbracket\hat{\tilde{\mathbf{x}}}_0^{\bar{o}}\rrbracket)) \subseteq P_o F_{k,0}(P^{-1}(\mathcal{B}_{\bar{\theta}_k}^{\infty}[\hat{c}_0^o] \times \llbracket\hat{\tilde{\mathbf{x}}}_0^{\bar{o}}\rrbracket))
		= P_o F_{k,0}(P^{-1}(\mathcal{B}_{\theta''_k}^{\infty}[\hat{c}_0^o] \times \llbracket\hat{\tilde{\mathbf{x}}}_0^{\bar{o}}\rrbracket)).
	\end{equation}

	\hrulefill
	
	\vspace*{4pt}
	
\end{figure*}

\begin{proof}
See \apxref{apx:Proof of lem:OIT-Inspired Invariance}, which is inspired by OIT.
\end{proof}

\begin{remark}\label{rek:OIT-Inspired Invariance}
From \lemref{lem:OIT-Inspired Invariance}, we know that even without the information of $\llbracket\hat{\tilde{\mathbf{x}}}_0^o\rrbracket$, we can enlarge $\theta_k$ until $P_o F_{k,0}(P^{-1}(\mathcal{B}_{\theta_k}^{\infty}[\hat{c}_0^o] \times \llbracket\hat{\tilde{\mathbf{x}}}_0^{\bar{o}}\rrbracket))$ becomes unchanged for $\theta_k \geq \bar{\theta}_k$.
As a result,
\begin{equation*}%\label{eqn:Implication of OIT-Inspired Invariance}
	P_o F_{k,0}(P^{-1}(\llbracket\tilde{\mathbf{x}}_0^o\rrbracket \times \llbracket\hat{\tilde{\mathbf{x}}}_0^{\bar{o}}\rrbracket)) \subseteq P_o F_{k,0}(P^{-1}(\mathcal{B}_{\bar{\theta}_k}^{\infty}[\hat{c}_0^o] \times \llbracket\hat{\tilde{\mathbf{x}}}_0^{\bar{o}}\rrbracket)),
\end{equation*}
which means $\mathcal{B}_{\bar{\theta}_k}^{\infty}[\hat{c}_0^o] \supseteq \llbracket\tilde{\mathbf{x}}_0^o\rrbracket$ always holds for $k \geq k_*$.
Based on this inspiration, we propose the OIT-inspired filtering framework in Algorithm~\ref{alg:OIT-Inspired Filtering}.
\end{remark}

\begin{algorithm}
\begin{footnotesize}
\caption{OIT-Inspired Filtering}\label{alg:OIT-Inspired Filtering}
\begin{algorithmic}[1]
\State  \textbf{Initialization:} Bounded $\llbracket\hat{\mathbf{x}}_0\rrbracket \subset \mathbb{R}^n$;\label{line:OIT-Inspired Filtering - Initialization}
\If {$k < \max\{\mu_o-1+n_{\lambda_0^o}, 1\} = k_*$}\label{line:OIT-Inspired Filtering - Start}
    \State  $\llbracket\hat{\mathbf{x}}_k| y_{0:k}\rrbracket \leftarrow$ Algorithm~\ref{alg:Classical Linear Set-Membership Filtering};\label{line:OIT-Inspired Filtering - Estimate from Classical SMF}
    \If {$\llbracket\hat{\mathbf{x}}_k| y_{0:k}\rrbracket = \emptyset$}\label{line:OIT-Inspired Filtering - Estimates Resetting - Start}
        \State  Choose a $\hat{\mathcal{T}}_0^o$ such that $\llbracket\hat{\mathbf{x}}_k| y_{0:k}\rrbracket = F_{k,0}(P^{-1}(\hat{\mathcal{T}}_0^o \times \llbracket\hat{\tilde{\mathbf{x}}}_0^{\bar{o}}\rrbracket))$ is non-empty and bounded;\label{line:OIT-Inspired Filtering - Estimates Resetting}
    \EndIf\label{line:OIT-Inspired Filtering - Estimates Resetting - End}
\Else
	\State  $\llbracket\hat{\mathbf{x}}_k| y_{0:k}\rrbracket \!=\! F_{k,0}(P^{-1}(\mathcal{B}_{\bar{\theta}_k}^{\infty}[\hat{c}_0^o] \times \llbracket\hat{\tilde{\mathbf{x}}}_0^{\bar{o}}\rrbracket))$;  \% Recursively\label{line:OIT-Inspired Filtering - Estimate Inspired by OIT}
\EndIf\label{line:OIT-Inspired Filtering - End}
\end{algorithmic}
\end{footnotesize}
\end{algorithm}

A line-by-line explanation of Algorithm~\ref{alg:OIT-Inspired Filtering} is as follows.
Line~\ref{line:OIT-Inspired Filtering - Initialization} initializes the algorithm.
Lines~\ref{line:OIT-Inspired Filtering - Start}-\ref{line:OIT-Inspired Filtering - End} give the filtering process at each $k \in \mathbb{N}_0$.
For $k < k_*$, the estimate $\llbracket\hat{\mathbf{x}}_k| y_{0:k}\rrbracket$ is identical to that of Algorithm~\ref{alg:Classical Linear Set-Membership Filtering} (see Line~\ref{line:OIT-Inspired Filtering - Estimate from Classical SMF} of Algorithm~\ref{alg:OIT-Inspired Filtering}), if it is not an empty set;
otherwise, it will be reset by Line~\ref{line:OIT-Inspired Filtering - Estimates Resetting}.
In Line~\ref{line:OIT-Inspired Filtering - Estimates Resetting}, we can choose $\hat{\mathcal{T}}_0^o = \mathcal{B}_{\theta}^{\infty}[0_{n_o}]$ with sufficiently large $\theta$, which is used in Algorithm~\ref{alg:OIT-CZ SMF}.
For $k \geq k_*$, the estimate $\llbracket\hat{\mathbf{x}}_k| y_{0:k}\rrbracket$ is determined by Line~\ref{line:OIT-Inspired Filtering - Estimate Inspired by OIT}.
Note that Line~\ref{line:OIT-Inspired Filtering - Estimate Inspired by OIT} can be implemented in a recursive manner by its definition in~\eqref{eqn:Filtering Map}.\footnote{Generally speaking, reduction methods are required to balance the accuracy and the complexity in Line~\ref{line:OIT-Inspired Filtering - Estimate Inspired by OIT} for specific filter designs.
But it is also possible to reduce the complexity without any accuracy loss (see the cases in \secref{sec:Classical and Stability-Guaranteed Filtering Frameworks}).}

\begin{theorem}[Stability of OIT-Inspired Filtering]\label{thm:Stability of OIT-Inspired Filtering}
If condition~(ii) in \thmref{thm:Stability Criterion} holds,
the filtering framework in Algorithm~\ref{alg:OIT-Inspired Filtering} is stable w.r.t. the initial condition.
\end{theorem}

\begin{proof}
See \apxref{apx:Proof of thm:Stability of OIT-Inspired Filtering}.
\end{proof}

\thmref{thm:Stability of OIT-Inspired Filtering} indicates that the OIT-inspired filtering framework given in Algorithm~\ref{alg:OIT-Inspired Filtering} does not rely on any information about $\llbracket\mathbf{x}_0\rrbracket$ to guarantee the well-posedness.

For Algorithm~\ref{alg:OIT-Inspired Filtering}, similar results in \propref{prop:Converse for Bounded Estimation Gap} and \corref{cor:Egregium} can also be derived, where the condition $\llbracket\hat{\tilde{\mathbf{x}}}_0^o\rrbracket \supseteq \llbracket\tilde{\mathbf{x}}_0^o\rrbracket$ is not needed any more.

\section{Stable and Fast Constrained Zonotopic SMF}\label{sec:A Stable and Fast Constrained Zonotopic SMF}

In this section, we develop a constrained zonotopic SMF under the new framework described in Algorithm~\ref{alg:OIT-Inspired Filtering}, where the OIT plays the pivotal role.
This SMF is not only with guaranteed stability but also with high efficiency and good accuracy.
We call it the OIT-inspired Constrained Zonotopic SMF (OIT-CZ SMF).

Before presenting the SMF, we introduce the constrained zonotope~\cite{ScottJ2016} in \defref{def:Extended Constrained Zonotope} with a small extension.

\begin{definition}[Extended Constrained Zonotope]\label{def:Extended Constrained Zonotope}
A set $\mathcal{Z} \subseteq \mathbb{R}^n$ is an (extended) constrained zonotope if there exists a quintuple $(\hat{G}, \hat{c}, \hat{A}, \hat{b}, \hat{h}) \in \mathbb{R}^{n\times n_g} \times \mathbb{R}^n \times \mathbb{R}^{n_c\times n_g} \times \mathbb{R}^{n_c} \times [0,\infty]^{n_g}$ such that $\mathcal{Z}$ is expressed by
\begin{equation*}%\label{eqn:Extended Constrained Zonotope}
\bigg\{\hat{G}\xi + \hat{c}\colon \hat{A} \xi = \hat{b},~\xi \in \prod_{j=1}^{n_g} \big[-\hat{h}^{(j)},\hat{h}^{(j)}\big]\bigg\} =: Z(\hat{G}, \hat{c}, \hat{A}, \hat{b}, \hat{h}),
\end{equation*}
where $\hat{h}^{(j)}$ is the $j$\textsuperscript{th} component of $\hat{h}$.
\end{definition}

In \defref{def:Extended Constrained Zonotope}, we slightly generalize the constrained zonotope in~\cite{ScottJ2016} by replacing $\|\xi\|_{\infty} \leq 1$, i.e., $\xi \in [-1,~1]^{n_g}$, with $\xi \in \prod_{j=1}^{n_g} \big[-\hat{h}^{(j)},\hat{h}^{(j)}\big]$.
The benefit is twofold:
(i)~we allow $\hat{h}^{(j)}$ to be infinity such that the posterior sets induced by unbounded prior sets (which is required by the SMFing framework in Algorithm~\ref{alg:OIT-Inspired Filtering}) can be fully described;
(ii)~the numerical stability of our proposed algorithm is improved.

Based on \defref{def:Extended Constrained Zonotope}, if $\llbracket\hat{\mathbf{x}}_0\rrbracket$, $\llbracket\mathbf{w}_k\rrbracket$, and $\llbracket\mathbf{v}_k\rrbracket$ are constrained zonotopes, the resulting $\llbracket\hat{\mathbf{x}}_k|y_{0:k-1}\rrbracket$ and $\llbracket\hat{\mathbf{x}}_k|y_{0:k}\rrbracket$ in Algorithm~\ref{alg:Classical Linear Set-Membership Filtering} are also constrained zonotopes without any approximations.
Specifically, by defining
\begin{equation}\label{eqn:Constrained Zonotopic Description}
\begin{split}
\llbracket\hat{\mathbf{x}}_k|y_{0:k-1}\rrbracket &= Z(\hat{G}_k^-, \hat{c}_k^-, \hat{A}_k^-, \hat{b}_k^-, \hat{h}_k^-),\\
\llbracket\hat{\mathbf{x}}_k|y_{0:k}\rrbracket &= Z(\hat{G}_k, \hat{c}_k, \hat{A}_k, \hat{b}_k, \hat{h}_k),\\
\llbracket\mathbf{w}_k\rrbracket &= Z(\hat{G}_{\mathbf{w}_k}, \hat{c}_{\mathbf{w}_k}, \hat{A}_{\mathbf{w}_k}, \hat{b}_{\mathbf{w}_k}, \hat{h}_{\mathbf{w}_k}),\\
\llbracket\mathbf{v}_k\rrbracket &= Z(\hat{G}_{\mathbf{v}_k}, \hat{c}_{\mathbf{v}_k}, \hat{A}_{\mathbf{v}_k}, \hat{b}_{\mathbf{v}_k}, \hat{h}_{\mathbf{v}_k}),
\end{split}
\end{equation}
the prediction step~\eqref{eqn:Prediction - Optimal Linear Set-Membership Filter} gives the exact $\llbracket\hat{\mathbf{x}}_k|y_{0:k-1}\rrbracket$ with
\begin{equation}\label{eqn:Prediction - CZ-SMF - Parameters}
\begin{split}
\!\!\!\!\!\hat{G}_k^- &\!=\!
\begin{bmatrix}
A \hat{G}_{k-1} & B \hat{G}_{\mathbf{w}_{k-1}}
\end{bmatrix},~
\hat{c}_k^- = A \hat{c}_{k-1} + B \hat{c}_{\mathbf{w}_{k-1}},\\
\!\!\!\!\!\hat{A}_k^- &\!=\!
\begin{bmatrix}
\hat{A}_{k-1} & 0 \\
0 & \hat{A}_{\mathbf{w}_{k-1}}
\end{bmatrix}\!,
\hat{b}_k^- \!=\!
\begin{bmatrix}
\hat{b}_{k-1} \\
\hat{b}_{\mathbf{w}_{k-1}}
\end{bmatrix}\!,
\hat{h}_k^- \!=\!
\begin{bmatrix}
\hat{h}_{k-1} \\
\hat{h}_{\mathbf{w}_{k-1}}
\end{bmatrix}\!,
\end{split}
\end{equation}
and the update step~\eqref{eqn:Update - Optimal Linear Set-Membership Filter} returns the exact $\llbracket\hat{\mathbf{x}}_k|y_{0:k}\rrbracket$ with
\begin{equation}\label{eqn:Update - CZ-SMF - Parameters}
\begin{split}
\hat{G}_k &=
\begin{bmatrix}
\hat{G}_k^- & 0
\end{bmatrix},~
\hat{c}_k = \hat{c}_k^-,~
\hat{h}_k =
\begin{bmatrix}
\hat{h}_k^- \\
\hat{h}_{\mathbf{v}_k}
\end{bmatrix},\\
\hat{A}_k &=
\begin{bmatrix}
\hat{A}_k^- & 0 \\
0 & \hat{A}_{\mathbf{v}_k}\\
C \hat{G}_k^- & \hat{G}_{\mathbf{v}_k}
\end{bmatrix},~
\hat{b}_k =
\begin{bmatrix}
\hat{b}_k^- \\
\hat{b}_{\mathbf{v}_k}\\
y_k - \hat{c}_{\mathbf{v}_k} - C \hat{c}_k^-
\end{bmatrix}.
\end{split}
\end{equation}
The proof of~\eqref{eqn:Prediction - CZ-SMF - Parameters} and~\eqref{eqn:Update - CZ-SMF - Parameters} is straightforward from~\cite{ScottJ2016}, by rewriting~\eqref{eqn:Update - Optimal Linear Set-Membership Filter} as $\llbracket\mathbf{x}_k|y_{0:k}\rrbracket = \{x_k \in \llbracket\mathbf{x}_k|y_{0:k-1}\rrbracket\colon C x_k \in \{y_k\} + \llbracket-\mathbf{v}_k\rrbracket\}$.

Now, we are ready to design the OIT-CZ SMF (see Algorithm~\ref{alg:OIT-CZ SMF}), where \propref{prop:Constrained Zonotopic Image under Filtering Map} plays an important role.

\begin{proposition}\label{prop:Constrained Zonotopic Image under Filtering Map}
The image of a constrained zonotope $\hat{\mathcal{Z}}_i^- = Z(\hat{G}_i^-, \hat{c}_i^-, \hat{A}_i^-, \hat{b}_i^-, \hat{h}_i^-)$ under the filtering map $F_{k,i}$ is
\begin{equation}\label{eqn:Constrained Zonotopic Image under Filtering Map}
\mathcal{Z}_k = F_{k,i} (\hat{\mathcal{Z}}_i^-) = Z(\hat{G}_k, \hat{c}_k, \hat{A}_k, \hat{b}_k, \hat{h}_k),
\end{equation}
where the parameters are given in~\eqref{eqn:Constrained Zonotopic Image under Filtering Map - Parameters} with $\hat{b}_{y_l} = y_l - \hat{c}_{\mathbf{v}_l} - C \hat{c}_l^-$ and $\hat{c}_l^- = A^{l-i} \hat{c}_i^- + \sum_{r = i}^{l - 1} A^{l-1-r} B \hat{c}_{\mathbf{w}_r}$ for $i \leq l \leq k$.
\end{proposition}

\begin{figure*}[!t]

\normalsize

\begin{equation}\label{eqn:Constrained Zonotopic Image under Filtering Map - Parameters}
\begin{split}
\hat{G}_k &=
\begin{bmatrix}
A^{k-i} \hat{G}_i^- & 0 & A^{k-i-1} B \hat{G}_{\mathbf{w}_i} & 0 & \ldots & B \hat{G}_{\mathbf{w}_{k-1}}  & 0
\end{bmatrix},\quad
\hat{c}_k = A^{k-i} \hat{c}_i^- + \sum_{r = i}^{k - 1} A^{k-1-r} B \hat{c}_{\mathbf{w}_r},\\
\hat{A}_k &=
\begin{bmatrix}
\hat{A}_i^- & 0 & 0 & 0 & \ldots & 0 & 0 \\
0 & \hat{A}_{\mathbf{v}_i} & 0 & 0 & \ldots & 0 & 0 \\
C\hat{G}_i^- & \hat{G}_{\mathbf{v}_i} & 0 & 0 & \ldots & 0 & 0 \\
0 & 0 & \hat{A}_{\mathbf{w}_i} & 0 & \ldots & 0 & 0 \\
0 & 0 & 0 & \hat{A}_{\mathbf{v}_{i+1}} & \ldots & 0 & 0 \\
C A \hat{G}_i^- & 0 & C B \hat{G}_{\mathbf{w}_i} & \hat{G}_{\mathbf{v}_{i+1}} & \ldots & 0 & 0 \\
\vdots & \vdots & \vdots & \vdots &  & \vdots & \vdots \\
0 & 0 & 0 & 0 & \ldots & \hat{A}_{\mathbf{w}_{k-1}} & 0 \\
0 & 0 & 0 & 0 & \ldots & 0 & \hat{A}_{\mathbf{v}_k} \\
C A^{k-i} \hat{G}_i^- & 0 & C A^{k-i-1} B \hat{G}_{\mathbf{w}_i} & 0 & \ldots & C B \hat{G}_{\mathbf{w}_{k-1}} & \hat{G}_{\mathbf{v}_k}
\end{bmatrix},\quad
\hat{b}_k =
\begin{bmatrix}
\hat{b}_i^- \\
\hat{b}_{\mathbf{v}_i} \\
\hat{b}_{y_i}\\
\hat{b}_{\mathbf{w}_i} \\
\hat{b}_{\mathbf{v}_{i+1}} \\
\hat{b}_{y_{i+1}}\\
\vdots \\
\hat{b}_{\mathbf{w}_{k-1}} \\
\hat{b}_{\mathbf{v}_k} \\
\hat{b}_{y_k}
\end{bmatrix},\quad
\hat{h}_k =
\begin{bmatrix}
\hat{h}_i^- \\
\hat{h}_{\mathbf{v}_i} \\
\hat{h}_{\mathbf{w}_i} \\
\hat{h}_{\mathbf{v}_{i+1}} \\
\vdots \\
\hat{h}_{\mathbf{w}_{k-1}} \\
\hat{h}_{\mathbf{v}_k}
\end{bmatrix}.
\end{split}
\end{equation}

\hrulefill

\vspace*{4pt}

\end{figure*}

\begin{proof}
	By setting $\hat{\mathcal{Z}}_i^- = \llbracket\hat{\mathbf{x}}_i|y_{0:i-1}\rrbracket$ and recursively using the prediction and update steps [i.e.,~\eqref{eqn:Prediction - CZ-SMF - Parameters} and~\eqref{eqn:Update - CZ-SMF - Parameters}] according to~\eqref{eqn:Filtering Map}, equation~\eqref{eqn:Constrained Zonotopic Image under Filtering Map} can be derived.
\end{proof}

\begin{algorithm}
\begin{footnotesize}
\caption{OIT-Inspired Constrained Zonotopic SMF (OIT-CZ SMF)}\label{alg:OIT-CZ SMF}
\begin{algorithmic}[1]
\State  \textbf{Initialization:} Bounded constrained zonotope $\llbracket\hat{\mathbf{x}}_0\rrbracket = Z(\hat{G}_0^-, \hat{c}_0^-, \hat{A}_0^-, \hat{b}_0^-, \hat{h}_0^-) \subset \mathbb{R}^n$, $\bar{\delta} \geq \max\{\mu_o-1+n_{\lambda_0^o}, 1\}$, $\varepsilon > 0$, $\Upsilon_{\infty} = \inf_{\gamma \in (\rho(\tilde{A}_{\bar{o}}), 1)} \frac{\max \{\gamma^{-k} \|\tilde{A}_{\bar{o}}^k\|_{\infty}\colon k \in \mathbb{N}_0\}}{1-\gamma}$;\label{line:OIT-CZ SMF - Initialization}
%\State  \textbf{Parameters:} Real number $\epsilon > 0$;\label{line:alg:ROIT-Aided Filter - Parameters}
\If {$k < \bar{\delta}$}\label{line:OIT-CZ SMF - Start}
    \State  $\mathcal{Z}_k = \llbracket\hat{\mathbf{x}}_k|y_{0:k}\rrbracket \leftarrow$ Algorithm~\ref{alg:Classical Linear Set-Membership Filtering} with~\eqref{eqn:Constrained Zonotopic Description}-\eqref{eqn:Update - CZ-SMF - Parameters};\label{line:OIT-CZ SMF - Optimal CZ-SMF}
    \If {$\mathcal{Z}_k = \emptyset$}\label{line:OIT-CZ SMF - Estimates Resetting - Start}
        \State  $\mathcal{Z}_k \leftarrow$~\eqref{eqn:Constrained Zonotopic Image under Filtering Map} with $i = 0$ and $\mathcal{Z}_0^- = P^{-1}\big(\mathcal{B}_{\theta}^{\infty}[0_{n_o}] \times \llbracket\hat{\tilde{\mathbf{x}}}_0^{\bar{o}}\rrbracket\big)$, where $\theta > 0$ is sufficiently large such that $\mathcal{Z}_k \neq \emptyset$;\label{line:OIT-CZ SMF - Estimates Resetting}
    \EndIf\label{line:OIT-CZ SMF - Estimates Resetting - End}
    \State  $\hat{\mathcal{T}}_k^{\bar{o}} \leftarrow Z(I_{n_o},\overline{c}_k^{\bar{o}},[~],[~],\overline{G}_k^{\bar{o}} 1_{n_o}) = \overline{\mathrm{IH}}(\widetilde{\mathcal{Z}}_k^{-,\bar{o}})$;\label{line:OIT-CZ SMF - Tobar - Case 1}
\Else
    \State  $\mathcal{Z}_k \leftarrow$~\eqref{eqn:Constrained Zonotopic Image under Filtering Map} with $i = k-\bar{\delta}$ and $\hat{\mathcal{Z}}_i^- = P^{-1} \big(\hat{\mathcal{T}}_{k-\bar{\delta}}^o \!\times\! \hat{\mathcal{T}}_{k-\bar{\delta}}^{\bar{o}}\big)$, where $\hat{\mathcal{T}}_{k-\bar{\delta}}^o = \mathcal{B}_{\bar{\theta}_k}^{\infty} \big[\mathrm{center}\big(P_o \overline{\mathrm{IH}} (\mathcal{Z}_{k-\bar{\delta}}^-)\big)\big]$ for $k-\bar{\delta} < \bar{\delta}$ and $\hat{\mathcal{T}}_{k-\bar{\delta}}^o = P_o \overline{\mathrm{IH}} (\mathcal{Z}_{k-\bar{\delta}}^-)$ for $k-\bar{\delta} \geq \bar{\delta}$;\label{line:OIT-CZ SMF - Estimate Inspired by OIT}
    \State  $Z(I_{n_o},\hat{c}_k^{\mathrm{in}},[~],[~],\hat{G}_k^{\mathrm{in}} 1_{n_o}) = \overline{\mathrm{IH}}(\tilde{A}_{21} \widetilde{\mathcal{Z}}_k^o \oplus \tilde{B}_{\bar{o}} \llbracket\mathbf{w}_k\rrbracket)$;\label{line:OIT-CZ SMF - Interval Hull of Unobservable Subsystem Input}
    \State  $\ell_k = \max\{\|\hat{G}_k^{\mathrm{in}}\|_{\infty}, \ell_{k-1}\}$, with $\ell_{\bar{\delta}-1} = 0$;\label{line:OIT-CZ SMF - Component Max}
    \State  $\hat{c}_{k+1}^{\bar{o}} = \tilde{A}_{\bar{o}} \hat{c}_k^{\bar{o}} + \hat{c}_k^{\mathrm{in}}$, with $\hat{c}_{\bar{\delta}}^{\bar{o}} = \mathrm{center}\big(\overline{\mathrm{IH}}(\widetilde{\mathcal{Z}}_{\bar{\delta}}^{\bar{o}})\big)$;\label{line:OIT-CZ SMF - c_obar}
    \State  $\hat{\mathcal{T}}_k^{\bar{o}} \!=\! \mathcal{B}_{\alpha_k}^{\infty}[\hat{c}_k^{\bar{o}}]$ with $\alpha_k \!=\! \frac{1}{2}\|A_{\bar{o}}^{k-\bar{\delta}}\|_{\infty} d_{\infty}\! (\widetilde{\mathcal{Z}}_{\bar{\delta}}^{\bar{o}}) + \!\Upsilon_{\infty} \ell_{k-1} \!+\! \varepsilon$;\label{line:OIT-CZ SMF - Tobar - Case 2}
\EndIf\label{line:OIT-CZ SMF - End}
\end{algorithmic}
\end{footnotesize}
\end{algorithm}

Algorithm~\ref{alg:OIT-CZ SMF} is under the framework of Algorithm~\ref{alg:OIT-Inspired Filtering}, and we provide the line-by-line explanation as follows.
\begin{itemize}
\item   Line~\ref{line:OIT-CZ SMF - Initialization} initializes Algorithm~\ref{alg:OIT-CZ SMF}, with the additional parameters $\bar{\delta} \geq \max\{\mu_o-1+n_{\lambda_0^o}, 1\}$ and $\varepsilon > 0$ that: a larger $\bar{\delta}$ leads to higher accuracy but increases the complexity (determined by Line~\ref{line:OIT-CZ SMF - Estimate Inspired by OIT});
    a smaller $\varepsilon$ makes the estimate corresponding to the unobservable system more accurate but brings slower convergence of~\eqref{eqn:Stability of OIT-CZ SMF - Finite-Time Inclusion} [indicated by Line~\ref{line:OIT-CZ SMF - Tobar - Case 2} and~\eqref{eqninpf:thm:Stability of OIT-CZ SMF - Well-Posedness - Step 2 - Distance Upper Bound 2}].
    Line~\ref{line:OIT-CZ SMF - Initialization} also gives an important constant $\Upsilon_{\infty}$ in estimating the unobservable state (see Line~\ref{line:OIT-CZ SMF - Estimate Inspired by OIT} with Line~\ref{line:OIT-CZ SMF - Tobar - Case 2}).\footnote{Note that $\max \{\gamma^{-k} \|\tilde{A}_{\bar{o}}^k\|_{\infty}\colon k \in \mathbb{N}_0\}$ can be calculated within finite steps (implied by the proof of \lemref{lem:Bound on Matrix Power Norm}).
    Thus, $\Upsilon_{\infty}$ can be computed by searching $\gamma$ over $(\rho(\tilde{A}_{\bar{o}}), 1)$.}
\item   Lines~\ref{line:OIT-CZ SMF - Optimal CZ-SMF}-\ref{line:OIT-CZ SMF - Tobar - Case 1} are for $k < \bar{\delta}$.
    Similarly to Lines~\ref{line:OIT-Inspired Filtering - Estimate from Classical SMF} and~\ref{line:OIT-Inspired Filtering - Estimates Resetting} of Algorithm~\ref{alg:OIT-Inspired Filtering}, Lines~\ref{line:OIT-CZ SMF - Optimal CZ-SMF} and~\ref{line:OIT-Inspired Filtering - Estimates Resetting} of Algorithm~\ref{alg:OIT-CZ SMF} give the estimate $\mathcal{Z}_k$:
    Line~\ref{line:OIT-CZ SMF - Optimal CZ-SMF} gives the estimate returned by the constrained zonotopic version of Algorithm~\ref{alg:Classical Linear Set-Membership Filtering}, i.e., with~\eqref{eqn:Constrained Zonotopic Description}-\eqref{eqn:Update - CZ-SMF - Parameters};
    in Line~\ref{line:OIT-CZ SMF - Estimates Resetting}, $\mathcal{B}_{\theta}^{\infty}[0_{n_o}]$ can be expressed by $Z(I_{n_o},0_{n_o},[~],[~],\alpha 1_{n_o})$ for improving the numerical stability, where $I_{n_o}$ is the identity matrix of size $n_o$ and $1_{n_o}$ is the $n_o$-dimensional all-ones column vector;
    one can double $\theta$ from $\theta = 1$ until $\mathcal{Z}_k \neq \emptyset$, and this can be done in finite steps.
    Line~\ref{line:OIT-CZ SMF - Tobar - Case 1} calculates the interval hull of $\widetilde{\mathcal{Z}}_k^{-,\bar{o}}$ by simply solving $2 n_o$ linear programmings, where $\overline{G}_k^{\bar{o}}$ and $\overline{c}_k^{\bar{o}}$ are the generator matrix and the center of the resulting interval hull, respectively, and $Z(I_{n_o},\overline{c}_k^{\bar{o}},[~],[~],\overline{G}_k^{\bar{o}} 1_{n_o})$ is employed to improve the numerical stability;
    note that $\widetilde{\mathcal{Z}}_k^{-,\bar{o}}$ is the projection of $\widetilde{\mathcal{Z}}_k^- = P \mathcal{Z}_k^-$ to the subspace w.r.t. $\tilde{x}_k^{\bar{o}}$, where $\mathcal{Z}_k^-$ is derived during the processing of calculating $\mathcal{Z}_k$ in Line~\ref{line:OIT-CZ SMF - Optimal CZ-SMF} or Line~\ref{line:OIT-CZ SMF - Estimates Resetting}.
\item   Lines~\ref{line:OIT-CZ SMF - Estimate Inspired by OIT}-\ref{line:OIT-CZ SMF - Tobar - Case 2} are for $k \geq \bar{\delta}$, where $\tilde{\mathcal{Z}}_k^o$ and $\tilde{\mathcal{Z}}_k^{\bar{o}}$ are the projections of $\widetilde{\mathcal{Z}}_k = P \mathcal{Z}_k$ to the subspaces w.r.t. $\tilde{x}_k^o$ and $\tilde{x}_k^{\bar{o}}$, respectively.
    Line~\ref{line:OIT-CZ SMF - Estimate Inspired by OIT} of Algorithm~\ref{alg:OIT-CZ SMF} gives the estimate $\mathcal{Z}_k$, which is a finite-horizon version of Line~\ref{line:OIT-Inspired Filtering - Estimate Inspired by OIT} of Algorithm~\ref{alg:OIT-Inspired Filtering} over the time window $[k-\bar{\delta},~k]$.
    In Line~\ref{line:OIT-CZ SMF - Estimate Inspired by OIT}, $\bar{\theta}_k$ is derived based on \lemref{lem:OIT-Inspired Invariance},\footnote{When utilizing \lemref{lem:OIT-Inspired Invariance}, one should regard $k - \bar{\delta}$, $\mathcal{B}_{\bar{\theta}_k}^{\infty} \big[\mathrm{center}\big(P_o \overline{\mathrm{IH}} (\mathcal{Z}_{k-\bar{\delta}}^-)\big)\big]$, and $\hat{\mathcal{T}}_{k-\bar{\delta}}^{\bar{o}}$ as $0$, $\mathcal{B}_{\bar{\theta}_k}^{\infty}[\hat{c}_0^o]$, and $\llbracket\hat{\tilde{\mathbf{x}}}_0^{\bar{o}}\rrbracket$, respectively.
    For $\bar{\theta}_k$, one can double $\theta_k$ from $\theta_k = 1$ until the equality in~\eqref{eqn:OIT-Inspired Invariance} holds, where the equality can be checked by calculating interval hulls.}
    and from Algorithm~\ref{alg:OIT-CZ SMF} we can observe that $\hat{\mathcal{T}}_{k-\bar{\delta}}^{\bar{o}}$ is calculated by Line~\ref{line:OIT-CZ SMF - Tobar - Case 1} (for $k-\bar{\delta} < \bar{\delta}$) and Lines~\ref{line:OIT-CZ SMF - Interval Hull of Unobservable Subsystem Input}-\ref{line:OIT-CZ SMF - Tobar - Case 2} (for $k-\bar{\delta} \geq \bar{\delta}$).\footnote{In Line~\ref{line:OIT-CZ SMF - Estimate Inspired by OIT}, we can replace $\mathcal{Z}_{k-\bar{\delta}}^-$ with $\mathcal{Z}_{k-\bar{\delta}}$ to improve the efficiency when $\overline{\mathrm{IH}} (\mathcal{Z}_{k-\bar{\delta}})$ is calculated (see \secref{sec:The Fast Constrained Zonotopic SMF}).}
    Line~\ref{line:OIT-CZ SMF - Interval Hull of Unobservable Subsystem Input} gives the interval hull of the ``input'' of the unobservable subsystem, where $\hat{c}_k^{\mathrm{in}}$ is the center of the resulting interval hull and $\hat{G}_k^{\mathrm{in}}$ is diagonal and positive semi-definite.
    In Line~\ref{line:OIT-CZ SMF - Component Max}, $\|\hat{G}_k^{\mathrm{in}}\|_{\infty}$ represents the maximum half-edge length of the interval hull derived by Line~\ref{line:OIT-CZ SMF - Interval Hull of Unobservable Subsystem Input}.
    Thus, $\ell_k$ records the greatest maximum half-edge length up to $k$, which determines the radius (in the sense of $\infty$-norm) of $\hat{\mathcal{T}}_{k-\bar{\delta}}^{\bar{o}}$ [see Line~\ref{line:OIT-CZ SMF - Tobar - Case 2}, where $d_{\infty}(\mathcal{S}) := \sup_{s,s' \in \mathcal{S}} \|s - s'\|_{\infty}$];
    for the center of $\hat{\mathcal{T}}_{k-\bar{\delta}}^{\bar{o}}$, it is calculated by Line~\ref{line:OIT-CZ SMF - c_obar}.
\end{itemize}

The following theorem describes not only the stability w.r.t. the initial condition but also two important properties of Algorithm~\ref{alg:OIT-CZ SMF}.
More specifically, for detectable $(A, C)$: the finite-time inclusion property fixes the ``outer boundedness breaking'' problem in the classical linear SMFing framework (see \rekref{rek:Stability Criterion});
the uniform boundedness of $\mathcal{Z}_k$ is guaranteed, i.e., the estimate cannot go unbounded as time elapses -- to the best of our knowledge it is the first time to propose a constrained-zonotopic SMF with rigorously proven uniform boundedness.

\begin{theorem}[Properties of OIT-CZ SMF]\label{thm:Stability of OIT-CZ SMF}
If $(A, C)$ is detectable,
the OIT-CZ SMF in Algorithm~\ref{alg:OIT-CZ SMF} is stable w.r.t. the initial condition.
Furthermore, there exists a $\underline{k} \geq 2\bar{\delta}$ such that
\begin{equation}\label{eqn:Stability of OIT-CZ SMF - Finite-Time Inclusion}
\mathcal{Z}_k \supseteq \llbracket\mathbf{x}_k| y_{0:k}\rrbracket,\quad k \geq \underline{k},
\end{equation}
which is the finite-time inclusion property of Algorithm~\ref{alg:OIT-CZ SMF}.
Finally, $\mathcal{Z}_k$ is uniformly bounded w.r.t. $k \in \mathbb{N}_0$.
\end{theorem}

\begin{proof}
See \apxref{apx:Proof of thm:Stability of OIT-CZ SMF}.
\end{proof}

Furthermore, Algorithm~\ref{alg:OIT-CZ SMF} has low computational complexity per step, especially when the system is observable:
the averaged complexity for $k \to \infty$ is determined by the case with $k \geq \bar{\delta}$ (i.e., Lines~\ref{line:OIT-CZ SMF - Estimate Inspired by OIT}-\ref{line:OIT-CZ SMF - Tobar - Case 2} in Algorithm~\ref{alg:OIT-CZ SMF});
for observable $(A, C)$, only Line~\ref{line:OIT-CZ SMF - Estimate Inspired by OIT} remains, and we can also set $\mathcal{Z}_{k-\bar{\delta}}^- = \mathbb{R}^n = \mathcal{B}_{\infty}^{\infty}[0_n]$;
from~\eqref{eqn:Constrained Zonotopic Image under Filtering Map - Parameters} we know that
In \secref{sec:Time Efficiency}, we show that Algorithm~\ref{alg:OIT-CZ SMF} has higher time efficiency compared with two most efficient constrained zonotopic SMFs in the toolbox CORA 2024~\cite{AlthoffM2024Manual}.

In terms of the accuracy, Algorithm~\ref{alg:OIT-CZ SMF} can be regarded as a reduction on the optimal estimate in Algorithm~\ref{alg:Classical Linear Set-Membership Filtering} with constrained zonotopic descriptions.
Different from the existing reduction methods (e.g.,~\cite{ScottJ2016}) based on geometric properties of constrained zonotopes, Algorithm~\ref{alg:OIT-CZ SMF} utilizes the properties of the dynamical system.
Thus, the reduction is in a long-term manner instead of a greedy/instanetous manner, which greatly overcomes the wrapping effect.
Therefore, the proposed OIT-CZ SMF has the desired performance improvement in terms of both complexity and accuracy.

\section{Numerical Examples}\label{sec:Numerical Examples}

\subsection{Classical and Stability-Guaranteed Frameworks}\label{sec:Classical and Stability-Guaranteed Filtering Frameworks}

In this subsection, we consider observable and detectable dynamical systems as illustrative examples to validate the theoretical results in \secref{sec:Stability Analysis of Optimal Linear Set-Membership Filtering Framework} and \secref{sec:Stability-Guaranteed Filtering Inspired by Observation-Information Tower}.

\begin{figure*}
\centering
\subfigure[]{\includegraphics [width=0.67\columnwidth,height=0.54\columnwidth]{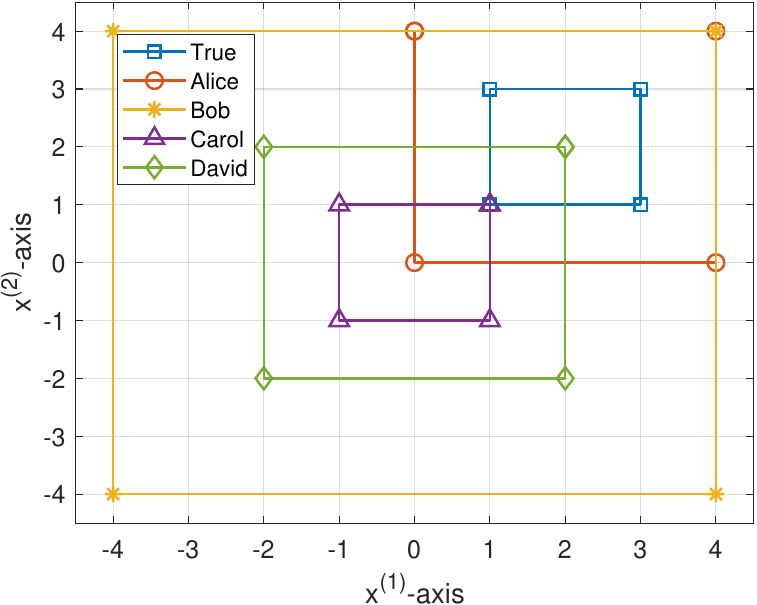}\label{fig:Illustrative example for observable system - A}}
\subfigure[]{\includegraphics [width=0.7\columnwidth,height=0.54\columnwidth]{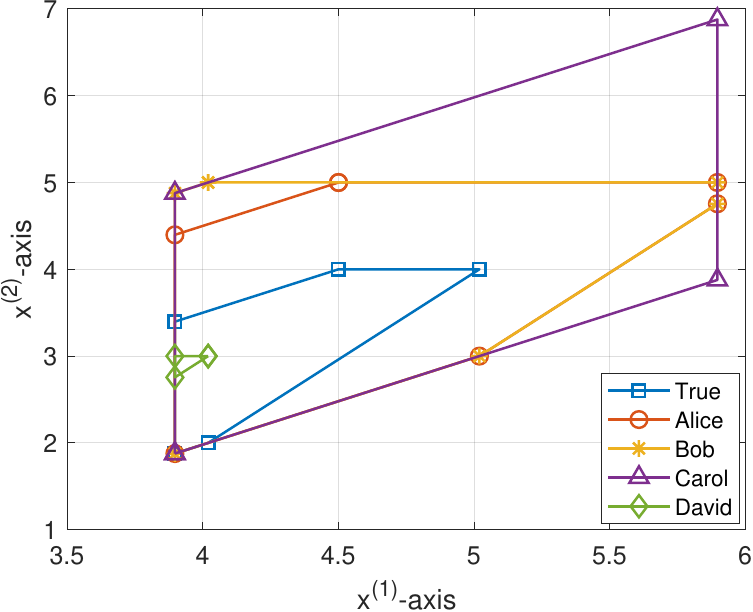}\label{fig:Illustrative example for observable system - B}}
\subfigure[]{\includegraphics [width=0.7\columnwidth,height=0.54\columnwidth]{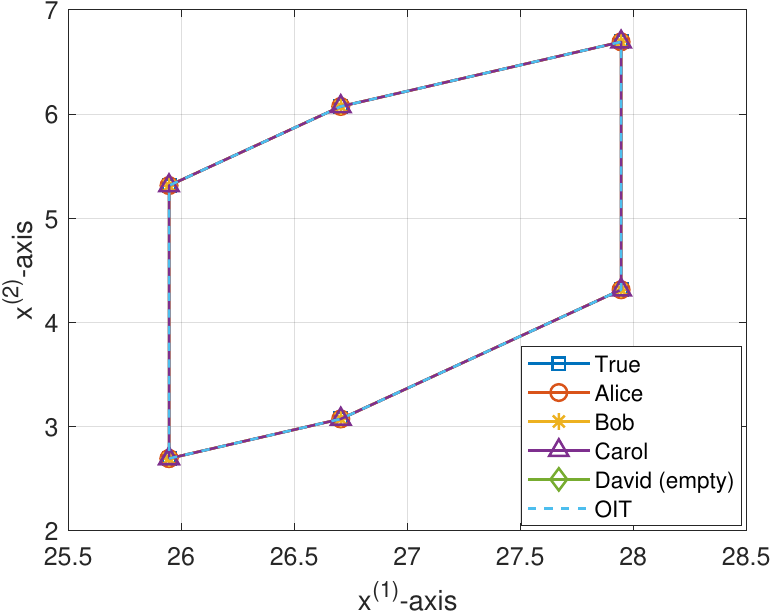}\label{fig:Illustrative example for observable system - C}}

\caption{Illustrative example for observable $(A, C)$:
(a) initial conditions and the true initial set at $k = 0$,
(b) posterior sets and the true set $\llbracket\mathbf{x}_1|y_{0:1}\rrbracket$ at $k = 1$,
(c) posterior sets, the true set $\llbracket\mathbf{x}_6|y_{0:6}\rrbracket$, and the OIT (see \defref{def:Observation-Information Tower}) at $k = 6$.
Alice and Bob know that by \propref{prop:Stability of Observable Systems}, the initial condition should include $\llbracket\mathbf{x}_0\rrbracket$ to safely make Algorithm~\ref{alg:Classical Linear Set-Membership Filtering} stable w.r.t. the initial condition:
Alice has a good knowledge of $\llbracket\mathbf{x}_0\rrbracket$ and she chooses $\llbracket\hat{\mathbf{x}}_0^{\mathrm{a}}\rrbracket = [0, 4] \times [0, 4]$ as the initial condition;
Bob does not know as much as Alice, but he can make sure that $\llbracket\mathbf{x}_0\rrbracket$ is within $[-4, 4] \times [-4, 4] = \llbracket\hat{\mathbf{x}}_0^{\mathrm{b}}\rrbracket$ [see (a)].
Carol knows that by \thmref{thm:Stability of OIT-Inspired Filtering}, Algorithm~\ref{alg:OIT-Inspired Filtering} is always stable w.r.t. the initial condition, and thus Carol chooses $\llbracket\hat{\mathbf{x}}_0^{\mathrm{c}}\rrbracket = [-1, 1] \times [-1, 1]$ as the initial condition [in fact, $\llbracket\hat{\mathbf{x}}_0^{\mathrm{c}}\rrbracket \not\supseteq \llbracket\mathbf{x}_0\rrbracket$, see (a)].
David chooses $\llbracket\hat{\mathbf{x}}_0^{\mathrm{d}}\rrbracket = [-2, 2] \times [-2, 2]$ as the initial condition while using Algorithm~\ref{alg:Classical Linear Set-Membership Filtering} [note that $\llbracket\hat{\mathbf{x}}_0^{\mathrm{d}}\rrbracket \not\supseteq \llbracket\mathbf{x}_0\rrbracket$, see (a)].
At $k = 1$, the estimates given by Alice, Bob, and Carol contain the true set, with $\llbracket\hat{\mathbf{x}}_1^{\mathrm{c}}|y_{0:1}\rrbracket \supset \llbracket\hat{\mathbf{x}}_1^{\mathrm{b}}|y_{0:1}\rrbracket \supset \llbracket\hat{\mathbf{x}}_1^{\mathrm{a}}|y_{0:1}\rrbracket \supset \llbracket\mathbf{x}_1|y_{0:1}\rrbracket$;
David provides the smallest estimate, but $\llbracket\hat{\mathbf{x}}_1^{\mathrm{d}}|y_{0:1}\rrbracket$ is inside $\llbracket\mathbf{x}_1|y_{0:1}\rrbracket$ (actually, the estimate becomes empty at $k = 2$, i.e., $\llbracket\hat{\mathbf{x}}_2^{\mathrm{d}}|y_{0:2}\rrbracket = \emptyset$).
At $k = 6$, except for $\llbracket\hat{\mathbf{x}}_6^{\mathrm{d}}|y_{0:6}\rrbracket = \emptyset$, all the posterior sets become the same (i.e., the estimation gap is $0$), and equal the OIT with $\delta = 2$ which gives a tight bound on the posterior sets.
}\label{fig:Illustrative example for observable system.}
\end{figure*}

\underline{Observable $(A, C)$:}
Consider a discretized second-order system described by~\eqref{eqn:State Equation} and~\eqref{eqn:Measurement Equation}, with parameters
\begin{multline}\label{eqn:System Paramters - Observable System}
A =
\begin{bmatrix}
1 & 1 \\
0 & 1
\end{bmatrix},\quad
B =
\begin{bmatrix}
0.5 \\
1
\end{bmatrix},\quad
C =
\begin{bmatrix}
1 & 0
\end{bmatrix},\\
\llbracket\mathbf{w}_k\rrbracket = [-1, 1],\quad
\llbracket\mathbf{v}_k\rrbracket = [-1, 1],
\end{multline}
which means $A$ is not Schur stable and $(A, C)$ is observable with $\mu = 2$.
The true initial set is $\llbracket\mathbf{x}_0\rrbracket = [1,3] \times [1,3]$.\footnote{The probability distributions of uncertain variables $\mathbf{x}_0,\mathbf{w}_{0:k},\mathbf{v}_{0:k}$ can be arbitrary for simulations.
In \secref{sec:Numerical Examples}, these uncertain variables are set to be uniformly distributed in their ranges.
The Matlab codes for all results in this paper are provided at  https://github.com/congyirui/Stability-of-Linear-SMF-2024.}

Assume that there are four designers Alice, Bob, Carol, and David to design SMFs for~\eqref{eqn:System Paramters - Observable System}, and the true initial set $\llbracket\mathbf{x}_0\rrbracket$ is unknown to them.
Alice, Bob, and David employ the classical filtering framework in Algorithm~\ref{alg:Classical Linear Set-Membership Filtering}, while Carol uses the OIT-inspired filtering framework in Algorithm~\ref{alg:OIT-Inspired Filtering}, where $P = I_2$.\footnote{Note that the exact solutions of Algorithm~\ref{alg:Classical Linear Set-Membership Filtering} and Algorithm~\ref{alg:OIT-Inspired Filtering} can be derived by employing halfspace-representation-based method with acceptable computational complexities.}
From~\figref{fig:Illustrative example for observable system.}, we can see that the difference of the initial conditions chosen by the first three designers are corrected by the measurements $y_{0:6}$, and the estimates converge to $\llbracket\mathbf{x}_6|y_{0:6}\rrbracket$;
thus, the estimation gaps of these three filters are $0$ for $k \geq 6$;
this implies the estimation gaps are bounded for $k \in \mathbb{N}_0$ which corroborates the results in \thmref{thm:Stability Criterion} and \thmref{thm:Stability of OIT-Inspired Filtering}.
In contrast, the initial condition chosen by David does not satisfy condition~(i) in \thmref{thm:Stability Criterion}, and the resulting estimate becomes empty as time elapses.

\underline{Detectable $(A, C)$:}
Consider the system with
\begin{multline}\label{eqn:System Paramters - Detectable System}
A =
\begin{bmatrix}
0.5 & 1 \\
0 & 1
\end{bmatrix},\quad
B =
\begin{bmatrix}
0.5 \\
1
\end{bmatrix},\quad
C =
\begin{bmatrix}
0 & 1
\end{bmatrix},\\
\llbracket\mathbf{w}_k\rrbracket = [-1, 1],\quad
\llbracket\mathbf{v}_k\rrbracket = [-1, 1],
\end{multline}
which implies $A$ is not Schur stable and $(A, C)$ is detectable with $\mu_o = 1$.
The true initial set is $\llbracket\mathbf{x}_0\rrbracket = [1,3] \times [1,3]$.

\begin{figure}
\centering
\includegraphics [width=0.8\columnwidth]{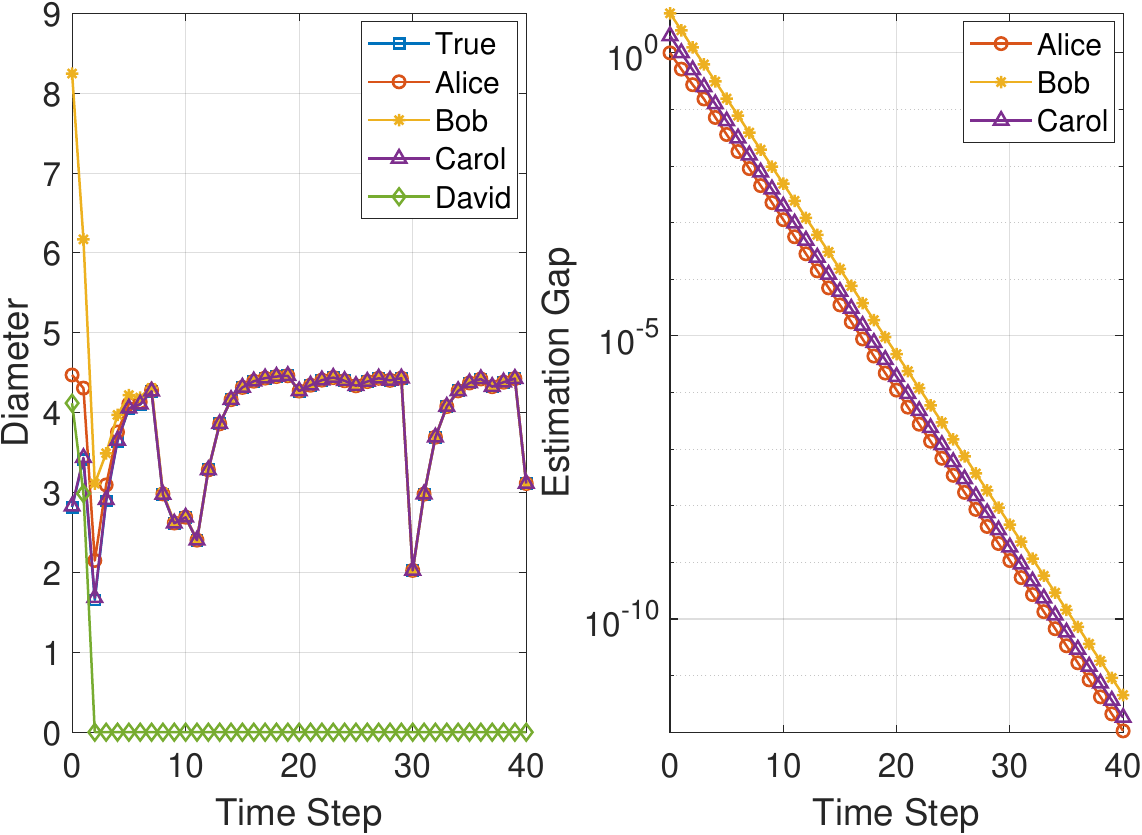}
\caption{Diameters and estimation gaps for detectable $(A, C)$ in one simulation run.
For the first three designers, $d(\llbracket\hat{\mathbf{x}}_k^{\mathrm{a}}|y_{0:k}\rrbracket)$, $d(\llbracket\hat{\mathbf{x}}_k^{\mathrm{b}}|y_{0:k}\rrbracket)$, and $d(\llbracket\hat{\mathbf{x}}_k^{\mathrm{c}}|y_{0:k}\rrbracket)$ are bounded and converge to $d(\llbracket\mathbf{x}_k|y_{0:k}\rrbracket)$;
the estimation gaps converge to $0$ exponentially fast.
For the last designer, $d(\llbracket\hat{\mathbf{x}}_k^{\mathrm{d}}|y_{0:k}\rrbracket) = 0$ from $k = 2$ (due to $\llbracket\hat{\mathbf{x}}_2^{\mathrm{d}}|y_{0:2}\rrbracket = \emptyset$).
}
\label{fig:Diameters and estimation gaps in detectable system.}
\end{figure}

The initial conditions chosen by Alice, Bob, Carol, and David are identical to those in \figref{fig:Illustrative example for observable system - A}, where Algorithm~\ref{alg:OIT-Inspired Filtering} is with $P = \left[\begin{smallmatrix} 0 & 1\\ -1 & 0\end{smallmatrix}\right]$.
\figref{fig:Diameters and estimation gaps in detectable system.} corroborates the theoretical results in \thmref{thm:Stability Criterion} and \thmref{thm:Stability of OIT-Inspired Filtering}, where the estimation gaps corresponding to non-empty estimates are bounded;
more specifically, they converge to $0$ exponentially fast.

\subsection{The Stable and Fast Constrained Zonotopic SMF}\label{sec:The Fast Constrained Zonotopic SMF}

Firstly, we use Monte Carlo simulation to test the OIT-CZ SMF in \secref{sec:Interval hull of the estimate};
then, we show the time efficiency of the OIT-CZ SMF in \secref{sec:Time Efficiency}.

\subsubsection{Interval hull of the estimate}\label{sec:Interval hull of the estimate}

In this part, we employ the OIT-CZ SMF to derive the interval hull of the estimate.
%, where the prior refinement technique in \rekref{rek:Prior Refinement} is used.

\begin{figure}
\centering
\subfigure[]{\includegraphics [width=0.75\columnwidth]{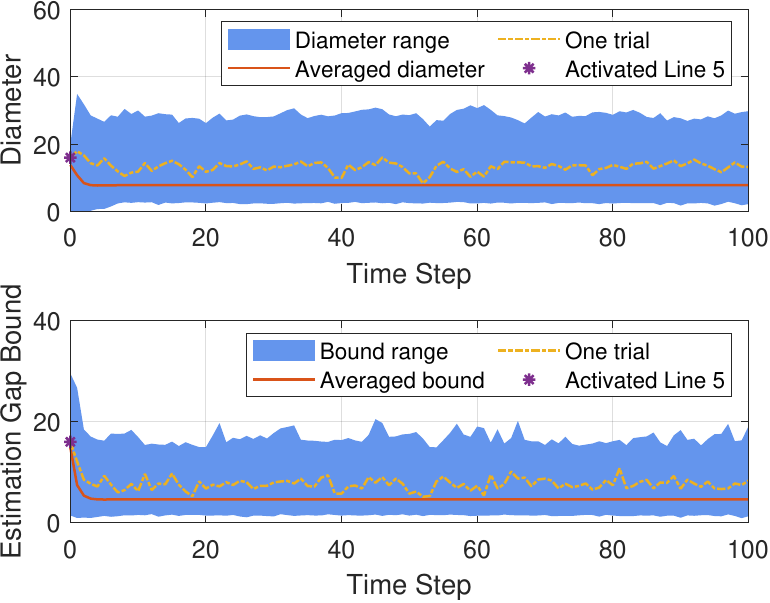}\label{fig:Diameters and bound on estimation gaps of OIT-CZ SMF for observable systems}}\\
\subfigure[]{\includegraphics [width=0.75\columnwidth]{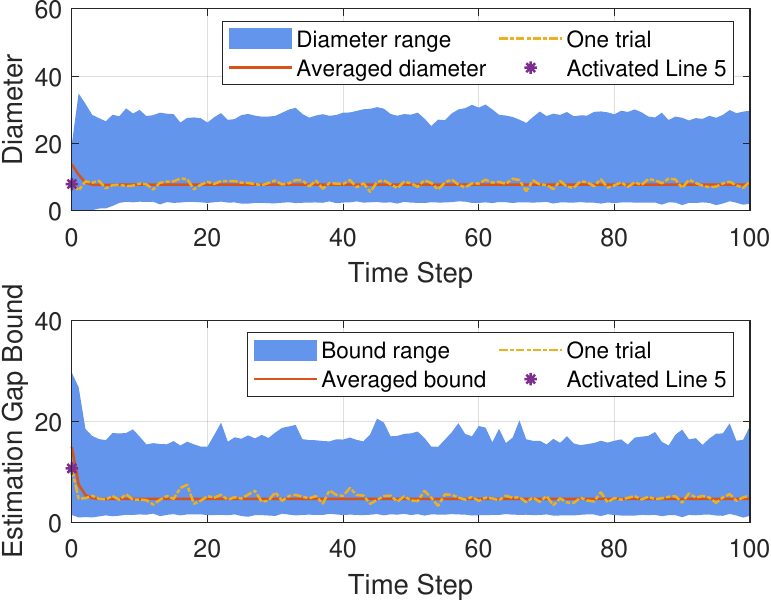}\label{fig:Diameters and bound on estimation gaps of OIT-CZ SMF for detectable systems}}
\caption{The diameter of the estimate $\mathcal{Z}_k$ and a bound on the estimation gap $d_k^{\mathrm{g}}(\mathcal{Z}_k)$ for OIT-CZ SMF, in the sense of $\infty$-norm.
(a) observable $(A, C)$,
(b) detectable $(A, C)$.
The bound on $d_k^{\mathrm{g}}(\mathcal{Z}_k)$ is $d_{\mathrm{H}}^{\infty}(\mathcal{Z}_k, \{x_k\}) := \sup_{\hat{x}_k \in \mathcal{Z}_k} \|\hat{x}_k - x_k\|_{\infty}$, which can be easily calculated by $\sup_{\hat{x}_k \in \overline{\mathrm{IH}}(\mathcal{Z}_k)} \|\hat{x}_k - x_k\|_{\infty}$.
}
\label{fig:Diameters and bound on estimation gaps of OIT-CZ SMF.}
\end{figure}

First, consider the randomly generated $(A, B, C)$ (by using \verb"drss" function in MATLAB) with observable $(A, C)$.
The process and measurement noises are with $\llbracket\mathbf{w}_k\rrbracket = [-1,1]^p$ and $\llbracket\mathbf{v}_k\rrbracket = [-1,1]^m$.
The true initial set is $\llbracket\mathbf{x}_0\rrbracket = [-10, 10]^n$, and the initial condition is $\llbracket\hat{\mathbf{x}}_0\rrbracket = [-10, 10]^n \oplus \{c\}$, where $c$ is randomly generated in $[-1, 1]^n$ for testing Algorithm~\ref{alg:OIT-CZ SMF}.
In the simulations, we set $n = 10$, $p = m \in \{5,\ldots,10\}$, and $\bar{\delta} = n - \mathrm{rank}(C) + 3 > \mu - 1$ (in Algorithm~\ref{alg:OIT-CZ SMF});
for each $p = m$, the simulations are conducted $1000$ times.
The results are shown in \figref{fig:Diameters and bound on estimation gaps of OIT-CZ SMF for observable systems}, where one of the simulation runs is highlighted by the (yellow) dash-dotted lines with the (purple) stars representing: $\mathcal{Z}_k = \emptyset$ holds in Line~\ref{line:OIT-CZ SMF - Estimates Resetting - Start} of Algorithm~\ref{alg:OIT-CZ SMF} and Line~\ref{line:OIT-CZ SMF - Estimates Resetting} derives a non-empty $\mathcal{Z}_k$ by resetting $\mathcal{Z}_0^-$.
We can see that the proposed OIT-CZ SMF is stable and uniformly bounded, which corroborates the theoretical results in \thmref{thm:Stability of OIT-CZ SMF}.

Second, consider the randomly generated $(A, B, C)$ (by using \verb"drss" function in MATLAB) with detectable $(A, C)$.
%Second, consider the detectable systems whose observable subsystem, or $(\tilde{A}_o, \tilde{B}_o, \tilde{C}_o)$, is generated by \verb"drss".
%
For the unobservable subsystem, $\tilde{A}_{\bar{o}}$ is a randomly created matrix with $\rho(\tilde{A}_{\bar{o}}) \leq 0.5$; each element in $\tilde{A}_{21}$ and $\tilde{B}_{\bar{o}}$ is randomly selected from $[0,~1]$.
The transformation matrix $P$ is a randomly derived orthogonal matrix such that by~\eqref{eqn:Observability Decomposition}: $A = P^{-1} \tilde{A} P$, $B = P^{-1} \tilde{B}$, $C = [\tilde{C}_o~0] P$ are finally obtained.
The sets $\llbracket\mathbf{w}_k\rrbracket$, $\llbracket\mathbf{v}_k\rrbracket$, $\llbracket\mathbf{x}_0\rrbracket$, and $\llbracket\hat{\mathbf{x}}_0\rrbracket$ the same as those in the simulations for observable $(A, C)$.
Also, we set $n = 10$, $n_o = p = m \in \{7,8,9\}$, $\bar{\delta} = n_o - \mathrm{rank}(\tilde{C}_o) + 3 > \mu_o - 1$, and $\varepsilon = 0.001$.
for each $n_o = p = m$, the simulations are conducted $1000$ times.
The results are shown in \figref{fig:Diameters and bound on estimation gaps of OIT-CZ SMF for detectable systems}, which validates the results in \thmref{thm:Stability of OIT-CZ SMF}.

\subsubsection{Time Efficiency}\label{sec:Time Efficiency}

In this part, we set\footnote{If $\llbracket\hat{\mathbf{x}}_0\rrbracket \neq \llbracket\mathbf{x}_0\rrbracket$, the classical algorithm can return an error due to the ill-posedness.} $\llbracket\hat{\mathbf{x}}_0\rrbracket = \llbracket\mathbf{x}_0\rrbracket$ and compare the computation time (w.r.t. the constrained zonotopic description) of the proposed OIT-CZ SMF and two classical constrained zonotopic SMFs;
these two classical SMFs are with the reduction methods \emph{girard} and \emph{combastel} in CORA 2024~\cite{AlthoffM2024Manual}, respectively.
Consider the randomly generated observable systems same as those in \secref{sec:Interval hull of the estimate}, but with $n = p = m \in \{10, 20, 30\}$.
We also set $\bar{\delta} = n - \mathrm{rank}(C) + 3$ in Algorithm~\ref{alg:OIT-CZ SMF}.
For these two classical SMFs, we set $n_c = 1$ and the degrees-of-freedom order~\cite{ScottJ2016} $o_d = (n_g - n_c) / n = 1$.
The simulations are conducted for $100$ runs over the time window $[0, 100]$ by using Matlab 2019b on a laptop with Intel Core i7-11800H@2.30GHz CPU, and the averaged computation time (per time step) is shown in \tabref{tab:Computation Time for the Constrained Zonotopic Description}.
Note that the diameters in the sense of $\infty$-norm for SMF (girard), SMF (combastel), and OIT-CZ SMF are respectively:
$14.8902$, $14.8902$, and $3.6878$ (when $n = 10$);
$29.3098$, $29.1371$, and $4.7596$ (when $n = 20$);
$37.2301$, $37.2301$, and $5.0911$ (when $n = 30$).
These results show that the OIT-CZ SMF achieves significantly higher accuracy with several orders of magnitude reduction in computation time, as compared to the classical SMFs.

\begin{table}[tb]
\caption{Computation Time (per Time Step) for the Constrained Zonotopic Description\label{tab:Computation Time for the Constrained Zonotopic Description}}
\centering
\begin{small}
\begin{tabular}{lccc}
\toprule
~& $n = 10$ & $n = 20$ & $n = 30$\\
\midrule
SMF (girard) & $0.5822$s & $3.8103$s & $11.9702$s\\
SMF (combastel) & $0.5809$s & $3.8014$s & $11.9546$s\\
OIT-CZ SMF & $0.0004$s & $0.0006$s & $0.0011$s\\
\bottomrule
\end{tabular}
\end{small}
\end{table}

\section{Conclusion}\label{sec:Conclusion}

In this paper, the stability of the SMFs w.r.t. the initial condition has been studied for linear time-invariant systems.
Specifically, the stability is given by the well-posedness and the bounded estimation gap.
Based on our proposed OIT, we have analyzed the stability of the classical linear SMFing framework, where an explicit sufficient condition has been given.
Then, we have provided a good necessary condition for the bounded estimation gap, which is very close to the sufficient condition.
To avoid unstable filter design resulted from improper initial conditions, the OIT-inspired filtering framework has been established to guaranteed the stability.
Under this new framework, we have developed the OIT-CZ SMF with guaranteed stability, uniform boundedness, high efficiency, and good accuracy.

\appendix

\section{Proof of \propref{prop:Set-Intersection-Based Outer Bound}}\label{apx:Proof of prop:Set-Intersection-Based Outer Bound}

We prove \propref{prop:Set-Intersection-Based Outer Bound} by induction.

\emph{Base case:} For $k = 0$, it follows from~\eqref{eqn:Update - Optimal Linear Set-Membership Filter},~\eqref{eqn:Observation-Information Set}, and~\eqref{eqn:State-Evolution Set} that $\llbracket\hat{\mathbf{x}}_k|y_{0:k}\rrbracket = \mathcal{O}_{0,0} \bigcap \mathcal{E}_0$, i.e.,~\eqref{eqn:Set-Intersection-Based Outer Bound} holds for $k = 0$.

\emph{Inductive step:} Assume~\eqref{eqn:Set-Intersection-Based Outer Bound} holds for any $k = k' \in \mathbb{N}_0$.
For $k = k'+1$, the prior set is derived by~\eqref{eqn:Prediction - Optimal Linear Set-Membership Filter} that
\begin{equation}\label{eqninpf:prop:Set-Intersection-Based Outer Bound - Prediction Step}
\begin{split}
&\llbracket\hat{\mathbf{x}}_{k'+1}| y_{0:k'}\rrbracket = A \llbracket\mathbf{x}_{k'}|y_{0:k'}\rrbracket \oplus B \llbracket\mathbf{w}_{k'}\rrbracket\\
&\subseteq A \bigg(\bigcap_{i=0}^{k'} \mathcal{O}_{k',i} \bigcap \mathcal{E}_{k'}\bigg) \oplus B \llbracket\mathbf{w}_{k'}\rrbracket\\
%&\stackrel{(a)}{=} \bigg(\bigcap_{i=0}^{k'} A \mathcal{O}_{k',i} \bigcap A \mathcal{E}_{k'}\bigg) + B \llbracket\mathbf{w}_{k'}\rrbracket\\
&\stackrel{(a)}{\subseteq} \bigcap_{i=0}^{k'} \left(A \mathcal{O}_{k',i} \oplus B \llbracket\mathbf{w}_{k'}\rrbracket\right) \bigcap \left(A \mathcal{E}_{k'} \oplus B \llbracket\mathbf{w}_{k'}\rrbracket\right)\\
&\stackrel{(b)}{=} \bigcap_{i=0}^{k'} \mathcal{O}_{k'+1,i} \bigcap \mathcal{E}_{k'+1},
\end{split}
\end{equation}
where
$(a)$ is established by~\eqref{eqn:Set Intersection Under Minkowski Sum}, and
$(b)$ is from the following two equations:
\begin{align}
A\mathcal{O}_{k',i} \oplus B \llbracket\mathbf{w}_{k'}\rrbracket &\stackrel{(c)}{=} \mathcal{O}_{k'+1,i},\label{eqninpf:prop:Set-Intersection-Based Outer Bound - 1}\\
A \mathcal{E}_{k'} \oplus B \llbracket\mathbf{w}_{k'}\rrbracket &\stackrel{(d)}{=} \mathcal{E}_{k'+1},\label{eqninpf:prop:Set-Intersection-Based Outer Bound - 2}
\end{align}
where $(c)$ and $(d)$ follow from $h^*\big(\sum_{i \in \mathcal{I}} \mathcal{S}_i\big) = \sum_{i \in \mathcal{I}} h^*(\mathcal{S}_i)$ for any linear map $h^*$.
By~\eqref{eqninpf:prop:Set-Intersection-Based Outer Bound - Prediction Step} and~\eqref{eqn:Update - Optimal Linear Set-Membership Filter}, the posterior set is outer bounded by
\begin{equation*}%\label{eqninpf:prop:Set-Intersection-Based Outer Bound - Update Step}
\begin{split}
&\llbracket\hat{\mathbf{x}}_{k'+1}| y_{0:k'+1}\rrbracket = \mathcal{X}_{k'+1}(C, y_{k'+1}, \llbracket\mathbf{v}_{k'+1}\rrbracket) \bigcap \llbracket\hat{\mathbf{x}}_{k'+1}| y_{0:k'}\rrbracket\\
&= \mathcal{O}_{k'+1,k'+1} \bigcap \llbracket\mathbf{x}_{k'+1}| y_{0:k'}\rrbracket \subseteq \bigcap_{i=0}^{k'+1} \mathcal{O}_{k'+1,i} \bigcap \mathcal{E}_{k'+1},
\end{split}
\end{equation*}
which implies~\eqref{eqn:Set-Intersection-Based Outer Bound} holds for $k = k'+1$.
Thus, \propref{prop:Set-Intersection-Based Outer Bound} is proven by induction.\hfill$\square$

\section{Proof of \thmref{thm:Uniform Boundedness of OIT}}\label{apx:Proof of thm:Uniform Boundedness of OIT}

To start with, we provide two lemmas as follows.

\begin{lemma}\textbf{\emph{(Uniform Boundedness Invariance Under Linear Map)}}\label{lem:Uniform Boundedness Invariance Under Linear Map}
$\forall l \in \mathcal{I}$, let the set $\mathcal{T}_{l,k} \subset \mathbb{R}^n$ be uniformly bounded w.r.t. $k \in \mathbb{N}_0$, and the linear map $h_l^*$ be independent of $k$.
Then, $\sum_{l \in \mathcal{I}} h_l^* (\mathcal{T}_{l,k})$ is uniformly bounded w.r.t. $k \in \mathbb{N}_0$.
\end{lemma}

\begin{proof}
Since $\forall l \in \mathcal{I}$, $\mathcal{T}_{l,k}$ is uniformly bounded w.r.t. $k \in \mathbb{N}_0$, there exists a $\bar{d}_l$ such that $d(\mathcal{T}_{l,k}) \leq \bar{d}_l,\quad \forall k \in \mathbb{N}_0$.
Thus, $\forall l \in \mathcal{I}$, the following holds for all $k \in \mathbb{N}_0$
\begin{equation}\label{eqninpf:lem:Uniform Boundedness Invariance Under Linear Map - 2}
\begin{split}
d(h_l^* (\mathcal{T}_{l,k})) &= \sup_{t, t' \in \mathcal{T}_{l,k}} \|h_l^*(t) - h_l^*(t')\|\\
&\leq \|h_l^*\| \sup_{t, t' \in \mathcal{T}_{l,k}} \|t - t'\| \leq \|h_l^*\| \bar{d}_l,
\end{split}
\end{equation}
where $\|h_l^*\|$ is the operator norm of $h_l^*$.
Now, we have $k \in \mathbb{N}_0$,
\begin{equation}\label{eqninpf:lem:Uniform Boundedness Invariance Under Linear Map - 3}
\!\!\!d\bigg(\sum_{l \in \mathcal{I}} h_l^* (\mathcal{T}_{l,k})\bigg) \stackrel{(e)}{\leq} \sum_{l \in \mathcal{I}} d(h_l^* (\mathcal{T}_{l,k})) \leq \sum_{l \in \mathcal{I}} \|h_l^*\| \bar{d}_l,
\end{equation}
where inequality $(e)$ can be easily obtained by using the following triangle inequality:
for the sets $\mathcal{R}_l$ ($l \in \mathcal{I}$),
\begin{multline}\label{eqninpf:lem:Uniform Boundedness Invariance Under Linear Map - Triangle Inequality}
d\bigg(\sum_{l \in \mathcal{I}} \mathcal{R}_l\bigg) = \sup_{r_l,r'_l \in \mathcal{R}_l, l \in \mathcal{I}} \|\sum_{l \in \mathcal{I}} (r_l - r'_l)\| \\\leq \sum_{l \in \mathcal{I}} \sup_{r_l,r'_l} \|r_l - r'_l\| = \sum_{l \in \mathcal{I}} d(\mathcal{R}_l).
\end{multline}
By~\eqref{eqninpf:lem:Uniform Boundedness Invariance Under Linear Map - 3}, $\sum_{l \in \mathcal{I}} h_l^* (\mathcal{T}_{l,k})$ is uniformly bounded.
\end{proof}

\begin{lemma}[Uniformly Bounded Intersection]\label{lem:Uniformly Bounded Intersection}
$\forall j \in \{0,\ldots,\delta\}$ ($\delta \geq 0$), let $\mathcal{S}_{j,k} \subset \mathbb{R}^n$ be any uniformly bounded set w.r.t. $k \in \mathbb{N}_0$.
Then,
\begin{equation}\label{eqn:Uniformly Bounded Intersection - 1}
\bigcap_{j=0}^{\delta} \big[\mathrm{ran}(\bar{D}_j) \oplus \mathcal{S}_{j,k}\big]
\end{equation}
is uniformly bounded w.r.t. $k \in \mathbb{N}_0$ [recall that $\mathrm{ran}(\bar{D}_j)$ returns the range space of matrix $\bar{D}_j \in \mathbb{R}^{n\times (n-m)}$], if
\begin{equation}\label{eqn:Uniformly Bounded Intersection - 2}
\mathrm{rank}\left(D^{\mathrm{T}}\right) = n,
\end{equation}
where $D = [D_0 \ldots D_{\delta}]$, and $D_j \in \mathbb{R}^{n\times m}$ ($j \in \{0,\ldots,\delta\}$) satisfies $\ker(D_j^{\mathrm{T}}) = \mathrm{ran}(\bar{D}_j)$.
\end{lemma}

\begin{proof}
This proof has three steps.
In the first step, we prove the following equality holds:
\begin{equation}\label{eqninpf:lem:Uniformly Bounded Intersection - 1}
\!\!\!\!\bigcap_{j=0}^{\delta} \big[\mathrm{ran}(\bar{D}_j) \oplus \mathcal{S}_{j,k}\big] = \!\bigcup_{s_k \in \mathcal{S}_k} \bigcap_{j=0}^{\delta} \big[\mathrm{ran}(\bar{D}_j) \oplus \{s_{j,k}\}\big],
\end{equation}
where $s_k = [s_{0,k}^{\mathrm{T}} \ldots s_{\delta,k}^{\mathrm{T}}]^{\mathrm{T}}$ and $\mathcal{S}_k := \mathcal{S}_{0,k}\times \cdots \times\mathcal{S}_{\delta,k}$ and $\times$ stands for the Cartesian product.
In the second step, we analyze $\bigcap_{j=0}^{\delta} \big[\mathrm{ran}(\bar{D}_j) \oplus \{s_{j,k}\}\big]$ from the perspective of solving linear equations.
Afterwards, we complete this proof in the third step.

\textit{Step~1:}
Since $\mathrm{ran}(\bar{D}_j) \oplus \mathcal{S}_{j,k} = \bigcup_{s_{j,k} \in \mathcal{S}_{j,k}} [\mathrm{ran}(\bar{D}_j) \oplus \{s_{j,k}\}] =: \bigcup_{s_{j,k} \in \mathcal{S}_{j,k}} \mathcal{T}_{j, s_{j,k}}$, we are readily to use distributive law of sets to get~\eqref{eqninpf:lem:Uniformly Bounded Intersection - 1}.
Specifically, we have\footnote{The readers can use mathematical induction to derive a rigorous proof. Due to the page limit, we omit it here.}
\begin{equation}\label{eqninpf:lem:Uniformly Bounded Intersection - 1 - Rewritten}
\bigcap_{j=0}^{\delta} \bigcup_{s_{j,k} \in \mathcal{S}_{j,k}} \mathcal{T}_{j, s_{j,k}} = \bigcup_{(s_{0,k},\ldots,s_{\delta,k})\in \mathcal{S}_k} \bigcap_{j=0}^{\delta} \mathcal{T}_{j, s_{j,k}},
\end{equation}
which means~\eqref{eqninpf:lem:Uniformly Bounded Intersection - 1} holds.

\textit{Step~2:}
$\forall s_k \in \mathcal{S}_k$, with $\ker(D_j^{\mathrm{T}}) = \mathrm{ran}(\bar{D}_j)$ we have
\begin{equation*}
\mathrm{ran}(\bar{D}_j) \oplus \{s_{j,k}\} = \left\{x\colon D_j^{\mathrm{T}} x = D_j^{\mathrm{T}} s_{j,k}\right\}, \quad j \in \{0,\ldots,{\delta}\},
\end{equation*}
which is the solution space of the linear equation $D_j^{\mathrm{T}} x = D_j^{\mathrm{T}} s_{j,k}$.
Thus, we get
\begin{equation}\label{eqninpf:lem:Uniformly Bounded Intersection - 3}
\bigcap_{j=0}^{\delta} \big[\mathrm{ran}(\bar{D}_j) \oplus \{s_{j,k}\}\big] = \left\{x\colon D^{\mathrm{T}} x = \bar{s}_k\right\},
\end{equation}
where $\bar{s}_k = [s_{0,k}^{\mathrm{T}}D_0 \ldots s_{\delta,k}^{\mathrm{T}}D_{\delta}]^{\mathrm{T}}$.
With~\eqref{eqn:Uniformly Bounded Intersection - 2}, we know that $D^{\mathrm{T}} x = \bar{s}_k$ has either a unique solution $(D^{\mathrm{T}})^+ \bar{s}_k$ or no solution, i.e., $\bigcap_{j=0}^{\delta} \big[\mathrm{ran}(\bar{D}_j) \oplus \{s_{j,k}\}\big]$ at most contains one element.
Hence, we have
\begin{equation}\label{eqninpf:lem:Uniformly Bounded Intersection - 3 - Rewritten}
\bigcap_{j=0}^{\delta} \big[\mathrm{ran}(\bar{D}_j) \oplus \{s_{j,k}\}\big] =
\begin{cases}
\left\{\left(D^{\mathrm{T}}\right)^+ \bar{s}_k\right\} & s_k \in \mathcal{S}_k^{\neq \emptyset},\\
\emptyset & \mathrm{otherwise},
\end{cases}
\end{equation}
where $\mathcal{S}_k^{\neq \emptyset} = \left\{s_k \in \mathcal{S}_k \colon \left\{x\colon D^{\mathrm{T}} x = \bar{s}_k\right\} \neq \emptyset\right\}$.

\textit{Step~3:}
With~\eqref{eqninpf:lem:Uniformly Bounded Intersection - 3 - Rewritten}, we can rewrite~\eqref{eqninpf:lem:Uniformly Bounded Intersection - 1} as follows
\begin{equation}\label{eqninpf:lem:Uniformly Bounded Intersection - 5}
\begin{split}
\bigcap_{j=0}^{\delta} \big[\mathrm{ran}(\bar{D}_j) \oplus \mathcal{S}_{j,k}\big]
&= \bigcup_{s_k\in \mathcal{S}_k^{\neq \emptyset}} \left\{\left(D^{\mathrm{T}}\right)^+ \bar{s}_k\right\}\\
&= \bigcup_{s_k\in \mathcal{S}_k^{\neq \emptyset}} \left\{H s_k\right\}
= H \mathcal{S}_k^{\neq \emptyset},
\end{split}
\end{equation}
where $H = \left(D^{\mathrm{T}}\right)^+ \mathrm{diag}\{D_0^{\mathrm{T}},\ldots,D_{\delta}^{\mathrm{T}}\}$.
As $\mathcal{S}_k^{\neq \emptyset} \subseteq \mathcal{S}_k$ is uniformly bounded\footnote{This is because $d(\mathcal{S}_k^{\neq \emptyset}) \leq d(\mathcal{S}_k) \leq \sqrt{\sum_{i=0}^{\delta} d^2(\mathcal{S}_{i,k})}$.} and $H$ is a linear map independent of $k$, $\bigcap_{j=0}^{\delta} \big[\mathrm{ran}(\bar{D}_j) \oplus \mathcal{S}_{j,k}\big]$ is uniformly bounded.
\end{proof}

Now, we prove \thmref{thm:Uniform Boundedness of OIT}.
With~\eqref{eqn:Specific Form of the Solution Set X}, the observation-information set in~\eqref{eqn:Observation-Information Set} can be rewritten as
\begin{multline}\label{eqn:Observation-Information Set - Rewritten}
\mathcal{O}_{k,i} = A^{k-i} \ker(C) \oplus A^{k-i} C^+ (\{y_i\} \oplus \llbracket-\mathbf{v}_i\rrbracket)\\ \oplus \sum_{r=i}^{k-1} A^{k-1-r} B \llbracket\mathbf{w}_r\rrbracket.
\end{multline}
Thus, with~\eqref{eqn:Observation-Information Set - Rewritten}, $j = i - k + \delta$, and $l = r - i + 1$, we can rewrite the OIT in \defref{def:Observation-Information Tower} as
\begin{equation}\label{eqn:Reduced Observation-Information Tower - Detailed Form}
\bigcap_{i=k-\delta}^k \mathcal{O}_{k,i} = \bigcap_{j=0}^{\delta} \left[A^{\delta-j} \ker(C) \oplus \mathcal{S}_{j,k}\right],
\end{equation}
where
\begin{multline}\label{eqninpf:thm:Uniform Boundedness of ROIT - Sjk}
\mathcal{S}_{j,k} = A^{\delta-j} C^+ (\{y_{k-\delta+j}\} \oplus \llbracket-\mathbf{v}_{k-\delta+j}\rrbracket) \\ \oplus \sum_{l=1}^{\delta-j} A^{\delta-j-l} B \llbracket\mathbf{w}_{l+k-\delta+j-1}\rrbracket.
\end{multline}
Let $\bar{C} \in \mathbb{R}^{(n - m) \times n}$ such that $\ker(C) = \mathrm{ran}(\bar{C}^{\mathrm{T}})$.
By $A^{\delta-j} \ker(C) = A^{\delta-j} \mathrm{ran}(\bar{C}^{\mathrm{T}}) = \mathrm{ran}(A^{\delta-j} \bar{C}^{\mathrm{T}}) =: \mathrm{ran}(\bar{D}_j)$,~\eqref{eqn:Reduced Observation-Information Tower - Detailed Form} can be rewritten as
\begin{equation}\label{eqn:Reduced Observation-Information Tower - Detailed Form - Rewritten}
\bigcap_{i=k-\delta}^k \mathcal{O}_{k,i} = \bigcap_{j=0}^{\delta} \left[\mathrm{ran}(\bar{D}_j) \oplus \mathcal{S}_{j,k}\right].
\end{equation}
Since $d(\llbracket\mathbf{w}_k\rrbracket) \leq d_w$ and $d(\llbracket\mathbf{v}_k\rrbracket) \leq d_v$, we know that $\{y_{k-\delta+j}\} \oplus \llbracket-\mathbf{v}_{k-\delta+j}\rrbracket$ and $\llbracket\mathbf{w}_{l+k-\delta+j-1}\rrbracket$ ($l \in \{1,\ldots,\delta-j\}$) are uniformly bounded w.r.t. $k \in \mathbb{N}_0$.
Hence, $\forall j \in \{0,\ldots,\delta\}$, by \lemref{lem:Uniform Boundedness Invariance Under Linear Map} $\mathcal{S}_{j,k}$ is uniformly bounded w.r.t. $k \in \mathbb{N}_0$.
Now, the precondition of \lemref{lem:Uniformly Bounded Intersection} is satisfied for~\eqref{eqn:Reduced Observation-Information Tower - Detailed Form - Rewritten}.
To guarantee the uniform boundedness of~\eqref{eqn:Reduced Observation-Information Tower - Detailed Form - Rewritten}, we only need to prove condition~\eqref{eqn:Uniformly Bounded Intersection - 2} with $\ker(D_j^{\mathrm{T}}) = \mathrm{ran}(\bar{D}_j)$ in \lemref{lem:Uniformly Bounded Intersection} holds, where $D_j^{\mathrm{T}} := C A^{j-\delta}$.
The proof has two steps.

\textit{Step~1:}
Prove $\ker(D_j^{\mathrm{T}}) = \mathrm{ran}(\bar{D}_j)$.
To be more specific, for non-singular $A$, we show that
\begin{equation}\label{eqninpf:thm:Uniform Boundedness of ROIT - Important Equality}
\ker(D_j^{\mathrm{T}}) = \ker(C A^{j-\delta}) = \mathrm{ran}(A^{\delta-j} \bar{C}^{\mathrm{T}}) = \mathrm{ran}(\bar{D}_j),
\end{equation}
as follows:
Firstly, we have
\begin{equation}\label{eqninpf:thm:Uniform Boundedness of ROIT - 2}
D_j^{\mathrm{T}} \bar{D}_j = C A^{j-\delta} A^{\delta-j} \bar{C}^{\mathrm{T}} = C \bar{C}^{\mathrm{T}} = 0,
\end{equation}
which implies $\mathrm{ran}(\bar{D}_j) \subseteq \ker(D_j^{\mathrm{T}})$.
Secondly, we check the rank of $\bar{D}_j$ as follows
\begin{multline}\label{eqninpf:thm:Uniform Boundedness of ROIT - Rank Check}
\mathrm{rank}(\bar{D}_j) = \mathrm{rank}(A^{\delta-j} \bar{C}^{\mathrm{T}}) = \mathrm{rank}(\bar{C}^{\mathrm{T}}) = n - \mathrm{rank}(C) \\= n - \mathrm{rank}(C A^{j-\delta}) = n - \mathrm{rank}(D_j^{\mathrm{T}}),
\end{multline}
which combined with~\eqref{eqninpf:thm:Uniform Boundedness of ROIT - 2} gives~\eqref{eqninpf:thm:Uniform Boundedness of ROIT - Important Equality}.

\textit{Step~2:}
Prove~\eqref{eqn:Uniformly Bounded Intersection - 2}.
Since $D_j^{\mathrm{T}} = C A^{j-\delta}$, we have $D^{\mathrm{T}} = O_{\delta}$, and the left-hand side of~\eqref{eqn:Uniformly Bounded Intersection - 2} equals
\begin{equation}\label{eqninpf:thm:Uniform Boundedness of ROIT - 3}
\mathrm{rank}\left(
O_{\delta}
\right) =
\mathrm{rank}\left(
\begin{bmatrix}
C \\
\vdots \\
C A^{\delta}
\end{bmatrix}
A^{-\delta} \right) \stackrel{(f)}{=} n,
\end{equation}
where $(f)$ is established by observable $(A, C)$ and $\delta \geq \mu - 1$ in \thmref{thm:Uniform Boundedness of OIT}.

Therefore, by \lemref{lem:Uniformly Bounded Intersection}, $\bigcap_{i=k-\delta}^k \mathcal{O}_{k,i}$ in~\eqref{eqn:Reduced Observation-Information Tower - Detailed Form - Rewritten} is uniformly bounded w.r.t. $k \geq \delta \geq \mu - 1$.

Next, we prove equation~\eqref{eqn:An Upper Bound on the Diameter of the OIT}.
With~\eqref{eqninpf:lem:Uniformly Bounded Intersection - 5} and $\mathcal{S}_k^{\neq \emptyset} \subseteq \mathcal{S}_k$, we can rewrite~\eqref{eqn:Reduced Observation-Information Tower - Detailed Form} as
\begin{equation}\label{eqninpf:thm:Uniform Boundedness of ROIT - 4}
\bigcap_{i=k-\delta}^k \mathcal{O}_{k,i} \subseteq O_{\delta}^+ \prod_{j=0}^{\delta} C A^{j-\delta} \mathcal{S}_{j,k},
\end{equation}
where $\prod$ is the Cartesian product of the sequence of $C A^{j-\delta} \mathcal{S}_{j,k}$ with the following form
\begin{equation*}
C C^+ (\{y_{k-\delta+j}\} \oplus \llbracket-\mathbf{v}_{k-\delta+j}\rrbracket) \oplus \sum_{l=1}^{\delta-j} C A^{-l} B \llbracket\mathbf{w}_{l+k-\delta+j-1}\rrbracket.
\end{equation*}
Thus, noticing that $\|C C^+\| = 1$, $\|O_{\delta}^+\| = 1/\sigma_{\min}(O_{\delta}^+)$, and $d(\{y_{k-\delta+j}\}) = 0$, we can get~\eqref{eqn:An Upper Bound on the Diameter of the OIT} from~\eqref{eqninpf:thm:Uniform Boundedness of ROIT - 4}.\hfill $\square$

\section{Proof of \thmref{thm:Stability Criterion}}\label{apx:Proof of thm:Stability Criterion}

Firstly, we consider the stability for observable $(A, C)$, and a sufficient condition is provided in \propref{prop:Stability of Observable Systems}.

\begin{proposition}[Stability of Observable Systems]\label{prop:Stability of Observable Systems}
For observable $(A, C)$ and bounded $\llbracket\hat{\mathbf{x}}_0\rrbracket \supseteq \llbracket\mathbf{x}_0\rrbracket$, the classical SMFing framework in Algorithm~\ref{alg:Classical Linear Set-Membership Filtering} is stable and $\llbracket\hat{\mathbf{x}}_k|y_{0:k}\rrbracket$ is uniformly bounded w.r.t. $k \in \mathbb{N}_0$.
\end{proposition}

\begin{proof}
We prove the well-posedness, the uniform boundedness of the estimate, and the boundedness of the estimation gap as follows.

\emph{\underline{Well-posedness:}}
Since $\llbracket\hat{\mathbf{x}}_0\rrbracket \supseteq \llbracket\mathbf{x}_0\rrbracket$, with~\eqref{eqn:Update - Optimal Linear Set-Membership Filter}, we have
\begin{multline}\label{eqninpf:prop:Stability of Observable Systems - 1}
\llbracket\hat{\mathbf{x}}_0|y_0\rrbracket = \mathcal{X}_0(C, y_0, \llbracket\mathbf{v}_0\rrbracket) \bigcap \llbracket\hat{\mathbf{x}}_0\rrbracket
\\\supseteq \mathcal{X}_0(C, y_0, \llbracket\mathbf{v}_0\rrbracket) \bigcap \llbracket\mathbf{x}_0\rrbracket
= \llbracket\mathbf{x}_0|y_0\rrbracket \stackrel{(g)}{\neq} \emptyset,
\end{multline}
i.e., $\llbracket\hat{\mathbf{x}}_0|y_0\rrbracket \neq \emptyset$, where $(g)$ follows from the fact that $y_0$ is generated by~\eqref{eqn:Measurement Equation} with at least one possible $x_0 \in \llbracket\mathbf{x}_0\rrbracket$.
With~\eqref{eqn:Prediction - Optimal Linear Set-Membership Filter}, we have $\llbracket\hat{\mathbf{x}}_1|y_0\rrbracket \supseteq \llbracket\mathbf{x}_1|y_0\rrbracket$, as $\llbracket\hat{\mathbf{x}}_0|y_0\rrbracket \supseteq \llbracket\mathbf{x}_0|y_0\rrbracket$.
Similarly to~\eqref{eqninpf:prop:Stability of Observable Systems - 1}, by~\eqref{eqn:Update - Optimal Linear Set-Membership Filter}, $\llbracket\hat{\mathbf{x}}_1|y_{0:1}\rrbracket \supseteq \llbracket\mathbf{x}_1|y_{0:1}\rrbracket \neq \emptyset$ holds.
Proceeding forward, we get $\llbracket\hat{\mathbf{x}}_k|y_{0:k}\rrbracket \supseteq \llbracket\mathbf{x}_k|y_{0:k}\rrbracket \neq \emptyset$ for $k \in \mathbb{N}_0$.

\emph{\underline{Uniformly bounded estimate:}}
Let $\bar{\mathbf{x}}_k = [(\bar{\mathbf{x}}_k^{\mathrm{a}})^\mathrm{T} (\bar{\mathbf{x}}_k^{\mathrm{b}})^\mathrm{T}]^\mathrm{T} = Q \mathbf{x}_k$ with a non-singular $Q \in \mathbb{R}^{n\times n}$ such that
\begin{equation}\label{eqn:Transformed System}
\begin{split}
\begin{bmatrix}
\bar{\mathbf{x}}_{k+1}^{\mathrm{a}} \\
\bar{\mathbf{x}}_{k+1}^{\mathrm{b}}
\end{bmatrix}
&=
\begin{bmatrix}
\bar{A}_{\mathrm{a}} & 0\\
0 & J_0
\end{bmatrix}
\begin{bmatrix}
\bar{\mathbf{x}}_k^{\mathrm{a}} \\
\bar{\mathbf{x}}_k^{\mathrm{b}}
\end{bmatrix}
+
\begin{bmatrix}
\bar{B}_{\mathrm{a}} \\
\bar{B}_{\mathrm{b}}
\end{bmatrix}
\mathbf{w}_k,\\
\mathbf{y}_k &= \bar{C}_{\mathrm{a}} \bar{\mathbf{x}}_k^{\mathrm{a}} + \bar{C}_{\mathrm{b}} \bar{\mathbf{x}}_k^{\mathrm{b}} + \mathbf{v}_k,
\end{split}
\end{equation}
where $\bar{A}_{\mathrm{a}}$ is non-singular and $J_0$ is a block diagonal matrix corresponding to all Jordan blocks associated with the eigenvalue $0$.
Thus, we have
\begin{equation}\label{eqninpf:prop:Stability of Observable Systems - Uniform Boundedness - Diameter}
\begin{split}
d&(\llbracket\hat{\mathbf{x}}_k|y_{0:k}\rrbracket) \leq \sup_{\hat{\bar{x}}_k,\hat{\bar{x}}'_k \in \llbracket\hat{\bar{\mathbf{x}}}_k|y_{0:k}\rrbracket} \|Q^{-1}\| \|\hat{\bar{x}}_k - \hat{\bar{x}}'_k\|\\
\stackrel{(h)}{\leq}& \sup_{\hat{\bar{x}}_k^{\mathrm{a}},\hat{\bar{x}}_k'^{\mathrm{a}} \in \llbracket\hat{\bar{\mathbf{x}}}_k^{\mathrm{a}}|y_{0:k}\rrbracket} \|Q^{-1}\| \sqrt{\|\hat{\bar{x}}_k^{\mathrm{a}} - \hat{\bar{x}}_k'^{\mathrm{a}}\|^2 + \bar{d}_{\mathrm{b}}^2},
\end{split}
\end{equation}
where $(h)$ follows from the uniform boundedness\footnote{Noticing that $J_0$ is a nilpotent matrix, we can derive the uniform boundedness of $\llbracket\hat{\bar{\mathbf{x}}}_k^{\mathrm{b}}\rrbracket \supseteq \llbracket\hat{\bar{\mathbf{x}}}_k^{\mathrm{b}}|y_{0:k}\rrbracket$ from $\bar{\mathbf{x}}_k^{\mathrm{b}} = J_0^k \bar{\mathbf{x}}_0^{\mathrm{b}} + \sum_{i = 0}^{k-1} J_0^{k-1-i} \bar{B}_{\mathrm{b}} \mathbf{w}_i$ and the boundedness of $\llbracket\hat{\mathbf{x}}_0\rrbracket \subset \mathbb{R}^n$.}
of $\llbracket\hat{\bar{\mathbf{x}}}_k^{\mathrm{b}}|y_{0:k}\rrbracket$ w.r.t. $k \in \mathbb{N}_0$ that: $\forall \hat{\bar{x}}_k^{\mathrm{b}},\hat{\bar{x}}_k'^{\mathrm{b}} \in \llbracket\hat{\bar{\mathbf{x}}}_k^{\mathrm{b}}|y_{0:k}\rrbracket$ there exists a $\bar{d}_{\mathrm{b}}$ such that $\|\hat{\bar{x}}_k^{\mathrm{b}} - \hat{\bar{x}}_k'^{\mathrm{b}}\| \leq \bar{d}_{\mathrm{b}}$.
Then, we only need to focus on the first subsystem of~\eqref{eqn:Transformed System}, i.e.,
\begin{equation}\label{eqninpf:prop:Stability of Observable Systems - Subsystem 1}
\begin{split}
\bar{\mathbf{x}}_{k+1}^{\mathrm{a}} &= \bar{A}_{\mathrm{a}} \bar{\mathbf{x}}_k^{\mathrm{a}} + \bar{B}_{\mathrm{a}} \mathbf{w}_k,\\
\mathbf{y}_k &= \bar{C}_{\mathrm{a}} \bar{\mathbf{x}}_k^{\mathrm{a}} + \bar{\mathbf{v}}_k,
\end{split}
\end{equation}
where $\bar{\mathbf{v}}_k = \bar{C}_{\mathrm{b}} \bar{\mathbf{x}}_k^{\mathrm{b}} + \mathbf{v}_k$ is an equivalent measurement noise [which is related to $\mathbf{w}_{0:k-1}$ due to~\eqref{eqn:Transformed System}] and $\llbracket\bar{\mathbf{v}}_k\rrbracket$ is uniformly bounded w.r.t. $k \in \mathbb{N}_0$.
Define $\overline{\mathcal{O}}_{k,i}^{\mathrm{a}} := \bar{A}_{\mathrm{a}}^{k-i} \overline{\mathcal{X}}_i^{\mathrm{a}}(\bar{C}_{\mathrm{a}},y_i,\llbracket \bar{\mathbf{v}}_i\rrbracket) \oplus \sum_{r=i}^{k-1} \bar{A}_{\mathrm{a}}^{k-1-r} \bar{B}_{\mathrm{a}} \llbracket\mathbf{w}_r\rrbracket$ with
\begin{equation*}%\label{eqn:Specific Form of the Solution Set X Bar}
\overline{\mathcal{X}}_i^{\mathrm{a}}(\bar{C}_{\mathrm{a}},y_i,\llbracket \bar{\mathbf{v}}_i\rrbracket) := \ker(\bar{C}_{\mathrm{a}}) \oplus \bar{C}_{\mathrm{a}}^+ (\left\{y_i\right\} \oplus \llbracket-\bar{\mathbf{v}}_i\rrbracket).
\end{equation*}
Note that $\overline{\mathcal{O}}_{k,i}^{\mathrm{a}}$ is an observation-information set in terms of~\eqref{eqninpf:prop:Stability of Observable Systems - Subsystem 1}.
Hence, we can derive $\llbracket\hat{\bar{\mathbf{x}}}_k^{\mathrm{a}}|y_{0:k}\rrbracket \subseteq \bigcap_{i=k-\delta}^{k} \overline{\mathcal{O}}_{k,i}^{\mathrm{a}}$.\footnote{Even though $\mathbf{w}_{0:k-1}$ and $\bar{\mathbf{v}}_k$ are related, we can assume the unrelatedness holds and employ Theorem~3 in~\cite{CongY2022TAC} to employ~\eqref{eqn:Set-Intersection-Based Outer Bound} and~\eqref{eqn:Introducing OIT} to get this result.}
Thus, for $k \geq \delta \geq \mu_{\mathrm{a}} - 1$, where $\mu_{\mathrm{a}}$ is the observability index w.r.t. $(\bar{A}_{\mathrm{a}},\bar{C}_{\mathrm{a}})$, we get
\begin{equation}\label{eqninpf:prop:Stability of Observable Systems - Uniform Boundedness - Diameter Upper Bound}
d(\llbracket\hat{\mathbf{x}}_k|y_{0:k}\rrbracket) \stackrel{\eqref{eqn:An Upper Bound on the Diameter of the OIT}}{\leq} \|Q^{-1}\| \sqrt{\bar{d}_{\delta}(\bar{A}_{\mathrm{a}}, \bar{B}_{\mathrm{a}}, \bar{C}_{\mathrm{a}})^2 + \bar{d}_{\mathrm{b}}^2},
\end{equation}
where $\bar{d}_{\delta}(\cdot, \cdot, \cdot)$ is defined in~\eqref{eqn:An Upper Bound on the Diameter of the OIT} with $d_v$ for the uniformly bounded equivalent noise $\bar{\mathbf{v}}_k$.
For $k < \delta$, bounded $\llbracket\hat{\mathbf{x}}_0\rrbracket$ lead to bounded $\llbracket\hat{\mathbf{x}}_k|y_{0:k}\rrbracket$ (determined by Algorithm~\ref{alg:Classical Linear Set-Membership Filtering}).
Thus $\llbracket\hat{\mathbf{x}}_k|y_{0:k}\rrbracket$ is uniformly bounded w.r.t. $k \in \mathbb{N}_0$.

\emph{\underline{Bounded estimation gap:}}
Due to the page limit\footnote{We recommend the readers to follow the ideas in~\eqref{eqninpf:thm:Stability Criterion - Boundedness of Estimation Gap - An Upper Bound on the Estimation Gap}-\eqref{eqninpf:thm:Stability Criterion - Boundedness of Estimation Gap - An Upper Bound on the Estimation Gap - Part 2 - UB} and derive a tighter bound.
%$d_k^{\mathrm{g}}(\llbracket\hat{\mathbf{x}}_k|y_{0:k}\rrbracket) \leq \|Q^{-1}\| \bar{d}_{\delta}(\bar{A}_{\mathrm{a}}, \bar{B}_{\mathrm{a}}, \bar{C}_{\mathrm{a}})$ for $k \geq \delta \geq \max\{\mu_1 - 1,1\}$.
%
However, the upper bound given in~\eqref{eqninpf:prop:Stability of Observable Systems - Upper Bound Estimation Gap - 1} is good enough to derive the boundedness of the estimation gap for observable systems.}, we provide the proof based on
\begin{equation}\label{eqninpf:prop:Stability of Observable Systems - Upper Bound Estimation Gap - 1}
\begin{split}
d_k^{\mathrm{g}}&(\llbracket\hat{\mathbf{x}}_k|y_{0:k}\rrbracket) \leq \sup_{\hat{x}_k \in \llbracket\hat{\mathbf{x}}_k|y_{0:k}\rrbracket\atop x_k \in \llbracket\mathbf{x}_k|y_{0:k}\rrbracket} \|\hat{x}_k - x_k\|,\\
&\stackrel{(j)}{\leq} \sup_{\hat{x}_k,\hat{x}'_k \in \llbracket\hat{\mathbf{x}}_k|y_{0:k}\rrbracket} \|\hat{x}_k - \hat{x}'_k\| = d(\llbracket\hat{\mathbf{x}}_k|y_{0:k}\rrbracket),
\end{split}
\end{equation}
where $(j)$ is from $\llbracket\hat{\mathbf{x}}_k|y_{0:k}\rrbracket \supseteq \llbracket\mathbf{x}_k|y_{0:k}\rrbracket$ (see the proof of well-posedness).
Therefore, for observable $(A, C)$, the boundedness of the estimation gap is established.
\end{proof}

Now, we prove the stability of Algorithm~\ref{alg:Classical Linear Set-Membership Filtering} w.r.t. the initial condition under conditions~(i) and~(ii) of \thmref{thm:Stability Criterion}.

\emph{\underline{Well-posedness:}}
Define $\widetilde{\mathcal{O}}_{k,i}^o := \tilde{A}_o^{k-i} \widetilde{\mathcal{X}}_i^o(\tilde{C}_o,y_i,\llbracket \mathbf{v}_i\rrbracket) \oplus \sum_{r=i}^{k-1} \tilde{A}_o^{k-1-r} \tilde{B}_o \llbracket\mathbf{w}_r\rrbracket$ with
\begin{equation*}
\begin{split}
\widetilde{\mathcal{X}}_i^o(\tilde{C}_o,y_i,\llbracket \mathbf{v}_i\rrbracket) &:= \left\{\tilde{x}_i^o\colon y_i = \tilde{C}_o \tilde{x}_i^o + v_i,~v_i \in \llbracket\mathbf{v}_i\rrbracket\right\}\\
&= \ker(\tilde{C}_o) \oplus \tilde{C}_o^+ (\left\{y_i\right\} \oplus \llbracket-\mathbf{v}_i\rrbracket).
\end{split}
\end{equation*}
Note that $\widetilde{\mathcal{O}}_{k,i}^o$ is an observation-information set in terms of the observable subsystem of~\eqref{eqn:Observability Decomposition}.
Then, we have
\begin{equation}\label{eqninpf:thm:Stability Criterion - Well-Posedness - k = 0}
\begin{split}
\llbracket\hat{\tilde{\mathbf{x}}}_0|y_0\rrbracket &= \llbracket P\hat{\mathbf{x}}_0|y_0\rrbracket = P \big(\mathcal{O}_{0,0} \bigcap \llbracket\hat{\mathbf{x}}_0\rrbracket\big)\\
&\stackrel{(k)}{=} P\mathcal{O}_{0,0} \bigcap \llbracket P\hat{\mathbf{x}}_0\rrbracket\\
&\stackrel{(l)}{=} \big(\widetilde{\mathcal{O}}_{0,0}^o \times \mathbb{R}^{n_{\bar{o}}}\big) \bigcap \llbracket\hat{\tilde{\mathbf{x}}}_0\rrbracket,\\
%&\stackrel{(c)}{\subseteq} \big(\widetilde{\mathcal{O}}_{0,0}^o \bigcap \llbracket\hat{\tilde{\mathbf{x}}}_0^o\rrbracket \big) \times \llbracket\hat{\tilde{\mathbf{x}}}_0^{\bar{o}}\rrbracket,
\end{split}
\end{equation}
where $(k)$ follows from the non-singularity of $P$;
$(l)$ is established by $P\mathcal{X}_0(C, y_0, \llbracket\mathbf{v}_0\rrbracket) = \{\tilde{x}_0\colon y_0 = C P^{-1} \tilde{x}_0 + v_0,~v_0 \in \llbracket\mathbf{v}_0\rrbracket\} = \{\tilde{x}_0\colon y_0 = [\tilde{C}_o~0] \tilde{x}_0 + v_0,~v_0 \in \llbracket\mathbf{v}_0\rrbracket\} = \{\tilde{x}_0^o\colon y_0 = \tilde{C}_o \tilde{x}_0^o + v_0,~v_0 \in \llbracket\mathbf{v}_0\rrbracket\} \times \{\tilde{x}_0^{\bar{o}} \in \mathbb{R}^{n_{\bar{o}}}\}$.
Substituting $\llbracket\hat{\tilde{\mathbf{x}}}_0\rrbracket = \llbracket\tilde{\mathbf{x}}_0\rrbracket$ into~\eqref{eqninpf:thm:Stability Criterion - Well-Posedness - k = 0} and noticing that $\llbracket\tilde{\mathbf{x}}_0|y_0\rrbracket \neq \emptyset$, we have
\begin{equation}\label{eqninpf:thm:Stability Criterion - Well-Posedness - k = 0 - Actual Range}
\llbracket\tilde{\mathbf{x}}_0|y_0\rrbracket = \big(\widetilde{\mathcal{O}}_{0,0}^o \times \mathbb{R}^{n_{\bar{o}}}\big) \bigcap \llbracket\tilde{\mathbf{x}}_0\rrbracket \neq \emptyset.
\end{equation}
This means $\exists \{\tilde{x}_0\} = \{\tilde{x}_0^o\} \times \{\tilde{x}_0^{\bar{o}}\}$ such that
\begin{equation*}%\label{eqninpf:thm:Stability Criterion - Well-Posedness - k = 0 - Non-empty Point}
\emptyset \neq \big(\widetilde{\mathcal{O}}_{0,0}^o \times \mathbb{R}^{n_{\bar{o}}}\big) \bigcap \big(\{\tilde{x}_0^o\} \times \{\tilde{x}_0^{\bar{o}}\}\big) = \big(\widetilde{\mathcal{O}}_{0,0}^o \bigcap \{\tilde{x}_0^o\} \big) \times \{\tilde{x}_0^{\bar{o}}\},
\end{equation*}
where $\tilde{x}_0^o \in \llbracket\tilde{\mathbf{x}}_0^o\rrbracket$.
Since $\llbracket\hat{\tilde{\mathbf{x}}}_0^o\rrbracket \supseteq \llbracket\tilde{\mathbf{x}}_0^o\rrbracket$ in condition~(i) holds, we get $\tilde{x}_0^o \in \llbracket\hat{\tilde{\mathbf{x}}}_0^o\rrbracket$.
Thus, $\exists \tilde{x}_0'^{\bar{o}}$ such that $\llbracket\hat{\tilde{\mathbf{x}}}_0\rrbracket \supseteq \{\tilde{x}_0^o\} \times \{\tilde{x}_0'^{\bar{o}}\}$, which together with~\eqref{eqninpf:thm:Stability Criterion - Well-Posedness - k = 0} gives
\begin{equation}\label{eqninpf:thm:Stability Criterion - Well-Posedness - k = 0 - Estimate}
\begin{split}
\llbracket\hat{\tilde{\mathbf{x}}}_0|y_0\rrbracket &= \big(\widetilde{\mathcal{O}}_{0,0}^o \times \mathbb{R}^{n_{\bar{o}}}\big) \bigcap \llbracket\hat{\tilde{\mathbf{x}}}_0\rrbracket\\
&\supseteq \big(\widetilde{\mathcal{O}}_{0,0}^o \bigcap \{\tilde{x}_0^o\} \big) \times \{\tilde{x}_0'^{\bar{o}}\} \neq \emptyset.
\end{split}
\end{equation}

From the first equalities of~\eqref{eqninpf:thm:Stability Criterion - Well-Posedness - k = 0 - Actual Range} and~\eqref{eqninpf:thm:Stability Criterion - Well-Posedness - k = 0 - Estimate}, we can derive $\llbracket\hat{\tilde{\mathbf{x}}}_0^o|y_0\rrbracket \supseteq \llbracket\tilde{\mathbf{x}}_0^o|y_0\rrbracket$, which combined with the observable subsystem in~\eqref{eqn:Observability Decomposition} yields $\llbracket\hat{\tilde{\mathbf{x}}}_1^o|y_0\rrbracket \supseteq \llbracket\tilde{\mathbf{x}}_1^o|y_0\rrbracket$.
Proceeding forward, we can obtain that $\llbracket\hat{\tilde{\mathbf{x}}}_k|y_{0:k}\rrbracket \neq \emptyset$ and $\llbracket\hat{\tilde{\mathbf{x}}}_k^o|y_{0:k}\rrbracket \supseteq \llbracket\tilde{\mathbf{x}}_k^o|y_{0:k}\rrbracket$ for $k \in \mathbb{N}_0$.

\emph{\underline{Boundedness of estimation gap:}}
From \defref{def:Estimation Gap}, the estimation gap $d_k^{\mathrm{g}}(\llbracket\hat{\mathbf{x}}_k|y_{0:k}\rrbracket) = d_{\mathrm{H}}(\llbracket\hat{\mathbf{x}}_k|y_{0:k}\rrbracket, \llbracket\mathbf{x}_k|y_{0:k}\rrbracket)$ can be upper bounded by
\begin{equation}\label{eqninpf:thm:Stability Criterion - Boundedness of Estimation Gap - An Upper Bound on the Estimation Gap}
d_k^{\mathrm{g}}(\llbracket\hat{\mathbf{x}}_k|y_{0:k}\rrbracket) \leq \|P^{-1}\| d_{\mathrm{H}}(\llbracket\hat{\tilde{\mathbf{x}}}_k|y_{0:k}\rrbracket, \llbracket\tilde{\mathbf{x}}_k|y_{0:k}\rrbracket).
\end{equation}
Noticing the structure of $d_{\mathrm{H}}(\mathcal{S},\mathcal{T})$ in~\eqref{eqn:Hausdorff Distance}, firstly we prove the upper boundedness of
\begin{equation}\label{eqninpf:thm:Stability Criterion - Boundedness of Estimation Gap - An Upper Bound on the Estimation Gap - Part 1}
\adjustlimits
\sup_{\hat{\tilde{x}}_k \in \llbracket\hat{\tilde{\mathbf{x}}}_k|y_{0:k}\rrbracket} \inf_{\tilde{x}_k \in \llbracket\tilde{\mathbf{x}}_k|y_{0:k}\rrbracket}\|\hat{\tilde{x}}_k - \tilde{x}_k\|.
\end{equation}

Applying \propref{prop:Stability of Observable Systems} to the observable subsystem of~\eqref{eqn:Observability Decomposition}, we get that $\forall \hat{\tilde{x}}_k^o \in \llbracket\hat{\tilde{\mathbf{x}}}_k^o|y_{0:k}\rrbracket$ and $\forall \tilde{x}_k^o \in \llbracket\tilde{\mathbf{x}}_k^o|y_{0:k}\rrbracket$, there exists a $\tilde{d}_o$ such that $\|\hat{\tilde{x}}_k^o - \tilde{x}_k^o\| \leq \tilde{d}_o$ for $k \in \mathbb{N}_0$.
Thus,~\eqref{eqninpf:thm:Stability Criterion - Boundedness of Estimation Gap - An Upper Bound on the Estimation Gap - Part 1} can be upper bounded by
\begin{equation}\label{eqninpf:thm:Stability Criterion - Boundedness of Estimation Gap - An Upper Bound on the Estimation Gap - Part 1 - UB}
\adjustlimits
\sup_{\hat{\tilde{x}}_k^{\bar{o}} \in \llbracket\hat{\tilde{\mathbf{x}}}_k^{\bar{o}}|y_{0:k}\rrbracket} \inf_{\tilde{x}_k^{\bar{o}} \in \llbracket\tilde{\mathbf{x}}_k^{\bar{o}}|y_{0:k}\rrbracket} \sqrt{\tilde{d}_{o}^2 + \|\hat{\tilde{x}}_k^{\bar{o}} - \tilde{x}_k^{\bar{o}}\|^2}.
\end{equation}
From the unobservable subsystem of~\eqref{eqn:Observability Decomposition}, we have
\begin{equation}\label{eqninpf:thm:Stability Criterion - Boundedness of Estimation Gap - Unobservable System}
\begin{split}
\llbracket\hat{\tilde{\mathbf{x}}}_k^{\bar{o}}&|y_{0:k}\rrbracket = \Big\llbracket\tilde{A}_{\bar{o}}^k \hat{\tilde{\mathbf{x}}}_0^{\bar{o}} + \sum_{i=0}^{k-1} \tilde{A}_{\bar{o}}^{k-1-i} (\tilde{A}_{21} \hat{\tilde{\mathbf{x}}}_i^o + \tilde{B}_{\bar{o}} \mathbf{w}_i)\Big|y_{0:k}\Big\rrbracket\\
&\stackrel{(m)}{\subseteq} \tilde{A}_{\bar{o}}^k \llbracket\hat{\tilde{\mathbf{x}}}_0^{\bar{o}}\rrbracket \oplus \sum_{i=0}^{k-1} \tilde{A}_{\bar{o}}^{k-1-i} (\tilde{A}_{21} \llbracket\hat{\tilde{\mathbf{x}}}_i^o|y_{0:i}\rrbracket \oplus \tilde{B}_{\bar{o}} \llbracket\mathbf{w}_i\rrbracket),
\end{split}
\end{equation}
where $(m)$ follows from $\llbracket\mathbf{a} + \mathbf{b}\rrbracket \subseteq \llbracket\mathbf{a}\rrbracket \oplus \llbracket\mathbf{b}\rrbracket$ and $\llbracket\mathbf{a}|b\rrbracket \subseteq \llbracket\mathbf{a}\rrbracket$.
Hence, $\forall \hat{\tilde{x}}_k^{\bar{o}} \in \llbracket\hat{\tilde{\mathbf{x}}}_k^{\bar{o}}|y_{0:k}\rrbracket$, $\exists \hat{\tilde{x}}_0^o \in \llbracket\hat{\tilde{\mathbf{x}}}_0^o\rrbracket,~\hat{\tilde{x}}_i^o \in \llbracket\hat{\tilde{\mathbf{x}}}_i^o|y_{0:i}\rrbracket,~w_i \in \llbracket\mathbf{w}_i\rrbracket~(i = 0,\ldots,k-1)$ such that
\begin{equation}\label{eqninpf:thm:Stability Criterion - Boundedness of Estimation Gap - Unobservable System - Equation 1}
\hat{\tilde{x}}_k^{\bar{o}} = \tilde{A}_{\bar{o}}^k \hat{\tilde{x}}_0^{\bar{o}} + \sum_{i=0}^{k-1} \tilde{A}_{\bar{o}}^{k-1-i} (\tilde{A}_{21} \hat{\tilde{x}}_i^o + \tilde{B}_{\bar{o}} w_i).
\end{equation}
Likewise, $\forall \tilde{x}_k^{\bar{o}} \in \llbracket\tilde{\mathbf{x}}_k^{\bar{o}}|y_{0:k},w_{0:k}\rrbracket \subseteq \llbracket\tilde{\mathbf{x}}_k^{\bar{o}}|y_{0:k}\rrbracket$, $\exists \tilde{x}_0^o \in \llbracket\tilde{\mathbf{x}}_0^o\rrbracket,~\tilde{x}_i^o \in \llbracket\tilde{\mathbf{x}}_i^o|y_{0:i}\rrbracket~(i = 0,\ldots,k-1)$ such that
\begin{equation}\label{eqninpf:thm:Stability Criterion - Boundedness of Estimation Gap - Unobservable System - Equation 2}
\tilde{x}_k^{\bar{o}} = \tilde{A}_{\bar{o}}^k \tilde{x}_0^{\bar{o}} + \sum_{i=0}^{k-1} \tilde{A}_{\bar{o}}^{k-1-i} (\tilde{A}_{21} \tilde{x}_i^o + \tilde{B}_{\bar{o}} w_i).
\end{equation}
Therefore,~\eqref{eqninpf:thm:Stability Criterion - Boundedness of Estimation Gap - An Upper Bound on the Estimation Gap - Part 1 - UB} is upper bounded by
\begin{equation}\label{eqninpf:thm:Stability Criterion - Boundedness of Estimation Gap - An Upper Bound on the Estimation Gap - Part 1 - UB 2}
\adjustlimits
\sup_{\hat{\tilde{x}}_k^{\bar{o}} \in \llbracket\hat{\tilde{\mathbf{x}}}_k^{\bar{o}}|y_{0:k}\rrbracket} \inf_{\tilde{x}_k^{\bar{o}} \in \llbracket\tilde{\mathbf{x}}_k^{\bar{o}}|y_{0:k},w_{0:k}\rrbracket} \sqrt{\tilde{d}_o^2 + \|\hat{\tilde{x}}_k^{\bar{o}} - \tilde{x}_k^{\bar{o}}\|^2},
\end{equation}
where [based on~\eqref{eqninpf:thm:Stability Criterion - Boundedness of Estimation Gap - Unobservable System - Equation 1} and~\eqref{eqninpf:thm:Stability Criterion - Boundedness of Estimation Gap - Unobservable System - Equation 2}]
\begin{equation}\label{eqninpf:thm:Stability Criterion - Boundedness of Estimation Gap - Equivalent Response}
\hat{\tilde{x}}_k^{\bar{o}} - \tilde{x}_k^{\bar{o}} = \tilde{A}_{\bar{o}}^k (\hat{\tilde{x}}_0^{\bar{o}} - \tilde{x}_0^{\bar{o}}) + \sum_{i=0}^{k-1} \tilde{A}_{21} \tilde{A}_{\bar{o}}^{k-1-i} (\hat{\tilde{x}}_i^o - \tilde{x}_i^o).
\end{equation}
By condition~(ii),~\eqref{eqninpf:thm:Stability Criterion - Boundedness of Estimation Gap - Equivalent Response}, and the uniform boundedness of $\llbracket\hat{\tilde{\mathbf{x}}}_k^o|y_{0:k}\rrbracket$ and $\llbracket\tilde{\mathbf{x}}_k^o|y_{0:k}\rrbracket$, there exists a $\bar{d}_{\bar{o}}$ such that
\begin{equation}\label{eqninpf:thm:Stability Criterion - Boundedness of Estimation Gap - Bound d_obar}
	\|\hat{\tilde{x}}_k^{\bar{o}} - \tilde{x}_k^{\bar{o}}\| \leq \bar{d}_{\bar{o}},
\end{equation}
which yields the boundedness of~\eqref{eqninpf:thm:Stability Criterion - Boundedness of Estimation Gap - An Upper Bound on the Estimation Gap - Part 1}.

Secondly, we analyze the boundedness of
\begin{equation}\label{eqninpf:thm:Stability Criterion - Boundedness of Estimation Gap - An Upper Bound on the Estimation Gap - Part 2}
\adjustlimits
\sup_{\tilde{x}_k \in \llbracket\tilde{\mathbf{x}}_k|y_{0:k}\rrbracket} \inf_{\hat{\tilde{x}}_k \in \llbracket\hat{\tilde{\mathbf{x}}}_k|y_{0:k}\rrbracket}\|\hat{\tilde{x}}_k - \tilde{x}_k\|
\end{equation}
in $d_{\mathrm{H}}(\llbracket\hat{\tilde{\mathbf{x}}}_k|y_{0:k}\rrbracket, \llbracket\tilde{\mathbf{x}}_k|y_{0:k}\rrbracket)$.
From $\llbracket\hat{\tilde{\mathbf{x}}}_k^o|y_{0:k}\rrbracket \supseteq \llbracket\tilde{\mathbf{x}}_k^o|y_{0:k}\rrbracket$ [guaranteed by condition~(i)],~\eqref{eqninpf:thm:Stability Criterion - Boundedness of Estimation Gap - An Upper Bound on the Estimation Gap - Part 2} can be rewritten as
\begin{equation}\label{eqninpf:thm:Stability Criterion - Boundedness of Estimation Gap - An Upper Bound on the Estimation Gap - Part 2 - UB}
\adjustlimits
\sup_{\tilde{x}_k^{\bar{o}} \in \llbracket\tilde{\mathbf{x}}_k^{\bar{o}}|y_{0:k}\rrbracket} \inf_{\hat{\tilde{x}}_k^{\bar{o}} \in \llbracket\hat{\tilde{\mathbf{x}}}_k^{\bar{o}}|y_{0:k}\rrbracket} \|\hat{\tilde{x}}_k^{\bar{o}} - \tilde{x}_k^{\bar{o}}\| \stackrel{\eqref{eqninpf:thm:Stability Criterion - Boundedness of Estimation Gap - Bound d_obar}}{\leq} \bar{d}_{\bar{o}}.
\end{equation}
Therefore,~\eqref{eqninpf:thm:Stability Criterion - Boundedness of Estimation Gap - An Upper Bound on the Estimation Gap - Part 2} is bounded for $k \in \mathbb{N}_0$.
To sum up, the estimation gap $d_k^{\mathrm{g}}(\llbracket\hat{\mathbf{x}}_k|y_{0:k}\rrbracket)$ is bounded w.r.t. $k \in \mathbb{N}_0$.\hfill$\square$

\section{Proof of \propref{prop:Converse for Bounded Estimation Gap}}\label{apx:Proof of prop:Converse for Bounded Estimation Gap}

The estimation gap can be lower bounded by
\begin{equation}\label{eqninpf:prop:Converse for Bounded Estimation Gap - Lower Bound on Estimation Gap}
\begin{split}
d_k^{\mathrm{g}}&(\llbracket\hat{\mathbf{x}}_k|y_{0:k}\rrbracket) \geq \|P\|^{-1} d_{\mathrm{H}}(\llbracket\hat{\tilde{\mathbf{x}}}_k|y_{0:k}\rrbracket, \llbracket\tilde{\mathbf{x}}_k|y_{0:k}\rrbracket)\\
&\stackrel{\eqref{eqn:Hausdorff Distance}}{\geq} \adjustlimits\sup_{\hat{\tilde{x}}_k \in \llbracket\hat{\tilde{\mathbf{x}}}_k|y_{0:k}\rrbracket} \inf_{\tilde{x}_k \in \llbracket\tilde{\mathbf{x}}_k|y_{0:k}\rrbracket}\|P\|^{-1}\|\hat{\tilde{x}}_k - \tilde{x}_k\|\\
&\geq \adjustlimits\sup_{\hat{\tilde{x}}_k \in \llbracket\hat{\tilde{\mathbf{x}}}_k|y_{0:k}\rrbracket} \inf_{\tilde{x}_k \in \llbracket\tilde{\mathbf{x}}_k|y_{0:k}\rrbracket}\|P\|^{-1}\|\hat{\tilde{x}}_k^{\bar{o}} - \tilde{x}_k^{\bar{o}}\|.
\end{split}
\end{equation}
From~\eqref{eqninpf:prop:Converse for Bounded Estimation Gap - Lower Bound on Estimation Gap} and~\eqref{eqninpf:thm:Stability Criterion - Boundedness of Estimation Gap - Equivalent Response}, we know:
if $\tilde{A}_{\bar{o}}$ is not marginally stable, then there exists a bounded $\llbracket\hat{\mathbf{x}}_0\rrbracket$ such that $\|\hat{\tilde{x}}_k^{\bar{o}} - \tilde{x}_k^{\bar{o}}\|$ in~\eqref{eqninpf:prop:Converse for Bounded Estimation Gap - Lower Bound on Estimation Gap} unboundedly increases with $k$.
Thus, for all bounded $\llbracket\hat{\mathbf{x}}_0\rrbracket \subset \mathbb{R}^n$, the boundedness of the estimation gap cannot be guaranteed when $\tilde{A}_{\bar{o}}$ is not marginally stable.
Therefore, $\tilde{A}_{\bar{o}}$ must be marginally stable to ensure the bounded $d_k^{\mathrm{g}}(\llbracket\hat{\mathbf{x}}_k|y_{0:k}\rrbracket)$.\hfill$\square$

\section{Proof of \corref{cor:Egregium}}\label{apx:Proof of cor:Egregium}

By \thmref{thm:Stability Criterion}, the classical SMFing framework with bounded $\llbracket\hat{\mathbf{x}}_0\rrbracket$ and $\llbracket\hat{\tilde{\mathbf{x}}}_0^o\rrbracket \supseteq \llbracket\tilde{\mathbf{x}}_0^o\rrbracket$ is stable w.r.t. the initial condition if $(A, C)$ is detectable.
Thus, we only focus on the uniform boundedness of the estimate.

From \propref{prop:Stability of Observable Systems}, $d(\llbracket\hat{\tilde{\mathbf{x}}}_k^o|y_{0:k}\rrbracket)$ is bounded.
Thus, with
\begin{equation*}
d(\llbracket\hat{\mathbf{x}}_k|y_{0:k}\rrbracket) = \|P^{-1}\|\sqrt{d(\llbracket\hat{\tilde{\mathbf{x}}}_k^o|y_{0:k}\rrbracket)^2 + d(\llbracket\hat{\tilde{\mathbf{x}}}_k^{\bar{o}}|y_{0:k}\rrbracket)^2},
\end{equation*}
we know that $\llbracket\hat{\mathbf{x}}_k|y_{0:k}\rrbracket$ is uniformly bounded w.r.t. $k \in \mathbb{N}_0$ if and only if $\llbracket\hat{\tilde{\mathbf{x}}}_k^{\bar{o}}|y_{0:k}\rrbracket$ is uniformly bounded w.r.t. $k \in \mathbb{N}_0$.
From~\eqref{eqninpf:thm:Stability Criterion - Boundedness of Estimation Gap - Unobservable System} with $\rho(\tilde{A}_{\bar{o}}) < 1$ [as $(A, C)$ is detectable], bounded $\llbracket\hat{\mathbf{x}}_0\rrbracket$, and uniformly bounded $\llbracket\hat{\tilde{\mathbf{x}}}_k^o|y_{0:k}\rrbracket$ and $\llbracket\mathbf{w}_k\rrbracket$, we can derive the uniform boundedness of $\llbracket\hat{\tilde{\mathbf{x}}}_k^{\bar{o}}|y_{0:k}\rrbracket$.
Therefore, $\llbracket\hat{\mathbf{x}}_k|y_{0:k}\rrbracket$ is uniformly bounded w.r.t. $k \in \mathbb{N}_0$ for detectable $(A, C)$.\hfill$\square$

\section{Proof of \lemref{lem:OIT-Inspired Invariance}}\label{apx:Proof of lem:OIT-Inspired Invariance}

Since for any $\mathcal{S} \subseteq \mathbb{R}^{n_o}$, $P_o F_{k,0}(P^{-1}(\mathcal{S} \times \llbracket\hat{\tilde{\mathbf{x}}}_0^{\bar{o}}\rrbracket))$ in~\eqref{eqn:OIT-Inspired Invariance} corresponds to the observable subsystem and the resulting set is independent of $\llbracket\hat{\tilde{\mathbf{x}}}_0^{\bar{o}}\rrbracket$, without loss of generality, we consider the pair $(A, C)$ is observable, i.e., $A = \tilde{A}_o$.
In that case,~\eqref{eqn:OIT-Inspired Invariance} becomes
\begin{equation}\label{eqninpf:lem:OIT-Inspired Invariance}
	F_{k,0}(\mathcal{B}_{\theta'_k}^{\infty}[\hat{c}_0^o]) \subseteq F_{k,0}(\mathcal{B}_{\bar{\theta}_k}^{\infty}[\hat{c}_0^o]) = F_{k,0}(\mathcal{B}_{\theta''_k}^{\infty}[\hat{c}_0^o]).
\end{equation}
Firstly, we prove~\eqref{eqninpf:lem:OIT-Inspired Invariance} for non-singular $A$.
Because $\mathcal{B}_{\theta'_k}^{\infty}[\hat{c}_0^o] \subseteq \mathcal{B}_{\bar{\theta}_k}^{\infty}[\hat{c}_0^o] \subseteq \mathcal{B}_{\theta''_k}^{\infty}[\hat{c}_0^o]$ for all $\theta'_k \leq \bar{\theta}_k \leq \theta''_k$, from \defref{def:Filtering Map} we have
\begin{equation}\label{eqninpf:lem:OIT-Inspired Invariance - Subseteq}
	F_{k,0}(\mathcal{B}_{\theta'_k}^{\infty}[\hat{c}_0^o]) \subseteq F_{k,0}(\mathcal{B}_{\bar{\theta}_k}^{\infty}[\hat{c}_0^o]) \subseteq F_{k,0}(\mathcal{B}_{\theta''_k}^{\infty}[\hat{c}_0^o]).
\end{equation}
By \thmref{thm:Uniform Boundedness of OIT}, $\forall k \geq \max\{\mu_o-1+n_{\lambda_0^o}, 1\}$, $\bigcap_{i=k-n_{\lambda_0^o}}^{k} \mathcal{O}_{k,i}$ is bounded, and $\exists \bar{\theta}_k \geq 0$ s.t. $\forall x_0 \in \mathbb{R}^n \setminus \mathcal{B}_{\bar{\theta}_k}^{\infty}[\hat{c}_0^o]$:
\begin{equation}\label{eqninpf:lem:OIT-Inspired Invariance - Equality}
	A^k x_0 + \sum_{i=0}^{k-1} A^{k-1-i} B w_i = x_k \notin \bigcap_{i=k-n_{\lambda_0^o}}^{k} \mathcal{O}_{k,i},
\end{equation}
which together with~\eqref{eqn:Set-Intersection-Based Outer Bound} and~\eqref{eqn:Introducing OIT} gives $F_{k,0}(\mathcal{B}_{\theta''_k}^{\infty}[\hat{c}_0^o]\setminus\mathcal{B}_{\bar{\theta}_k}^{\infty}[\hat{c}_0^o]) = \emptyset$ for $\theta''_k \geq \bar{\theta}_k$.
Thus, $F_{k,0}(\mathcal{B}_{\bar{\theta}_k}^{\infty}[\hat{c}_0^o]) = F_{k,0}(\mathcal{B}_{\theta''_k}^{\infty}[\hat{c}_0^o])$, and with~\eqref{eqninpf:lem:OIT-Inspired Invariance - Subseteq} we get~\eqref{eqninpf:lem:OIT-Inspired Invariance}.

When $A$ is singular, from~\eqref{eqn:Transformed System} we have $\bar{\mathbf{x}}_k^{\mathrm{b}} = \sum_{i = 0}^{k-1} J_0^{k-1-i} \bar{B}_{\mathrm{b}} \mathbf{w}_i$ for $k \geq n_{\lambda_0^o}$, which is not affected by the choice of $\mathcal{B}_{\theta_k}^{\infty}[\hat{c}_0^o]$;
for the subsystem w.r.t. $\bar{\mathbf{x}}_k^{\mathrm{a}}$, the proof is the same as that of~\eqref{eqninpf:lem:OIT-Inspired Invariance}, where the measurement equation is in~\eqref{eqninpf:prop:Stability of Observable Systems - Subsystem 1} with the equivalent noise $\bar{\mathbf{v}}_k = \bar{C}_{\mathrm{b}} \sum_{i = 0}^{k-1} J_0^{k-1-i} \bar{B}_{\mathrm{b}} \mathbf{w}_i + \mathbf{v}_k$ for $k \geq n_{\lambda_0^o}$.\hfill$\square$

\section{Proof of \thmref{thm:Stability of OIT-Inspired Filtering}}\label{apx:Proof of thm:Stability of OIT-Inspired Filtering}

\emph{\underline{Well-posedness:}}
For $k < k_*$, Line~\ref{line:OIT-Inspired Filtering - Estimates Resetting} resets every empty $\llbracket\hat{\mathbf{x}}_k| y_{0:k}\rrbracket$ to a non-empty set.
For $k \geq k_*$, the estimate $\llbracket\hat{\mathbf{x}}_k| y_{0:k}\rrbracket$ in Line~\ref{line:OIT-Inspired Filtering - Estimate Inspired by OIT} is non-empty, which is guaranteed by $\llbracket\hat{\tilde{\mathbf{x}}}_0^o\rrbracket = \mathcal{B}_{\bar{\theta}_k}^{\infty}[\hat{c}_0^o] \supseteq \llbracket\tilde{\mathbf{x}}_0^o\rrbracket$ (see \rekref{rek:OIT-Inspired Invariance}) and the well-posedness part [see~\eqref{eqninpf:thm:Stability Criterion - Well-Posedness - k = 0}-\eqref{eqninpf:thm:Stability Criterion - Well-Posedness - k = 0 - Estimate}] in the proof of \thmref{thm:Stability Criterion}.
Thus, $\llbracket\hat{\mathbf{x}}_k| y_{0:k}\rrbracket \neq \emptyset$ holds for $k \in \mathbb{N}_0$.

\emph{\underline{Boundedness of estimation gap:}}
For $k < k_*$, bounded $\llbracket\mathbf{x}_0\rrbracket$ with Lines~\ref{line:OIT-Inspired Filtering - Estimate from Classical SMF} and~\ref{line:OIT-Inspired Filtering - Estimates Resetting} indicates $d_k^{\mathrm{g}}(\llbracket\hat{\mathbf{x}}_k|y_{0:k}\rrbracket) < \infty$.
For $k \geq k_*$, using the same techniques in~\eqref{eqninpf:thm:Stability Criterion - Boundedness of Estimation Gap - An Upper Bound on the Estimation Gap}-\eqref{eqninpf:thm:Stability Criterion - Boundedness of Estimation Gap - An Upper Bound on the Estimation Gap - Part 2 - UB} [note that $\llbracket\hat{\tilde{\mathbf{x}}}_0^o\rrbracket = \mathcal{B}_{\bar{\theta}_k}^{\infty}[\hat{c}_0^o] \supseteq \llbracket\tilde{\mathbf{x}}_0^o\rrbracket$ ensures the preconditon $\llbracket\hat{\mathbf{x}}_0\rrbracket \supseteq \llbracket\mathbf{x}_0\rrbracket$ in \propref{prop:Stability of Observable Systems} when it is applied to the observable system to derive~\eqref{eqninpf:thm:Stability Criterion - Boundedness of Estimation Gap - An Upper Bound on the Estimation Gap - Part 1 - UB}], we can obtain the boundedness of $d_k^{\mathrm{g}}(\llbracket\hat{\mathbf{x}}_k|y_{0:k}\rrbracket)$.
Therefore, the estimation gap $d_k^{\mathrm{g}}(\llbracket\hat{\mathbf{x}}_k|y_{0:k}\rrbracket)$ is bounded for $k \in \mathbb{N}_0$.\hfill$\square$

\section{Proof of \thmref{thm:Stability of OIT-CZ SMF}}\label{apx:Proof of thm:Stability of OIT-CZ SMF}

Firstly, we provide \lemref{lem:Bound on Matrix Power Norm} for bounding the norm of matrix power, which is a variant of Lemma~3 in~\cite{CongY2021TAC}.

\begin{lemma}[Bound on Norm of Matrix Power]\label{lem:Bound on Matrix Power Norm}
Given $F \in \mathbb{R}^{n \times n}$ with $\rho(F) < 1$, for $\gamma \in (\rho(F), 1)$, $\|F^k\|_{\infty} \leq \beta_{\gamma}(F) \gamma^k$ holds for all $k \in \mathbb{N}_0$, where $\beta_{\gamma}(F) = \max \{\gamma^{-k} \|F^k\|_{\infty}\colon k \in \mathbb{N}_0\}$.
\end{lemma}

\begin{proof}
$\forall \gamma \in (\rho(F), 1)$, let $\tilde{F} = \gamma^{-1} F$, and we have $\rho(\tilde{F}) = \rho(F)/\gamma < 1$.
Since $\lim_{k\to\infty} \|\tilde{F}^k\|_{\infty} = 0$, we can find a $\underline{k} \in \mathbb{N}_0$ such that for all $k \geq \underline{k}$, the following holds
\begin{align}\label{eqninpf:lem:Bound on Matrix Power Norm - 1}
\gamma^{-k} \|F^k\|_{\infty} = \|\gamma^{-k} F^k\|_{\infty} = \|\tilde{F}^k\|_{\infty} < 1,
\end{align}
which implies $\|F^k\|_{\infty} < \gamma^{k}$.
It together with $\beta_{\gamma}(F) = \max \{1, \gamma^{-k} \|F^k\|_{\infty}~(1 \leq k \leq \underline{k})\} = \max \{\gamma^{-k} \|F^k\|_{\infty}\colon k \in \mathbb{N}_0\}$\footnote{Note that $\gamma^{-0} \|F^0\|_{\infty} = 1$ and $\gamma^{-k} \|F^k\|_{\infty} < 1$ for $k > \underline{k}$.} gives $\|F^k\|_{\infty} \leq \beta_{\gamma}(F) \gamma^k$ for all $k \in \mathbb{N}_0$.
\end{proof}

When $(A, C)$ is detectable, the uniform boundedness of the estimate and the boundedness of the estimation gap can be readily derived based on the results in \secref{sec:Stability-Guaranteed Filtering Inspired by Observation-Information Tower}, when we replace $F_{k,0}$ and $\llbracket\hat{\tilde{\mathbf{x}}}_0^{\bar{o}}\rrbracket$ with $F_{k,k-\bar{\delta}}$ and $\hat{\mathcal{T}}_{k-\bar{\delta}}^{\bar{o}}$, respectively.
Thus, we only focus on the well-posedness and finite-time inclusion property of Algorithm~\ref{alg:OIT-CZ SMF}.

\emph{\underline{Well-posedness:}}
Similarly to the well-posedness part in \apxref{apx:Proof of thm:Stability of OIT-Inspired Filtering} (Lines~\ref{line:OIT-CZ SMF - Estimates Resetting} and~\ref{line:OIT-CZ SMF - Estimate Inspired by OIT} in Algorithm~\ref{alg:OIT-CZ SMF} play the same role as Lines~\ref{line:OIT-Inspired Filtering - Estimates Resetting} and~\ref{line:OIT-Inspired Filtering - Estimate Inspired by OIT} in Algorithm~\ref{alg:OIT-Inspired Filtering}), we have $\mathcal{Z}_k \neq \emptyset$ for $k \in \mathbb{N}_0$ and
\begin{equation}\label{eqninpf:thm:Stability of OIT-CZ SMF - Well-Posedness - Observable Subsystem Inclusion Property}
\widetilde{\mathcal{Z}}_k^o \supseteq \llbracket\tilde{\mathbf{x}}_k^o|y_{0:k}\rrbracket,\quad k \geq \bar{\delta}.
\end{equation}
The only part needs highlighting is:
for Line~\ref{line:OIT-CZ SMF - Estimate Inspired by OIT}, when $\bar{\delta} \leq k < 2\bar{\delta}$, \lemref{lem:OIT-Inspired Invariance} guarantees $\hat{\mathcal{T}}_{k-\bar{\delta}}^o = \mathcal{B}_{\bar{\theta}_k}^{\infty} \big[\mathrm{center}\big(P_o \overline{\mathrm{IH}} (\mathcal{Z}_{k-\bar{\delta}}^-)\big)\big] \supseteq \llbracket\tilde{\mathbf{x}}_{k-\bar{\delta}}^o|y_{0:k-\bar{\delta}-1}\rrbracket$;
when $k \geq 2\bar{\delta}$, $\hat{\mathcal{T}}_{k-\bar{\delta}}^o = P_o \overline{\mathrm{IH}} (\mathcal{Z}_{k-\bar{\delta}}^-) \supseteq \llbracket\tilde{\mathbf{x}}_{k-\bar{\delta}}^o|y_{0:k-\bar{\delta}-1}\rrbracket$;
thus,
\begin{equation}\label{eqninpf:thm:Stability of OIT-CZ SMF - Well-Posedness - Inclusion}
	\hat{\mathcal{T}}_k^o \supseteq \llbracket\tilde{\mathbf{x}}_k^o|y_{0:k-1}\rrbracket \supseteq \llbracket\tilde{\mathbf{x}}_k^o|y_{0:k}\rrbracket, \quad k \in \mathbb{N}_0.
\end{equation}

\emph{\underline{Boundedness of estimation gap:}}
When $(A, C)$ is detectable, condition~(ii) in \thmref{thm:Stability Criterion} holds.
Then, the proof is similar to that of \thmref{thm:Stability of OIT-Inspired Filtering}.

\emph{\underline{Finite-time inclusion:}}
The proof includes two steps.
In the first step, we show that
\begin{equation}\label{eqninpf:thm:Stability of OIT-CZ SMF - Well-Posedness - Step 1 - 1}
\llbracket\tilde{\mathbf{x}}_k|y_{0:k}\rrbracket = \llbracket P \mathbf{x}_k|y_{0:k}\rrbracket \subseteq \llbracket\tilde{\mathbf{x}}_k^o|y_{0:k}\rrbracket \times \mathcal{T}_k^{\bar{o}}
\end{equation}
holds for $k \in \mathbb{N}_0$, where
\begin{equation}\label{eqninpf:thm:Stability of OIT-CZ SMF - Well-Posedness - Step 1 - 2}
\!\!\!\!\!\mathcal{T}_k^{\bar{o}} \!:=\! \tilde{A}_{\bar{o}}^k \llbracket\tilde{\mathbf{x}}_0^{\bar{o}}\rrbracket \oplus \sum_{i=0}^{k-1} \tilde{A}_{\bar{o}}^{k-1-i} (\tilde{A}_{21} \llbracket\tilde{\mathbf{x}}_i^o|y_{0:i}\rrbracket \oplus \tilde{B}_{\bar{o}} \llbracket\mathbf{w}_i\rrbracket).
\end{equation}
In the second step, the finite-time inclusion property~\eqref{eqn:Stability of OIT-CZ SMF - Finite-Time Inclusion} is proven based on~\eqref{eqninpf:thm:Stability of OIT-CZ SMF - Well-Posedness - Step 1 - 1}.

\textit{Step~1:}
%
%Similarly to the proof of~\eqref{eqninpf:prop:Observability Decomposition Based OIT Bound - 1}
%
We use the mathematical induction on~\eqref{eqninpf:thm:Stability of OIT-CZ SMF - Well-Posedness - Step 1 - 1}.

\emph{Base~case:}
For $k = 0$, we have $\llbracket\tilde{\mathbf{x}}_0|y_0\rrbracket \subseteq \llbracket\tilde{\mathbf{x}}_0^o|y_0\rrbracket \times \llbracket\tilde{\mathbf{x}}_0^{\bar{o}}|y_0\rrbracket = \llbracket\tilde{\mathbf{x}}_0^o|y_0\rrbracket \times \mathcal{T}_0^{\bar{o}}$.
Thus,~\eqref{eqninpf:thm:Stability of OIT-CZ SMF - Well-Posedness - Step 1 - 1} holds for $k = 0$.

\emph{Inductive step:}
Assume~\eqref{eqninpf:thm:Stability of OIT-CZ SMF - Well-Posedness - Step 1 - 1} holds for any $k = l \in \mathbb{N}_0$.
For $k = l+1$, with~\eqref{eqn:Observability Decomposition}
we have
\begin{multline}\label{eqninpf:thm:Stability of OIT-CZ SMF - Well-Posedness - Step 1 - Inductive Step - 1}
\llbracket\tilde{\mathbf{x}}_{l+1}|y_{0:l+1}\rrbracket \subseteq (\tilde{A}_o \llbracket\tilde{\mathbf{x}}_l^o|y_{0:l}\rrbracket \oplus \tilde{B}_o \llbracket\mathbf{w}_l\rrbracket) \\ \times
(\tilde{A}_{\bar{o}} \mathcal{T}_l^{\bar{o}} \oplus \tilde{A}_{21} \llbracket\tilde{\mathbf{x}}_l^o|y_{0:l}\rrbracket \oplus \tilde{B}_{\bar{o}} \llbracket\mathbf{w}_l\rrbracket).
\end{multline}
Noticing that $\mathcal{T}_{l+1}^{\bar{o}} = \tilde{A}_{\bar{o}} \mathcal{T}_l^{\bar{o}} \oplus \tilde{A}_{21} \llbracket\tilde{\mathbf{x}}_l^o|y_{0:l}\rrbracket \oplus \tilde{B}_{\bar{o}} \llbracket\mathbf{w}_l\rrbracket$ and
\begin{equation}\label{eqninpf:thm:Stability of OIT-CZ SMF - Well-Posedness - Step 1 - Inductive Step - 2}
\llbracket\tilde{\mathbf{x}}_{l+1}|y_{0:l+1}\rrbracket \subseteq \llbracket\tilde{\mathbf{x}}_{l+1}^o|y_{0:l+1}\rrbracket \times \llbracket\tilde{\mathbf{x}}_{l+1}^{\bar{o}}|y_{0:l+1}\rrbracket,
\end{equation}
we have [as $(\mathcal{T}_1 \times \mathcal{T}_2) \bigcap (\mathcal{T}_3 \times \mathcal{T}_4) = (\mathcal{T}_1 \bigcap \mathcal{T}_3) \times (\mathcal{T}_2 \bigcap \mathcal{T}_4)$]
\begin{multline}\label{eqninpf:thm:Stability of OIT-CZ SMF - Well-Posedness - Step 1 - Inductive Step - 3}
\llbracket\tilde{\mathbf{x}}_{l+1}|y_{0:l+1}\rrbracket \subseteq \Big[(\tilde{A}_o \llbracket\tilde{\mathbf{x}}_l^o|y_{0:l}\rrbracket \oplus \tilde{B}_o \llbracket\mathbf{w}_l\rrbracket) \\ \bigcap \llbracket\tilde{\mathbf{x}}_{l+1}^o|y_{0:l+1}\rrbracket \Big] \times
\big(\mathcal{T}_{l+1}^{\bar{o}} \bigcap \llbracket\tilde{\mathbf{x}}_{l+1}^{\bar{o}}|y_{0:l+1}\rrbracket \big).
\end{multline}
It implies $\llbracket\tilde{\mathbf{x}}_{l+1}|y_{0:l+1}\rrbracket \subseteq \llbracket\tilde{\mathbf{x}}_{l+1}^o|y_{0:l+1}\rrbracket \times \mathcal{T}_{l+1}^{\bar{o}}$, i.e.,~\eqref{eqninpf:thm:Stability of OIT-CZ SMF - Well-Posedness - Step 1 - 1} holds for $k = l+1$.
Thus,~\eqref{eqninpf:thm:Stability of OIT-CZ SMF - Well-Posedness - Step 1 - 1} is proven by induction.

\textit{Step~2:}
We split the RHS of~\eqref{eqninpf:thm:Stability of OIT-CZ SMF - Well-Posedness - Step 1 - 2} as $\mathcal{T}_k^{\bar{o}} = \mathcal{T}_{k,1}^{\bar{o}} + \mathcal{T}_{k,2}^{\bar{o}}$:
\begin{align}
%\mathcal{T}_k^{\bar{o}} &= \mathcal{T}_{k,1}^{\bar{o}} + \mathcal{T}_{k,2}^{\bar{o}}\label{eqninpf:thm:Stability of OIT-CZ SMF - Well-Posedness - Step 2 - Tk},\\
\!\!\!\!\!\mathcal{T}_{k,1}^{\bar{o}} \!&=\! \tilde{A}_{\bar{o}}^k \llbracket\tilde{\mathbf{x}}_0^{\bar{o}}\rrbracket \!\oplus \!\!\sum_{i=0}^{\bar{\delta}-1} \!\tilde{A}_{\bar{o}}^{k-1-i} (\tilde{A}_{21} \llbracket\tilde{\mathbf{x}}_i^o|y_{0:i}\rrbracket \!\oplus\! \tilde{B}_{\bar{o}} \llbracket\mathbf{w}_i\rrbracket),\label{eqninpf:thm:Stability of OIT-CZ SMF - Well-Posedness - Step 2 - Tk1}\\
\!\!\!\!\!\mathcal{T}_{k,2}^{\bar{o}} \!&=\! \sum_{i=\bar{\delta}}^{k-1} \tilde{A}_{\bar{o}}^{k-1-i} (\tilde{A}_{21} \llbracket\tilde{\mathbf{x}}_i^o|y_{0:i}\rrbracket \oplus \tilde{B}_{\bar{o}} \llbracket\mathbf{w}_i\rrbracket).\label{eqninpf:thm:Stability of OIT-CZ SMF - Well-Posedness - Step 2 - Tk2}
\end{align}

Then, we also spilt $\mathcal{B}_{\alpha_k}^{\infty}[\hat{c}_k^{\bar{o}}]$ in Line~\ref{line:OIT-CZ SMF - Tobar - Case 2} into two parts:
\begin{equation}\label{eqninpf:thm:Stability of OIT-CZ SMF - Well-Posedness - Step 2 - Bk}
\hat{\mathcal{T}}_k^{\bar{o}} = \mathcal{B}_{\alpha_k}^{\infty}[\hat{c}_k^{\bar{o}}] = \mathcal{B}_{\alpha_{k,1}}^{\infty}[\hat{c}_{k,1}^{\bar{o}}] \oplus \mathcal{B}_{\alpha_{k,2}}^{\infty}[\hat{c}_{k,2}^{\bar{o}}],
\end{equation}
where $\alpha_{k,1} = \frac{1}{2} \|A_{\bar{o}}^{k-\bar{\delta}}\|_{\infty} d_{\infty} (\llbracket\hat{\tilde{\mathbf{x}}}_{\bar{\delta}}^{\bar{o}}|y_{0:\bar{\delta}}\rrbracket) + \varepsilon$, $\alpha_{k,2} = \Upsilon_{\infty} \ell_{k-1}$, $\hat{c}_{k,1}^{\bar{o}} = \tilde{A}_{\bar{o}}^{k-\bar{\delta}} \hat{c}_{\bar{\delta}}^{\bar{o}}$, and $\hat{c}_{k,2}^{\bar{o}} = \sum_{i=\bar{\delta}}^{k-1} \tilde{A}_{\bar{o}}^{k-1-i} \hat{c}_i^{\mathrm{in}}$ such that $\alpha_k = \alpha_{k,1} + \alpha_{k,2}$ and $\hat{c}_k^{\bar{o}} = \hat{c}_{k,1}^{\bar{o}} + \hat{c}_{k,2}^{\bar{o}}$.
Next, we prove
\begin{equation}\label{eqninpf:thm:Stability of OIT-CZ SMF - Well-Posedness - Step 2 - Tk1 subseteq Bk1}
\mathcal{T}_{k,1}^{\bar{o}} \subseteq \mathcal{B}_{\alpha_{k,1}}^{\infty}[\hat{c}_{k,1}^{\bar{o}}],\quad k \geq \underline{k}',
\end{equation}
for sufficiently large $\underline{k}' \geq \bar{\delta}$, and
\begin{equation}\label{eqninpf:thm:Stability of OIT-CZ SMF - Well-Posedness - Step 2 - Tk2 subseteq Bk2}
\mathcal{T}_{k,2}^{\bar{o}} \subseteq \mathcal{B}_{\alpha_{k,2}}^{\infty}[\hat{c}_{k,2}^{\bar{o}}],\quad k \geq \bar{\delta},
\end{equation}
such that
\begin{equation}\label{eqninpf:thm:Stability of OIT-CZ SMF - Well-Posedness - Step 2 - Tk subseteq Bk}
\mathcal{T}_k^{\bar{o}} = \mathcal{T}_{k,1}^{\bar{o}} \oplus \mathcal{T}_{k,2}^{\bar{o}} \subseteq \mathcal{B}_{\alpha_k}^{\infty}[\hat{c}_k^{\bar{o}}] = \hat{\mathcal{T}}_k^{\bar{o}},\quad k \geq \underline{k}'.
\end{equation}

For~\eqref{eqninpf:thm:Stability of OIT-CZ SMF - Well-Posedness - Step 2 - Tk1 subseteq Bk1}, $\forall t_k \in \mathcal{T}_{k,1}^{\bar{o}}$, its distance from the center of $\mathcal{B}_{\alpha_{k,1}}^{\infty}[\hat{c}_{k,1}^{\bar{o}}]$ is upper bounded by
\begin{equation}\label{eqninpf:thm:Stability of OIT-CZ SMF - Well-Posedness - Step 2 - Distance Upper Bound}
\|t_k - \hat{c}_{k,1}^{\bar{o}}\|_{\infty} \leq \|t_k\|_{\infty} + \|\tilde{A}_{\bar{o}}^{k-\bar{\delta}} \hat{c}_{\bar{\delta}}^{\bar{o}}\|_{\infty},\quad k \geq \bar{\delta}.
\end{equation}
Since~\eqref{eqninpf:thm:Stability of OIT-CZ SMF - Well-Posedness - Step 2 - Tk1} implies $\lim_{k\to\infty} \|t_k\|_{\infty} = 0$, we can further derive that for $\varepsilon > 0$ given in Line~\ref{line:OIT-CZ SMF - Initialization}, $\exists \underline{k}' \geq \bar{\delta}$ such that
\begin{equation}\label{eqninpf:thm:Stability of OIT-CZ SMF - Well-Posedness - Step 2 - Distance Upper Bound 2}
\|t_k - \hat{c}_{k,1}^{\bar{o}}\|_{\infty} \leq \varepsilon \leq \alpha_{k,1},\quad k \geq \underline{k}',
\end{equation}
which means $t_k \in \mathcal{B}_{\alpha_{k,1}}^{\infty}[\hat{c}_{k,1}^{\bar{o}}]$, i.e.,~\eqref{eqninpf:thm:Stability of OIT-CZ SMF - Well-Posedness - Step 2 - Tk1 subseteq Bk1} holds.

For~\eqref{eqninpf:thm:Stability of OIT-CZ SMF - Well-Posedness - Step 2 - Tk2 subseteq Bk2}, as~\eqref{eqninpf:thm:Stability of OIT-CZ SMF - Well-Posedness - Observable Subsystem Inclusion Property} holds, $\mathcal{T}_{k,2}^{\bar{o}}$ in~\eqref{eqninpf:thm:Stability of OIT-CZ SMF - Well-Posedness - Step 2 - Tk2} is bounded by
\begin{equation}\label{eqninpf:thm:Stability of OIT-CZ SMF - Well-Posedness - Step 2 - Tk2 - Outer Bound}
\begin{split}
\mathcal{T}_{k,2}^{\bar{o}} &\subseteq \sum_{i=\bar{\delta}}^{k-1} \tilde{A}_{\bar{o}}^{k-1-i} (\tilde{A}_{21} \widetilde{\mathcal{Z}}_i^o \oplus \tilde{B}_{\bar{o}} \llbracket\mathbf{w}_i\rrbracket)\\
&\subseteq \sum_{i=\bar{\delta}}^{k-1} \tilde{A}_{\bar{o}}^{k-1-i} \overline{\mathrm{IH}}(\tilde{A}_{21} \widetilde{\mathcal{Z}}_i^o \oplus \tilde{B}_{\bar{o}} \llbracket\mathbf{w}_i\rrbracket)\\
&\stackrel{(n)}{\subseteq} \sum_{i=\bar{\delta}}^{k-1} \tilde{A}_{\bar{o}}^{k-1-i} \mathcal{B}_{\ell_i}^{\infty}[\hat{c}_i^{\mathrm{in}}] \stackrel{(o)}{\subseteq} \mathcal{B}_{\alpha_{k,2}}^{\infty}[\hat{c}_{k,2}^{\bar{o}}],
\end{split}
\end{equation}
where $(n)$ follows from Lines~\ref{line:OIT-CZ SMF - Interval Hull of Unobservable Subsystem Input}-\ref{line:OIT-CZ SMF - Component Max} and the fact that $2 \|\hat{G}_i^{\mathrm{in}}\|_{\infty}$ gives the maximum edge length of the interval hull.
Recall that $\alpha_{k,2} = \Upsilon_{\infty} \ell_{k-1}$ and $\hat{c}_{k,2}^{\bar{o}} = \sum_{i=\bar{\delta}}^{k-1} \tilde{A}_{\bar{o}}^{k-1-i} \hat{c}_i^{\mathrm{in}}$, and
observe that
\begin{align*}%\label{eqninpf:thm:Stability of OIT-CZ SMF - Well-Posedness - Step 2 - Bound on Sum}
&\sum_{i=\bar{\delta}}^{k-1} \tilde{A}_{\bar{o}}^{k-1-i} \mathcal{B}_{\ell_i}^{\infty}[\hat{c}_i^{\mathrm{in}}] \subseteq \sum_{i=\bar{\delta}}^{k-1} \mathcal{B}_{\|\tilde{A}_{\bar{o}}^{k-1-i}\|_{\infty}\ell_i}^{\infty}[\hat{c}_i^{\mathrm{in}}],\\
&\sum_{i=\bar{\delta}}^{k-1} \|\tilde{A}_{\bar{o}}^{k-1-i}\|_{\infty} \stackrel{\mathrm{\lemref{lem:Bound on Matrix Power Norm}}}{\leq} \!\!\inf_{\gamma \in (\rho(\tilde{A}_{\bar{o}}), 1)}\!\! \frac{\beta_{\gamma}(\tilde{A}_{\bar{o}})}{1-\gamma} = \Upsilon_{\infty}.
\end{align*}
We can further derive $(o)$ in~\eqref{eqninpf:thm:Stability of OIT-CZ SMF - Well-Posedness - Step 2 - Tk2 - Outer Bound}, i.e.,~\eqref{eqninpf:thm:Stability of OIT-CZ SMF - Well-Posedness - Step 2 - Tk2 subseteq Bk2} holds.

Thus,~\eqref{eqninpf:thm:Stability of OIT-CZ SMF - Well-Posedness - Step 2 - Tk1 subseteq Bk1} and~\eqref{eqninpf:thm:Stability of OIT-CZ SMF - Well-Posedness - Step 2 - Tk2 subseteq Bk2} give~\eqref{eqninpf:thm:Stability of OIT-CZ SMF - Well-Posedness - Step 2 - Tk subseteq Bk}.

With~\eqref{eqninpf:thm:Stability of OIT-CZ SMF - Well-Posedness - Inclusion},~\eqref{eqninpf:thm:Stability of OIT-CZ SMF - Well-Posedness - Step 1 - 1}, and~\eqref{eqninpf:thm:Stability of OIT-CZ SMF - Well-Posedness - Step 2 - Tk subseteq Bk} we get $\llbracket\mathbf{x}_k| y_{0:k}\rrbracket \subseteq P^{-1} \big(\hat{\mathcal{T}}_k^o \times \hat{\mathcal{T}}_k^{\bar{o}}\big)$ for $k \geq \underline{k}'$.
Setting $\underline{k} = \underline{k}' + \bar{\delta}$, we can obtain for $k \geq \underline{k} \geq 2\bar{\delta}$,
\begin{equation}\label{eqninpf:thm:Stability of OIT-CZ SMF - Well-Posedness - Step 2 - An Outer Bound on Actual Range}
\llbracket\mathbf{x}_{k-\bar{\delta}}| y_{0:k-\bar{\delta}}\rrbracket \subseteq P^{-1} \big(\hat{\mathcal{T}}_{k-\bar{\delta}}^o \times \hat{\mathcal{T}}_{k-\bar{\delta}}^{\bar{o}}\big).
\end{equation}
Therefore, we have
\begin{multline}\label{eqninpf:thm:Stability of OIT-CZ SMF - Finite-Time Inclusion}
	\llbracket\mathbf{x}_k| y_{0:k}\rrbracket \stackrel{\eqref{eqn:Filtering Map}}{=} F_{k,k-\bar{\delta}}\big(\llbracket\mathbf{x}_{k-\bar{\delta}}| y_{0:k-\bar{\delta}}\rrbracket\big)\\
	\stackrel{\eqref{eqninpf:thm:Stability of OIT-CZ SMF - Well-Posedness - Step 2 - An Outer Bound on Actual Range}}{\subseteq} F_{k,k-\bar{\delta}}\big( P^{-1} \big(\hat{\mathcal{T}}_{k-\bar{\delta}}^o \times \hat{\mathcal{T}}_{k-\bar{\delta}}^{\bar{o}}\big)\big) \stackrel{\mathrm{Line}~\ref{line:OIT-CZ SMF - Estimate Inspired by OIT}}{=} \mathcal{Z}_k
\end{multline}
for $k \geq \underline{k}$, i.e.,~\eqref{eqn:Stability of OIT-CZ SMF - Finite-Time Inclusion} holds.

\emph{\underline{Uniformly bounded estimate:}}
From Algorithm~\ref{alg:OIT-CZ SMF}, the uniform boundedness of $\mathcal{Z}_k$ only depends on $k \geq \bar{\delta}$.
By Line~\ref{line:OIT-CZ SMF - Estimate Inspired by OIT}, $\mathcal{Z}_k^o = P_o F_{k,k-\bar{\delta}}\big( P^{-1} \big(\hat{\mathcal{T}}_{k-\bar{\delta}}^o \times \hat{\mathcal{T}}_{k-\bar{\delta}}^{\bar{o}}\big)\big)$ holds for $k \geq \bar{\delta}$, which combined with \propref{prop:Stability of Observable Systems} and~\eqref{eqninpf:thm:Stability of OIT-CZ SMF - Well-Posedness - Inclusion} guarantees the uniform boundedness of $\mathcal{Z}_k^o$ w.r.t. $k \geq \bar{\delta}$.
Then, similarly to \corref{cor:Egregium}, we can get the uniform boundedness of $\mathcal{Z}_k$ w.r.t. $k \in \mathbb{N}_0$.\hfill$\square$

\end{document}